\documentclass[12pt]{amsart}

\usepackage[all]{xy}
\usepackage{amsmath}
\usepackage{amsfonts}
\usepackage{amssymb}
\usepackage{amscd}
\usepackage{amsthm}
\usepackage{latexsym}
\usepackage{amsbsy}
\usepackage{graphicx}
\usepackage{epstopdf}

\input xypic

\usepackage{color}

\newtheorem{thm}{Theorem}[section]
\newtheorem{prop}[thm]{Proposition}

\newtheorem{lem}[thm]{Lemma}

\newtheorem{rem}[thm]{Remark}

\numberwithin{equation}{section}

\def\bG{{\mathbb G}}

\def\bL{{\mathbb L}}

\def\A{{\mathbb A}}
\def\C{{\mathbb C}}

\renewcommand{\H}{{\mathbb H}}

\renewcommand{\P}{{\mathbb P}}
\def\Q{{\mathbb Q}}
\def\R{{\mathbb R}}
\def\Z{{\mathbb Z}}
\def\K{{\mathbb K}}

\def\m{{\mathfrak m}}

\def\cutint{{\int \!\!\!\!\!\! -}}

\def\cD{{\mathcal D}}

\def\cM{{\mathcal M}}

\def\cP{{\mathcal P}}
\def\cQ{{\mathcal Q}}
\def\cR{{\mathcal R}}
\def\cS{{\mathcal S}}
\def\cT{{\mathcal T}}

\def\cV{{\mathcal V}}

\def\cX{{\mathcal X}}

\def\cZ{{\mathcal Z}}

\def\Spec{{\rm Spec}}

\def\Tr{{\rm Tr}}

\title[Periods and Motives in Robertson--Walker spacetimes]{Periods and motives in the spectral action of 
Robertson--Walker spacetimes}
\author{Farzad Fathizadeh and Matilde Marcolli}
\address{Division of Physics, Mathematics, and Astronomy, California Institute of Technology, 1200 E California Blvd, Pasadena, CA 91125, USA}
\email{farzadf@caltech.edu}
\email{matilde@caltech.edu}

\begin{document}
\maketitle

\begin{abstract}
We show that, when considering the scaling factor as an affine variable, the coefficients of the 
asymptotic expansion of the spectral action on a (Euclidean) Robertson-Walker spacetime 
are periods of mixed Tate motives, involving relative motives of complements of unions of
hyperplanes and quadric hypersurfaces 
and divisors given by unions of coordinate hyperplanes.  
\end{abstract}

\tableofcontents

\section{Introduction}

Over the past decade, Grothendieck's theory of motives has come to 
play an increasingly important role in theoretical physics. While the existence of
a relation between motives and periods of algebraic varieties and computations in 
high-energy physics might have seemed surprising and unexpected, the existence of underlying
motivic structures in quantum field theory has now been widely established, see for
instance \cite{BEK}, \cite{BrSch}, \cite{CoMa}, \cite{Mar}. Typically, periods and motives occur in
quantum field theory in the perturbative approach, through the asymptotic expansion in Feynman 
diagrams, where in the terms of the asymptotic expansion the renormalized Feynman integrals 
are identified with periods of certain hypersurface complements. 
The nature of the motive of the hypersurface constraints the
class of numbers that can occur as periods.  Similarly, a large body of
recent work on amplitudes in $N=4$ Supersymmetric Yang--Mills has uncovered 
another setting where the connection to periods and motives plays an important role,
see \cite{Ampl}, \cite{Gon1}, \cite{Gon2}. 

\smallskip

In this paper, we present another surprising instance of the occurrences of periods and
motives in theoretical physics, this time in a model of (modified) gravity based on the spectral action functional
of \cite{CCact}. The situation is somewhat similar to the one seen in the quantum field theory
setting, with some important differences. As in the QFT framework, we deal with an asymptotic
expansion, which in our case is given by the large energy expansion of the spectral action
functional. We show in this paper that, in the case of (Euclidean) Robertson--Walker spacetimes,
the terms of the asymptotic expansion of the spectral action functional can be expressed as
periods of mixed Tate motives, given by complements of quadric hypersurfaces. An important
difference, with respect to the case of a scalar massless quantum field theory of \cite{BEK}, is
that here we need to consider only one quadric hypersurface for each term of the expansion,
whereas in the quantum field theory case one has to deal with the much more complicated
motive of a union of quadric hypersurfaces, associated to the edges of the Feynman graph. 
On the other hand, the algebraic differential form that is integrated on a semi-algebraic set
in the hypersurface complement is much more complicated in the spectral action case considered
here, than in the quantum field theory case: the terms in the algebraic differential form
arise from the computation, via pseudo-differential calculus, of a parametrix for the
square of the Dirac operator on the Robertson--Walker spacetime, after a suitable change
of variables in the integral. While the explicit expression of the differential form, even for
the simplest cases of the coefficients $a_2$ and $a_4$ can take up several pages, the
structure of the terms can be understood, as we explain in the following sections,
and the domain of definition is, in the case of the $a_{2n}$ term, the complement of 
a union of two hyperplanes and a 
quadric hypersurfaces defined by a family of quadrics $Q_{\alpha, 2n}$ in an affine 
space $\A^{2n+3}$. 

\smallskip

In Section~\ref{RWsec} we describe our choice of coordinates, and the resulting form of 
the pseudodifferential symbol of the square of the Dirac operator on a (Euclidean) 
Robertson-Walker metric. In Section~\ref{WodSec}, we describe briefly how the Seeley-DeWitt
coefficients of the heat kernel expansion can be computed in terms of Wodzicki residues,
by taking products with auxiliary tori with flat metrics. Section~\ref{a2Sec} gives the explicit
computation of the $a_2$ term, showing that, before integrating in the time variable, and
treating the scaling factor as an affine parameter $\alpha \in \A^1\smallsetminus\{ 0 \}=\bG_m$,
one can write the resulting integral as a period obtained by integrating an algebraic (over $\Q$) 
differential form over a $\Q$-semi-algebraic set. The differential form is 
defined on the complement in $\A^5$ of a union of two hyperplanes and the quadric determined
by the vanishing of the quadratic form $Q_{\alpha, 2}=u_1^2 +\alpha^{-2} (u_2^2 + u_3^2 + u_4^2)$.
The $\Q$-semialgebraic set in this hypersurface complement has boundary
contained in a divisor given by a union of coordinate hyperplanes. Although the boundary
divisor and the hypersurface intersect nontrivially, all the integrals are convergent and
we do not have a renormalization problem, unlike what happens in the quantum field theory setting.
In Section~\ref{a4Sec} we
prove an analogous result for the $a_4$ term, with the very lengthy full expression of the
$a_4$ term given in the appendix. In Section~\ref{a2nSec}, using an inductive argument
and the result of the previous cases, we prove that the terms $a_{2n}$ can all be identified
(prior to time-integration) with periods of motives of complements of quadric hypersurfaces
obtained from a family of quadrics
$$ Q_{\alpha, 2n} = u_1^2 + \alpha^{-2} (u_2^2+u_3^2+u_4^2) + u_5^2 + \cdots + u_{2n+2}^2.  $$
The algebraic differential forms depend on $2n$ auxiliary affine parameters $\alpha_1,\ldots,\alpha_{2n}$,
which correspond to the time derivatives of the scaling factor of the Robertson--Walker metric. 
In Section \ref{MotSec} we analyze more explicitly the motive, showing that, over a quadratic field
extension $\Q(\sqrt{-1})$ where the quadrics become isotropic, it is a mixed Tate motive, while over $\Q$ it is 
a form of a Tate motive in the sense of \cite{Rost}, \cite{Vishik1}, \cite{Vishik2}. We compute explicitly,
by a simple inductive argument, the class in the Grothendieck ring of the relevant hypersurface 
complement.

\smallskip

\subsection{The spectral model of gravity}

The {\em spectral action} functional, introduced in \cite{CCact} is a regularized trace of the
Dirac operator $D$ given by
$$ \cS(\Lambda) = \Tr(f(D/\Lambda))= \sum_{\lambda \in \text{Spec}(D)}\text{Mult}(\lambda)f(\lambda/\Lambda),$$
where the test function $f$ is a smooth even rapidly decaying function, which should be thought of as a 
smooth approximation to a cutoff function. The parameter $\Lambda>0$ is an energy scale. One
of the main advantages of this action functional is that it is not only defined for smooth compact
Riemannian spin manifolds, but also for a more general class of geometric objects that include
the noncommutative analogs of Riemannian manifolds, finitely summable spectral triples, see \cite{CoS3}.
In particular, the spectral action functional applied to almost commutative geometries (products of 
manifolds and finite noncommutative spaces) is used as a method to generate particle physics
models with varying possible matter sectors depending on the finite geometry and with
matter coupled to gravity, see \cite{WvS} for a recent overview. It was shown in \cite{CCact} that,
in the case of commutative and almost commutative geometries, the spectral action functional
has an asymptotic expansion for large energy $\Lambda$, 
$$ \Tr (f(D/\Lambda))\sim\,\sum_{\beta \in \Sigma^+_{ST}}\,f_\beta\,\Lambda^\beta \,\,  
 \cutint |D|^{-\beta} \,\, +\,f(0)\,\zeta_{D}(0) + \cdots, $$
where the coefficients depend
on momenta $f_\beta=\,\int_{0}^{\infty}f(v)\,v^{\beta-1}\,dv$ and Taylor coefficients 
of the test function $f$ and on residues 
$$  \cutint |D|^{-\beta} =  \frac{1}{2} {\rm Res}_{s=\beta}  \,\, \zeta_D(s) $$
at poles of the zeta function $\zeta_D(s)$ of the
Dirac operator. The leading terms of the asymptotic expansion recover the usual local terms
of an action functional for gravity, the Einstein-Hilbert action with cosmological term, with
additional modified gravity terms given by Weyl conformal gravity and Gauss-Bonnet gravity.
In the case of an almost commutative geometry the leading terms of the asymptotic
expansion also determine the Lagrangian of the resulting particle physics model. The spectral
action on ordinary manifold, as an action functional of modified gravity, was applied to cosmological 
models, see \cite{Mar2} for an overview. In the manifold case, the Mellin transform relation
between zeta function and trace of the heat kernel expresses the coefficients of the spectral
action expansion in terms of the Seeley-DeWitt coefficients $a_{2n}$ of the heat kernel expansion,
$$ \Tr(e^{-t D^2}) \sim_{t\to 0+}\,\,\, t^{-m/2} \sum_{n=0}^\infty a_{2n}(D^2) \,t^n . $$
Pseudodifferential calculus techniques and the parametrix method can then be applied to the 
computation of the symbol and the Seeley-DeWitt coefficients. The resulting computations
can easily become intractable, but a computationally more efficient method introduced in \cite{FFMRationality},
based on Wodzicki residues and products by auxiliary flat tori can be applied to make the problem 
more easily tractable. 

\smallskip

In the case of the (Euclidean) Robertson-Walker spacetimes, it was conjectured in \cite{CC}
and proved in \cite{FGK} that all the terms in the expansion of the spectral action are
polynomials with rational coefficients in the scaling factor and its derivatives.
This rationality result suggests the existence of an underlying arithmetic structure.
In the case of the Bianchi IX metrics, a similar rationality result was proved in \cite{FFMRationality}
and the underlying arithmetic structure was analyzed in \cite{FFM2} for the Bianchi IX gravitational
instantons, in terms of modular forms. Here we consider the case of the Robertson-Walker spacetimes
and we look for arithmetic structures in the expansion of the spectral action in terms of 
periods and motives. A similar motivic analysis of the Bianchi IX case will be carried out in forthcoming work.

\section{Robertson-Walker metric and the Dirac operator}\label{RWsec}

We consider the Robertson-Walker metric with the expansion factor $a(t)$, 
\[
ds^2 = dt^2 + a(t)^2 d\sigma^2, 
\]
where $d\sigma^2$ is the round metric on the 3-dimensional sphere $\mathbb{S}^3$. Using the 
Hopf coordinates for $\mathbb{S}^3$, we use the local chart
\[
x = (t, \eta, \phi_1, \phi_2) 
\mapsto
(t, \sin \eta \cos \phi_1, \sin \eta \sin \phi_2, \cos \eta \cos \phi_1, \cos \eta \sin \phi_2 ),  
\]
\[
0 < \eta < \frac{\pi}{2}, 
\qquad 
0 < \phi_1 < 2 \pi, 
\qquad 
0 < \phi_2 < 2 \pi.   
\]
In this coordinate system, the Robertson-Walker metric is written as 
\[
ds^2 = dt^2 + a(t)^2 \left ( d \eta^2 + \sin^2 (\eta)  \, d\phi_1^2 + \cos^2 (\eta) \, d \phi_2^2 \right ),  
\]
or alternatively we write:
\[
(g_{\mu \nu}) = 
\left(
\begin{array}{cccc}
 1 & 0 & 0 & 0 \\
 0 & a(t)^2 & 0 & 0 \\
 0 & 0 & a(t)^2 \sin ^2(\eta ) & 0 \\
 0 & 0 & 0 & a(t)^2 \cos ^2(\eta ) \\
\end{array}
\right), 
\]
with 
\[
(g^{\mu \nu}) = (g_{\mu \nu})^{-1}= 
\left(
\begin{array}{cccc}
 1 & 0 & 0 & 0 \\
 0 & \frac{1}{a(t)^2} & 0 & 0 \\
 0 & 0 & \frac{\csc ^2(\eta )}{a(t)^2} & 0 \\
 0 & 0 & 0 & \frac{\sec ^2(\eta )}{a(t)^2} \\
\end{array}
\right). 
\]

\smallskip
\subsection{Pseudodifferential symbol}
One can write the local expression for the Dirac operator $D$, as in \S 2 of \cite{FGK},
and one finds that the pseudodifferential symbol $\sigma_{D}(x, \xi)$ of  D is 
given by the sum $q_1(x, \xi)+q_0(x, \xi)$, where, using the notation 
$\xi = (\xi_1, \xi_2, \xi_3, \xi_4) \in \mathbb{R}^4$ for an element of the cotangent fibre 
$T_x^* M \simeq \mathbb{R}^4$ at the point $x = (t, \eta, \phi_1, \phi_2)$, we have 
\begin{eqnarray}
 q_1(x, \xi) &=& 
\left(
\begin{array}{cccc}
 0 & 0 & \frac{i \sec (\eta ) \xi _4}{a(t)}-\xi _1 & \frac{i \xi _2}{a(t)}+\frac{\csc (\eta ) \xi _3}{a(t)} \\
 0 & 0 & \frac{i \xi _2}{a(t)}-\frac{\csc (\eta ) \xi _3}{a(t)} & -\xi _1-\frac{i \sec (\eta ) \xi _4}{a(t)} \\
 -\xi _1-\frac{i \sec (\eta ) \xi _4}{a(t)} & -\frac{i \xi _2}{a(t)}-\frac{\csc (\eta ) \xi _3}{a(t)} & 0 & 0 \\
 \frac{\csc (\eta ) \xi _3}{a(t)}-\frac{i \xi _2}{a(t)} & \frac{i \sec (\eta ) \xi _4}{a(t)}-\xi _1 & 0 & 0 \\
\end{array}
\right), \nonumber \\
&& \nonumber \\
&& \nonumber \\
q_0 (\xi) &=& 
\left(
\begin{array}{cccc}
 0 & 0 & \frac{3 i a'(t)}{2 a(t)} & \frac{\cot (\eta )-\tan (\eta )}{2 a(t)} \\
 0 & 0 & \frac{\cot (\eta )-\tan (\eta )}{2 a(t)} & \frac{3 i a'(t)}{2 a(t)} \\
 \frac{3 i a'(t)}{2 a(t)} & \frac{\tan (\eta )-\cot (\eta )}{2 a(t)} & 0 & 0 \\
 \frac{\tan (\eta )-\cot (\eta )}{2 a(t)} & \frac{3 i a'(t)}{2 a(t)} & 0 & 0 \\
\end{array}
\right). \nonumber \\
\end{eqnarray}
These matrices can be used to find the pseudodifferential symbol of the 
square of the Dirac operator: 
\[
\sigma_{D^2}(x, \xi) = p_2(x, \xi) + p_1(x, \xi) + p_0(x, \xi), 
\]
where, denoting the $4 \times 4$ identity matrix by $I_{4 \times 4}$, 
we have:  
\[
p_2(x, \xi) 
= 
q_1(x, \xi) \, q_1(x, \xi) 
= 
\left( \sum g^{\mu \nu} \xi_\mu \xi_\nu \right) I_{4 \times 4} 
\]
\[
= \left (   \xi _1^2+\frac{\xi _2^2}{a(t)^2}+\frac{\csc ^2(\eta ) \xi _3^2}{a(t)^2}+\frac{\sec ^2(\eta ) \xi _4^2}{a(t)^2} \right ) I_{4 \times 4}, 
\]
\[
p_1(x, \xi)
= 
q_0(x, \xi) \, q_1(x, \xi)+ q_1(x, \xi) \, q_0(x, \xi) + 
\sum _{j=1}^4 -i \frac{\partial q_1}{\partial \xi _j}(x, \xi) \, \frac{\partial q_1}{\partial x_j} (x, \xi), 
\]
\[
p_0(x, \xi) = 
q_0(x, \xi) \, q_0(x, \xi) +\sum _{j=1}^4 -i \frac{\partial q_1}{\partial \xi _j}(x, \xi) \, 
\frac{\partial q_0}{\partial x_j}(x, \xi). 
\]

\section{Heat expansion and the Wodzicki residue}\label{WodSec}

It is in general computationally difficult to obtain explicit expressions for the
Seeley--DeWitt coefficients of the heat kernel expansions, even for nicely
homogeneous and isotropic metrics like the Friedmann--Robertson--Walker case.
A computationally more efficient method was introduced in
\cite{FFMRationality}, based on products with auxiliary flat tori and Wodzicki residues.
We apply it here to calculate  
the coefficients $a_{2n}$ that appear in the small time heat kernel expansion 
\begin{equation} \label{HeatExp}
\text{Trace}(e^{-sD^2})\,\, \sim_{s \to 0^+}\,\, a_0 s^{-2} + a_2 s^{-1} + a_4 + a_6 s + a_8s^2+ \cdots .  
\end{equation}
In fact, it is proved in \cite{FFMRationality} that, for any non-negative even integer $r$, 
we have 
\begin{equation} \label{HeatCoefResidue}
a_{2+r} = \frac{1}{2^5\, \pi^{4+r/2}} \textnormal{Res}(\Delta^{-1}),
\end{equation}
where 
\[
\Delta = D^2 \otimes 1 + 1 \otimes \Delta_{\mathbb{T}^r},
\]
in which $\Delta_{\mathbb{T}^r}$ is the flat Laplacian on 
the $r$-dimensional torus $\mathbb{T}^r = \left ( \mathbb{R}/\mathbb{Z} \right )^r$. 
Here, the linear functional $\text{Res}$ defined on the algebra of classical 
pseudodifferential operators is the Wodzicki residue, which is defined as follows. Assume that 
the dimension of the manifold is $m$, and that the symbol of a classical pseudodifferential 
operator is given in a local chart $U$ by 
\[
\sigma (x, \xi) 
\sim 
\sum_{j=0}^\infty \sigma_{d-j} (x, \xi) \qquad (\xi \to \infty),
\]
where each $\sigma_{d-j} : U \times \left ( \mathbb{R}^m \setminus \{ 0\}\right ) \to M_r(\mathbb{C})$ 
is positively homogeneous of order $d-j$ in $\xi$. Then one needs to consider the 1-density 
defined by 
\begin{equation} \label{WodResDensity}
\textnormal{wres}_x P_\sigma 
= 
\left ( \int_{|\xi |=1} 
\textrm{tr} \left (\sigma_{-m}(x, \xi)  \right ) |\sigma_{\xi, \, m-1} | 
\right ) |dx^0 \wedge dx^1 \wedge \cdots \wedge dx^{m-1}|, 
\end{equation}
in which $\sigma_{\xi, \, m-1} $ is the volume form of the unit sphere $|\xi|=1$ in the 
cotangent fibre $\mathbb{R}^m \simeq T_x^* M$:
\[
\sigma_{\xi, \, m-1} = \sum_{j=1}^m (-1)^{j-1} \xi_j \, d\xi_1 
\wedge \cdots \wedge {\widehat d \xi_j} \wedge \cdots \wedge d \xi_m. 
\]
The Wodzicki residue of the pseudodifferential operator $P_\sigma$ associated 
with the symbol $\sigma$ is by definition the integral of the above 1-density 
associated to $\sigma$: 
\begin{equation} \label{WodResFormula}
\text{Res} \left ( P_\sigma \right) = \int_{M} \textnormal{wres}_x P_\sigma. 
\end{equation}
One can find a detailed  discussion of the Wodzicki residue in Chapter 7 of \cite{GVFbook} 
and references therein.

\medskip
\section{The $a_2$ term and quadric surfaces in $\P^3$}\label{a2Sec}

We will refer here to the form of the $a_2$ term before performing the integration in 
the time variable $t$. As we show below, this is the part that is naturally expressible
in terms of periods of motives.

\begin{thm}\label{a2compute}
The form $b_{-4}(x, \xi)$ derived from $\textnormal{tr}(\sigma_{-4}(x, \xi) )$ computing the 
$a_2$ term of the heat kernel expansion of $D^2$ is a rational differential form
$$ \Omega = f \, \widetilde\sigma_3, $$
in affine coordinates $(u_0,u_1,u_2,u_3,u_4)\in \A^5$, $\alpha\in \bG_m$, and $(\alpha_1,\alpha_2)\in \A^2$
where the functions $$ f (u_0,u_1,u_2,u_3,u_4,\alpha, \alpha_1, \alpha_2)= f_{(\alpha_1,\alpha_2)}(u_0,u_1,u_2,u_3,u_4,\alpha) $$ are 
$\Q$-linear combinations of rational functions of the form
$$ \frac{P(u_0,u_1,u_2,u_3,u_4,\alpha, \alpha_1, \alpha_2)}
{\alpha^{2r} u_0^k (1-u_0)^m (u_1^2 + \alpha^{-2} (u_2^2 + u_3^2 + u_4^2))^\ell}, $$
where
$$ P(u_0,u_1,u_2,u_3,u_4,\alpha, \alpha_1, \alpha_2) = P_{(\alpha_1, \alpha_2)}(u_0,u_1,u_2,u_3,u_4,\alpha) $$
are polynomials in $\Q[u_0,u_1,u_2,u_3,u_4,\alpha, \alpha_1, \alpha_2]$ and 
where $r$, $k$, $m$ and $\ell$ are non-negative integers, and with $\widetilde\sigma_3=\widetilde\sigma_3(u_0,u_1,u_2,u_3,u_4)$ the 
algebraic differential form
$$  \frac{1}{2} \Big ( u_1  \, du_0 \, du_2 \, du_3 \, du_4 
- u_2  \, du_0\, du_1 \, du_3 \, du_4 + u_3 \, d u_0 \, du_1 \, du_2 \, du_4 
-  u_4   \, d u_0 \, du_1 \, du_2 \, du_3 \Big ). $$
The forms $\Omega^\alpha=\Omega^\alpha_{(\alpha_1,\alpha_2)}$ obtained by restricting 
the above to a fixed value of 
$\alpha\in \A^1\smallsetminus \{ 0 \}$ are a two parameter family 
of algebraic differential forms on the algebraic variety over $\Q$
given by the complement in $\A^5$ of the union of two affine hyperplanes $H_0=\{ u_0=0 \}$
and $H_1=\{ u_0=1 \}$ and the hypersurface $\widehat{CZ}_\alpha$ defined by 
the vanishing of the quadratic form
\begin{equation}\label{Qalpha}
 Q_{\alpha, 2} = u_1^2 + \alpha^{-2} (u_2^2 + u_3^2 + u_4^2). 
\end{equation} 
\end{thm}

\proof We use the formula \eqref{HeatCoefResidue} in the special case of $r=0$ 
to calculate the term $a_2$ appearing in the heat kernel expansion \eqref{HeatExp}. 
In this case we have 
\[
a_{2} = \frac{1}{2^5\, \pi^{4}} \, \textnormal{Res} \left ( (D^2)^{-1} \right ),
\]
where $(D^2)^{-1}$ denotes the parametrix of $D^2$. In order to use the formula 
\eqref{WodResDensity}, since the dimension of the 
manifold is 4, we need to calculate the term $\sigma_{-4}(x, \xi)$ that is homogeneous 
of order $-4$ in the expansion of the symbol of $(D^2)^{-1}$. By performing symbolic 
calculations we find that: 
\[
\textnormal{tr}(\sigma_{-4}(x, \xi) ) = 
\]
\begin{center}
\begin{math}
\frac{32 \cot ^2(\eta ) \xi _3^4 \csc ^4(\eta )}{a(t)^6 \left(\xi _1^2+\frac{\xi _2^2}{a(t)^2}+\frac{\csc ^2(\eta ) \xi _3^2}{a(t)^2}+\frac{\sec ^2(\eta ) \xi _4^2}{a(t)^2}\right){}^4}+\frac{32 \xi _2^2 \xi _3^2 \csc ^4(\eta )}{a(t)^6 \left(\xi _1^2+\frac{\xi _2^2}{a(t)^2}+\frac{\csc ^2(\eta ) \xi _3^2}{a(t)^2}+\frac{\sec ^2(\eta ) \xi _4^2}{a(t)^2}\right){}^4}+\frac{32 \xi _3^4 a'(t)^2 \csc ^4(\eta )}{a(t)^6 \left(\xi _1^2+\frac{\xi _2^2}{a(t)^2}+\frac{\csc ^2(\eta ) \xi _3^2}{a(t)^2}+\frac{\sec ^2(\eta ) \xi _4^2}{a(t)^2}\right){}^4}-\frac{8 \xi _3^2 \csc ^4(\eta )}{a(t)^4 \left(\xi _1^2+\frac{\xi _2^2}{a(t)^2}+\frac{\csc ^2(\eta ) \xi _3^2}{a(t)^2}+\frac{\sec ^2(\eta ) \xi _4^2}{a(t)^2}\right){}^3}-\frac{192 \xi _1^2 \xi _3^4 a'(t)^2 \csc ^4(\eta )}{a(t)^6 \left(\xi _1^2+\frac{\xi _2^2}{a(t)^2}+\frac{\csc ^2(\eta ) \xi _3^2}{a(t)^2}+\frac{\sec ^2(\eta ) \xi _4^2}{a(t)^2}\right){}^5}-\frac{384 \cot (\eta ) \xi _1 \xi _2 \xi _3^4 a'(t) \csc ^4(\eta )}{a(t)^7 \left(\xi _1^2+\frac{\xi _2^2}{a(t)^2}+\frac{\csc ^2(\eta ) \xi _3^2}{a(t)^2}+\frac{\sec ^2(\eta ) \xi _4^2}{a(t)^2}\right){}^5}-\frac{192 \cot ^2(\eta ) \xi _2^2 \xi _3^4 \csc ^4(\eta )}{a(t)^8 \left(\xi _1^2+\frac{\xi _2^2}{a(t)^2}+\frac{\csc ^2(\eta ) \xi _3^2}{a(t)^2}+\frac{\sec ^2(\eta ) \xi _4^2}{a(t)^2}\right){}^5}-\frac{384 \sec (\eta ) \xi _1 \xi _2 \xi _3^2 \xi _4^2 a'(t) \csc ^3(\eta )}{a(t)^7 \left(\xi _1^2+\frac{\xi _2^2}{a(t)^2}+\frac{\csc ^2(\eta ) \xi _3^2}{a(t)^2}+\frac{\sec ^2(\eta ) \xi _4^2}{a(t)^2}\right){}^5}+\frac{64 \cot ^2(\eta ) \xi _2^2 \xi _3^2 \csc ^2(\eta )}{a(t)^6 \left(\xi _1^2+\frac{\xi _2^2}{a(t)^2}+\frac{\csc ^2(\eta ) \xi _3^2}{a(t)^2}+\frac{\sec ^2(\eta ) \xi _4^2}{a(t)^2}\right){}^4}+\frac{16 \cot (\eta ) \cot (2 \eta ) \xi _3^2 \csc ^2(\eta )}{a(t)^4 \left(\xi _1^2+\frac{\xi _2^2}{a(t)^2}+\frac{\csc ^2(\eta ) \xi _3^2}{a(t)^2}+\frac{\sec ^2(\eta ) \xi _4^2}{a(t)^2}\right){}^3}+\frac{384 \sec ^2(\eta ) \xi _2^2 \xi _3^2 \xi _4^2 \csc ^2(\eta )}{a(t)^8 \left(\xi _1^2+\frac{\xi _2^2}{a(t)^2}+\frac{\csc ^2(\eta ) \xi _3^2}{a(t)^2}+\frac{\sec ^2(\eta ) \xi _4^2}{a(t)^2}\right){}^5}+\frac{64 \xi _2^2 \xi _3^2 a'(t)^2 \csc ^2(\eta )}{a(t)^6 \left(\xi _1^2+\frac{\xi _2^2}{a(t)^2}+\frac{\csc ^2(\eta ) \xi _3^2}{a(t)^2}+\frac{\sec ^2(\eta ) \xi _4^2}{a(t)^2}\right){}^4}+\frac{4 \xi _3^2 a'(t)^2 \csc ^2(\eta )}{a(t)^4 \left(\xi _1^2+\frac{\xi _2^2}{a(t)^2}+\frac{\csc ^2(\eta ) \xi _3^2}{a(t)^2}+\frac{\sec ^2(\eta ) \xi _4^2}{a(t)^2}\right){}^3}+\frac{64 \sec ^2(\eta ) \xi _3^2 \xi _4^2 a'(t)^2 \csc ^2(\eta )}{a(t)^6 \left(\xi _1^2+\frac{\xi _2^2}{a(t)^2}+\frac{\csc ^2(\eta ) \xi _3^2}{a(t)^2}+\frac{\sec ^2(\eta ) \xi _4^2}{a(t)^2}\right){}^4}+\frac{48 \cot (\eta ) \xi _1 \xi _2 \xi _3^2 a'(t) \csc ^2(\eta )}{a(t)^5 \left(\xi _1^2+\frac{\xi _2^2}{a(t)^2}+\frac{\csc ^2(\eta ) \xi _3^2}{a(t)^2}+\frac{\sec ^2(\eta ) \xi _4^2}{a(t)^2}\right){}^4}+\frac{8 \xi _3^2 a''(t) \csc ^2(\eta )}{a(t)^3 \left(\xi _1^2+\frac{\xi _2^2}{a(t)^2}+\frac{\csc ^2(\eta ) \xi _3^2}{a(t)^2}+\frac{\sec ^2(\eta ) \xi _4^2}{a(t)^2}\right){}^3}-\frac{\csc ^2(\eta )}{a(t)^2 \left(\xi _1^2+\frac{\xi _2^2}{a(t)^2}+\frac{\csc ^2(\eta ) \xi _3^2}{a(t)^2}+\frac{\sec ^2(\eta ) \xi _4^2}{a(t)^2}\right){}^2}-\frac{12 \cot ^2(\eta ) \xi _3^2 \csc ^2(\eta )}{a(t)^4 \left(\xi _1^2+\frac{\xi _2^2}{a(t)^2}+\frac{\csc ^2(\eta ) \xi _3^2}{a(t)^2}+\frac{\sec ^2(\eta ) \xi _4^2}{a(t)^2}\right){}^3}-\frac{32 \xi _1^2 \xi _3^2 a''(t) \csc ^2(\eta )}{a(t)^3 \left(\xi _1^2+\frac{\xi _2^2}{a(t)^2}+\frac{\csc ^2(\eta ) \xi _3^2}{a(t)^2}+\frac{\sec ^2(\eta ) \xi _4^2}{a(t)^2}\right){}^4}-\frac{48 \xi _1^2 \xi _3^2 a'(t)^2 \csc ^2(\eta )}{a(t)^4 \left(\xi _1^2+\frac{\xi _2^2}{a(t)^2}+\frac{\csc ^2(\eta ) \xi _3^2}{a(t)^2}+\frac{\sec ^2(\eta ) \xi _4^2}{a(t)^2}\right){}^4}-\frac{96 \cot (2 \eta ) \xi _1 \xi _2 \xi _3^2 a'(t) \csc ^2(\eta )}{a(t)^5 \left(\xi _1^2+\frac{\xi _2^2}{a(t)^2}+\frac{\csc ^2(\eta ) \xi _3^2}{a(t)^2}+\frac{\sec ^2(\eta ) \xi _4^2}{a(t)^2}\right){}^4}-\frac{96 \cot (\eta ) \cot (2 \eta ) \xi _2^2 \xi _3^2 \csc ^2(\eta )}{a(t)^6 \left(\xi _1^2+\frac{\xi _2^2}{a(t)^2}+\frac{\csc ^2(\eta ) \xi _3^2}{a(t)^2}+\frac{\sec ^2(\eta ) \xi _4^2}{a(t)^2}\right){}^4}-\frac{64 \sec ^2(\eta ) \xi _3^2 \xi _4^2 \csc ^2(\eta )}{a(t)^6 \left(\xi _1^2+\frac{\xi _2^2}{a(t)^2}+\frac{\csc ^2(\eta ) \xi _3^2}{a(t)^2}+\frac{\sec ^2(\eta ) \xi _4^2}{a(t)^2}\right){}^4}-\frac{384 \xi _1^2 \xi _2^2 \xi _3^2 a'(t)^2 \csc ^2(\eta )}{a(t)^6 \left(\xi _1^2+\frac{\xi _2^2}{a(t)^2}+\frac{\csc ^2(\eta ) \xi _3^2}{a(t)^2}+\frac{\sec ^2(\eta ) \xi _4^2}{a(t)^2}\right){}^5}-\frac{384 \sec ^2(\eta ) \xi _1^2 \xi _3^2 \xi _4^2 a'(t)^2 \csc ^2(\eta )}{a(t)^6 \left(\xi _1^2+\frac{\xi _2^2}{a(t)^2}+\frac{\csc ^2(\eta ) \xi _3^2}{a(t)^2}+\frac{\sec ^2(\eta ) \xi _4^2}{a(t)^2}\right){}^5}-\frac{384 \cot (\eta ) \xi _1 \xi _2^3 \xi _3^2 a'(t) \csc ^2(\eta )}{a(t)^7 \left(\xi _1^2+\frac{\xi _2^2}{a(t)^2}+\frac{\csc ^2(\eta ) \xi _3^2}{a(t)^2}+\frac{\sec ^2(\eta ) \xi _4^2}{a(t)^2}\right){}^5}+\frac{384 \sec ^3(\eta ) \xi _1 \xi _2 \xi _3^2 \xi _4^2 a'(t) \csc (\eta )}{a(t)^7 \left(\xi _1^2+\frac{\xi _2^2}{a(t)^2}+\frac{\csc ^2(\eta ) \xi _3^2}{a(t)^2}+\frac{\sec ^2(\eta ) \xi _4^2}{a(t)^2}\right){}^5}+\frac{32 \csc ^2(2 \eta ) \xi _2^2}{a(t)^4 \left(\xi _1^2+\frac{\xi _2^2}{a(t)^2}+\frac{\csc ^2(\eta ) \xi _3^2}{a(t)^2}+\frac{\sec ^2(\eta ) \xi _4^2}{a(t)^2}\right){}^3}+\frac{32 \sec ^4(\eta ) \xi _2^2 \xi _4^2}{a(t)^6 \left(\xi _1^2+\frac{\xi _2^2}{a(t)^2}+\frac{\csc ^2(\eta ) \xi _3^2}{a(t)^2}+\frac{\sec ^2(\eta ) \xi _4^2}{a(t)^2}\right){}^4}+\frac{32 \sec ^4(\eta ) \xi _4^4 \tan ^2(\eta )}{a(t)^6 \left(\xi _1^2+\frac{\xi _2^2}{a(t)^2}+\frac{\csc ^2(\eta ) \xi _3^2}{a(t)^2}+\frac{\sec ^2(\eta ) \xi _4^2}{a(t)^2}\right){}^4}+\frac{64 \sec ^2(\eta ) \xi _2^2 \xi _4^2 \tan ^2(\eta )}{a(t)^6 \left(\xi _1^2+\frac{\xi _2^2}{a(t)^2}+\frac{\csc ^2(\eta ) \xi _3^2}{a(t)^2}+\frac{\sec ^2(\eta ) \xi _4^2}{a(t)^2}\right){}^4}+\frac{32 \xi _2^4 a'(t)^2}{a(t)^6 \left(\xi _1^2+\frac{\xi _2^2}{a(t)^2}+\frac{\csc ^2(\eta ) \xi _3^2}{a(t)^2}+\frac{\sec ^2(\eta ) \xi _4^2}{a(t)^2}\right){}^4}+\frac{32 \sec ^4(\eta ) \xi _4^4 a'(t)^2}{a(t)^6 \left(\xi _1^2+\frac{\xi _2^2}{a(t)^2}+\frac{\csc ^2(\eta ) \xi _3^2}{a(t)^2}+\frac{\sec ^2(\eta ) \xi _4^2}{a(t)^2}\right){}^4}+\frac{4 \xi _2^2 a'(t)^2}{a(t)^4 \left(\xi _1^2+\frac{\xi _2^2}{a(t)^2}+\frac{\csc ^2(\eta ) \xi _3^2}{a(t)^2}+\frac{\sec ^2(\eta ) \xi _4^2}{a(t)^2}\right){}^3}+\frac{4 \sec ^2(\eta ) \xi _4^2 a'(t)^2}{a(t)^4 \left(\xi _1^2+\frac{\xi _2^2}{a(t)^2}+\frac{\csc ^2(\eta ) \xi _3^2}{a(t)^2}+\frac{\sec ^2(\eta ) \xi _4^2}{a(t)^2}\right){}^3}+\frac{64 \sec ^2(\eta ) \xi _2^2 \xi _4^2 a'(t)^2}{a(t)^6 \left(\xi _1^2+\frac{\xi _2^2}{a(t)^2}+\frac{\csc ^2(\eta ) \xi _3^2}{a(t)^2}+\frac{\sec ^2(\eta ) \xi _4^2}{a(t)^2}\right){}^4}+\frac{3 a'(t)^2}{a(t)^2 \left(\xi _1^2+\frac{\xi _2^2}{a(t)^2}+\frac{\csc ^2(\eta ) \xi _3^2}{a(t)^2}+\frac{\sec ^2(\eta ) \xi _4^2}{a(t)^2}\right){}^2}+\frac{96 \cot (2 \eta ) \sec ^2(\eta ) \xi _2^2 \xi _4^2 \tan (\eta )}{a(t)^6 \left(\xi _1^2+\frac{\xi _2^2}{a(t)^2}+\frac{\csc ^2(\eta ) \xi _3^2}{a(t)^2}+\frac{\sec ^2(\eta ) \xi _4^2}{a(t)^2}\right){}^4}+\frac{384 \sec ^4(\eta ) \xi _1 \xi _2 \xi _4^4 \tan (\eta ) a'(t)}{a(t)^7 \left(\xi _1^2+\frac{\xi _2^2}{a(t)^2}+\frac{\csc ^2(\eta ) \xi _3^2}{a(t)^2}+\frac{\sec ^2(\eta ) \xi _4^2}{a(t)^2}\right){}^5}+\frac{384 \sec ^2(\eta ) \xi _1 \xi _2^3 \xi _4^2 \tan (\eta ) a'(t)}{a(t)^7 \left(\xi _1^2+\frac{\xi _2^2}{a(t)^2}+\frac{\csc ^2(\eta ) \xi _3^2}{a(t)^2}+\frac{\sec ^2(\eta ) \xi _4^2}{a(t)^2}\right){}^5}+\frac{8 \xi _2^2 a''(t)}{a(t)^3 \left(\xi _1^2+\frac{\xi _2^2}{a(t)^2}+\frac{\csc ^2(\eta ) \xi _3^2}{a(t)^2}+\frac{\sec ^2(\eta ) \xi _4^2}{a(t)^2}\right){}^3}+\frac{8 \sec ^2(\eta ) \xi _4^2 a''(t)}{a(t)^3 \left(\xi _1^2+\frac{\xi _2^2}{a(t)^2}+\frac{\csc ^2(\eta ) \xi _3^2}{a(t)^2}+\frac{\sec ^2(\eta ) \xi _4^2}{a(t)^2}\right){}^3}+\frac{6 a''(t)}{a(t) \left(\xi _1^2+\frac{\xi _2^2}{a(t)^2}+\frac{\csc ^2(\eta ) \xi _3^2}{a(t)^2}+\frac{\sec ^2(\eta ) \xi _4^2}{a(t)^2}\right){}^2}-\frac{\sec ^2(\eta )}{a(t)^2 \left(\xi _1^2+\frac{\xi _2^2}{a(t)^2}+\frac{\csc ^2(\eta ) \xi _3^2}{a(t)^2}+\frac{\sec ^2(\eta ) \xi _4^2}{a(t)^2}\right){}^2}-\frac{4}{a(t)^2 \left(\xi _1^2+\frac{\xi _2^2}{a(t)^2}+\frac{\csc ^2(\eta ) \xi _3^2}{a(t)^2}+\frac{\sec ^2(\eta ) \xi _4^2}{a(t)^2}\right){}^2}-\frac{24 \xi _1^2 a''(t)}{a(t) \left(\xi _1^2+\frac{\xi _2^2}{a(t)^2}+\frac{\csc ^2(\eta ) \xi _3^2}{a(t)^2}+\frac{\sec ^2(\eta ) \xi _4^2}{a(t)^2}\right){}^3}-\frac{12 \xi _1^2 a'(t)^2}{a(t)^2 \left(\xi _1^2+\frac{\xi _2^2}{a(t)^2}+\frac{\csc ^2(\eta ) \xi _3^2}{a(t)^2}+\frac{\sec ^2(\eta ) \xi _4^2}{a(t)^2}\right){}^3}-\frac{16 \cot (2 \eta ) \xi _1 \xi _2 a'(t)}{a(t)^3 \left(\xi _1^2+\frac{\xi _2^2}{a(t)^2}+\frac{\csc ^2(\eta ) \xi _3^2}{a(t)^2}+\frac{\sec ^2(\eta ) \xi _4^2}{a(t)^2}\right){}^3}-\frac{16 \cot ^2(2 \eta ) \xi _2^2}{a(t)^4 \left(\xi _1^2+\frac{\xi _2^2}{a(t)^2}+\frac{\csc ^2(\eta ) \xi _3^2}{a(t)^2}+\frac{\sec ^2(\eta ) \xi _4^2}{a(t)^2}\right){}^3}-\frac{8 \sec ^4(\eta ) \xi _4^2}{a(t)^4 \left(\xi _1^2+\frac{\xi _2^2}{a(t)^2}+\frac{\csc ^2(\eta ) \xi _3^2}{a(t)^2}+\frac{\sec ^2(\eta ) \xi _4^2}{a(t)^2}\right){}^3}-\frac{12 \sec ^2(\eta ) \xi _4^2 \tan ^2(\eta )}{a(t)^4 \left(\xi _1^2+\frac{\xi _2^2}{a(t)^2}+\frac{\csc ^2(\eta ) \xi _3^2}{a(t)^2}+\frac{\sec ^2(\eta ) \xi _4^2}{a(t)^2}\right){}^3}-\frac{16 \cot (2 \eta ) \sec ^2(\eta ) \xi _4^2 \tan (\eta )}{a(t)^4 \left(\xi _1^2+\frac{\xi _2^2}{a(t)^2}+\frac{\csc ^2(\eta ) \xi _3^2}{a(t)^2}+\frac{\sec ^2(\eta ) \xi _4^2}{a(t)^2}\right){}^3}-\frac{32 \xi _1^2 \xi _2^2 a''(t)}{a(t)^3 \left(\xi _1^2+\frac{\xi _2^2}{a(t)^2}+\frac{\csc ^2(\eta ) \xi _3^2}{a(t)^2}+\frac{\sec ^2(\eta ) \xi _4^2}{a(t)^2}\right){}^4}-\frac{32 \sec ^2(\eta ) \xi _1^2 \xi _4^2 a''(t)}{a(t)^3 \left(\xi _1^2+\frac{\xi _2^2}{a(t)^2}+\frac{\csc ^2(\eta ) \xi _3^2}{a(t)^2}+\frac{\sec ^2(\eta ) \xi _4^2}{a(t)^2}\right){}^4}-\frac{48 \xi _1^2 \xi _2^2 a'(t)^2}{a(t)^4 \left(\xi _1^2+\frac{\xi _2^2}{a(t)^2}+\frac{\csc ^2(\eta ) \xi _3^2}{a(t)^2}+\frac{\sec ^2(\eta ) \xi _4^2}{a(t)^2}\right){}^4}-\frac{48 \sec ^2(\eta ) \xi _1^2 \xi _4^2 a'(t)^2}{a(t)^4 \left(\xi _1^2+\frac{\xi _2^2}{a(t)^2}+\frac{\csc ^2(\eta ) \xi _3^2}{a(t)^2}+\frac{\sec ^2(\eta ) \xi _4^2}{a(t)^2}\right){}^4}-\frac{96 \cot (2 \eta ) \xi _1 \xi _2^3 a'(t)}{a(t)^5 \left(\xi _1^2+\frac{\xi _2^2}{a(t)^2}+\frac{\csc ^2(\eta ) \xi _3^2}{a(t)^2}+\frac{\sec ^2(\eta ) \xi _4^2}{a(t)^2}\right){}^4}-\frac{96 \cot (2 \eta ) \sec ^2(\eta ) \xi _1 \xi _2 \xi _4^2 a'(t)}{a(t)^5 \left(\xi _1^2+\frac{\xi _2^2}{a(t)^2}+\frac{\csc ^2(\eta ) \xi _3^2}{a(t)^2}+\frac{\sec ^2(\eta ) \xi _4^2}{a(t)^2}\right){}^4}-\frac{48 \sec ^2(\eta ) \xi _1 \xi _2 \xi _4^2 \tan (\eta ) a'(t)}{a(t)^5 \left(\xi _1^2+\frac{\xi _2^2}{a(t)^2}+\frac{\csc ^2(\eta ) \xi _3^2}{a(t)^2}+\frac{\sec ^2(\eta ) \xi _4^2}{a(t)^2}\right){}^4}-\frac{192 \xi _1^2 \xi _2^4 a'(t)^2}{a(t)^6 \left(\xi _1^2+\frac{\xi _2^2}{a(t)^2}+\frac{\csc ^2(\eta ) \xi _3^2}{a(t)^2}+\frac{\sec ^2(\eta ) \xi _4^2}{a(t)^2}\right){}^5}-\frac{192 \sec ^4(\eta ) \xi _1^2 \xi _4^4 a'(t)^2}{a(t)^6 \left(\xi _1^2+\frac{\xi _2^2}{a(t)^2}+\frac{\csc ^2(\eta ) \xi _3^2}{a(t)^2}+\frac{\sec ^2(\eta ) \xi _4^2}{a(t)^2}\right){}^5}-\frac{384 \sec ^2(\eta ) \xi _1^2 \xi _2^2 \xi _4^2 a'(t)^2}{a(t)^6 \left(\xi _1^2+\frac{\xi _2^2}{a(t)^2}+\frac{\csc ^2(\eta ) \xi _3^2}{a(t)^2}+\frac{\sec ^2(\eta ) \xi _4^2}{a(t)^2}\right){}^5}-\frac{192 \sec ^4(\eta ) \xi _2^2 \xi _4^4 \tan ^2(\eta )}{a(t)^8 \left(\xi _1^2+\frac{\xi _2^2}{a(t)^2}+\frac{\csc ^2(\eta ) \xi _3^2}{a(t)^2}+\frac{\sec ^2(\eta ) \xi _4^2}{a(t)^2}\right){}^5}.
\end{math}
\end{center}

One can see easily that each term in this expression that has an odd power of $\xi_j$ in 
its numerator will finally integrate to $0$ when we perform the  integration of the 
1-density, using formulas \eqref{WodResDensity} and \eqref{WodResFormula}. Therefore 
we can reduce to considering only the following expression in our calculation, which is obtained 
from the above expression for $\textnormal{tr}(\sigma_{-4}(x, \xi) ) $ after eliminating 
any term that has an odd exponent in the numerator: 
\begin{equation} \label{bminus4}
b_{-4}(x, \xi) = b_{-4}(t, \eta, \xi) =  
\end{equation}
\begin{center}
\begin{math}
\frac{32 \cot ^2(\eta ) \xi _3^4 \csc ^4(\eta )}{a(t)^6 \left(\xi _1^2+\frac{\xi _2^2}{a(t)^2}+\frac{\csc ^2(\eta ) \xi _3^2}{a(t)^2}+\frac{\sec ^2(\eta ) \xi _4^2}{a(t)^2}\right){}^4}+\frac{32 \xi _2^2 \xi _3^2 \csc ^4(\eta )}{a(t)^6 \left(\xi _1^2+\frac{\xi _2^2}{a(t)^2}+\frac{\csc ^2(\eta ) \xi _3^2}{a(t)^2}+\frac{\sec ^2(\eta ) \xi _4^2}{a(t)^2}\right){}^4}+\frac{32 \xi _3^4 a'(t)^2 \csc ^4(\eta )}{a(t)^6 \left(\xi _1^2+\frac{\xi _2^2}{a(t)^2}+\frac{\csc ^2(\eta ) \xi _3^2}{a(t)^2}+\frac{\sec ^2(\eta ) \xi _4^2}{a(t)^2}\right){}^4}-\frac{8 \xi _3^2 \csc ^4(\eta )}{a(t)^4 \left(\xi _1^2+\frac{\xi _2^2}{a(t)^2}+\frac{\csc ^2(\eta ) \xi _3^2}{a(t)^2}+\frac{\sec ^2(\eta ) \xi _4^2}{a(t)^2}\right){}^3}-\frac{192 \xi _1^2 \xi _3^4 a'(t)^2 \csc ^4(\eta )}{a(t)^6 \left(\xi _1^2+\frac{\xi _2^2}{a(t)^2}+\frac{\csc ^2(\eta ) \xi _3^2}{a(t)^2}+\frac{\sec ^2(\eta ) \xi _4^2}{a(t)^2}\right){}^5}-\frac{192 \cot ^2(\eta ) \xi _2^2 \xi _3^4 \csc ^4(\eta )}{a(t)^8 \left(\xi _1^2+\frac{\xi _2^2}{a(t)^2}+\frac{\csc ^2(\eta ) \xi _3^2}{a(t)^2}+\frac{\sec ^2(\eta ) \xi _4^2}{a(t)^2}\right){}^5}+\frac{64 \cot ^2(\eta ) \xi _2^2 \xi _3^2 \csc ^2(\eta )}{a(t)^6 \left(\xi _1^2+\frac{\xi _2^2}{a(t)^2}+\frac{\csc ^2(\eta ) \xi _3^2}{a(t)^2}+\frac{\sec ^2(\eta ) \xi _4^2}{a(t)^2}\right){}^4}+\frac{16 \cot (\eta ) \cot (2 \eta ) \xi _3^2 \csc ^2(\eta )}{a(t)^4 \left(\xi _1^2+\frac{\xi _2^2}{a(t)^2}+\frac{\csc ^2(\eta ) \xi _3^2}{a(t)^2}+\frac{\sec ^2(\eta ) \xi _4^2}{a(t)^2}\right){}^3}+\frac{384 \sec ^2(\eta ) \xi _2^2 \xi _3^2 \xi _4^2 \csc ^2(\eta )}{a(t)^8 \left(\xi _1^2+\frac{\xi _2^2}{a(t)^2}+\frac{\csc ^2(\eta ) \xi _3^2}{a(t)^2}+\frac{\sec ^2(\eta ) \xi _4^2}{a(t)^2}\right){}^5}+\frac{64 \xi _2^2 \xi _3^2 a'(t)^2 \csc ^2(\eta )}{a(t)^6 \left(\xi _1^2+\frac{\xi _2^2}{a(t)^2}+\frac{\csc ^2(\eta ) \xi _3^2}{a(t)^2}+\frac{\sec ^2(\eta ) \xi _4^2}{a(t)^2}\right){}^4}+\frac{4 \xi _3^2 a'(t)^2 \csc ^2(\eta )}{a(t)^4 \left(\xi _1^2+\frac{\xi _2^2}{a(t)^2}+\frac{\csc ^2(\eta ) \xi _3^2}{a(t)^2}+\frac{\sec ^2(\eta ) \xi _4^2}{a(t)^2}\right){}^3}+\frac{64 \sec ^2(\eta ) \xi _3^2 \xi _4^2 a'(t)^2 \csc ^2(\eta )}{a(t)^6 \left(\xi _1^2+\frac{\xi _2^2}{a(t)^2}+\frac{\csc ^2(\eta ) \xi _3^2}{a(t)^2}+\frac{\sec ^2(\eta ) \xi _4^2}{a(t)^2}\right){}^4}+\frac{8 \xi _3^2 a''(t) \csc ^2(\eta )}{a(t)^3 \left(\xi _1^2+\frac{\xi _2^2}{a(t)^2}+\frac{\csc ^2(\eta ) \xi _3^2}{a(t)^2}+\frac{\sec ^2(\eta ) \xi _4^2}{a(t)^2}\right){}^3}-\frac{\csc ^2(\eta )}{a(t)^2 \left(\xi _1^2+\frac{\xi _2^2}{a(t)^2}+\frac{\csc ^2(\eta ) \xi _3^2}{a(t)^2}+\frac{\sec ^2(\eta ) \xi _4^2}{a(t)^2}\right){}^2}-\frac{12 \cot ^2(\eta ) \xi _3^2 \csc ^2(\eta )}{a(t)^4 \left(\xi _1^2+\frac{\xi _2^2}{a(t)^2}+\frac{\csc ^2(\eta ) \xi _3^2}{a(t)^2}+\frac{\sec ^2(\eta ) \xi _4^2}{a(t)^2}\right){}^3}-\frac{32 \xi _1^2 \xi _3^2 a''(t) \csc ^2(\eta )}{a(t)^3 \left(\xi _1^2+\frac{\xi _2^2}{a(t)^2}+\frac{\csc ^2(\eta ) \xi _3^2}{a(t)^2}+\frac{\sec ^2(\eta ) \xi _4^2}{a(t)^2}\right){}^4}-\frac{48 \xi _1^2 \xi _3^2 a'(t)^2 \csc ^2(\eta )}{a(t)^4 \left(\xi _1^2+\frac{\xi _2^2}{a(t)^2}+\frac{\csc ^2(\eta ) \xi _3^2}{a(t)^2}+\frac{\sec ^2(\eta ) \xi _4^2}{a(t)^2}\right){}^4}-\frac{96 \cot (\eta ) \cot (2 \eta ) \xi _2^2 \xi _3^2 \csc ^2(\eta )}{a(t)^6 \left(\xi _1^2+\frac{\xi _2^2}{a(t)^2}+\frac{\csc ^2(\eta ) \xi _3^2}{a(t)^2}+\frac{\sec ^2(\eta ) \xi _4^2}{a(t)^2}\right){}^4}-\frac{64 \sec ^2(\eta ) \xi _3^2 \xi _4^2 \csc ^2(\eta )}{a(t)^6 \left(\xi _1^2+\frac{\xi _2^2}{a(t)^2}+\frac{\csc ^2(\eta ) \xi _3^2}{a(t)^2}+\frac{\sec ^2(\eta ) \xi _4^2}{a(t)^2}\right){}^4}-\frac{384 \xi _1^2 \xi _2^2 \xi _3^2 a'(t)^2 \csc ^2(\eta )}{a(t)^6 \left(\xi _1^2+\frac{\xi _2^2}{a(t)^2}+\frac{\csc ^2(\eta ) \xi _3^2}{a(t)^2}+\frac{\sec ^2(\eta ) \xi _4^2}{a(t)^2}\right){}^5}-\frac{384 \sec ^2(\eta ) \xi _1^2 \xi _3^2 \xi _4^2 a'(t)^2 \csc ^2(\eta )}{a(t)^6 \left(\xi _1^2+\frac{\xi _2^2}{a(t)^2}+\frac{\csc ^2(\eta ) \xi _3^2}{a(t)^2}+\frac{\sec ^2(\eta ) \xi _4^2}{a(t)^2}\right){}^5}+\frac{32 \csc ^2(2 \eta ) \xi _2^2}{a(t)^4 \left(\xi _1^2+\frac{\xi _2^2}{a(t)^2}+\frac{\csc ^2(\eta ) \xi _3^2}{a(t)^2}+\frac{\sec ^2(\eta ) \xi _4^2}{a(t)^2}\right){}^3}+\frac{32 \sec ^4(\eta ) \xi _2^2 \xi _4^2}{a(t)^6 \left(\xi _1^2+\frac{\xi _2^2}{a(t)^2}+\frac{\csc ^2(\eta ) \xi _3^2}{a(t)^2}+\frac{\sec ^2(\eta ) \xi _4^2}{a(t)^2}\right){}^4}+\frac{32 \sec ^4(\eta ) \xi _4^4 \tan ^2(\eta )}{a(t)^6 \left(\xi _1^2+\frac{\xi _2^2}{a(t)^2}+\frac{\csc ^2(\eta ) \xi _3^2}{a(t)^2}+\frac{\sec ^2(\eta ) \xi _4^2}{a(t)^2}\right){}^4}+\frac{64 \sec ^2(\eta ) \xi _2^2 \xi _4^2 \tan ^2(\eta )}{a(t)^6 \left(\xi _1^2+\frac{\xi _2^2}{a(t)^2}+\frac{\csc ^2(\eta ) \xi _3^2}{a(t)^2}+\frac{\sec ^2(\eta ) \xi _4^2}{a(t)^2}\right){}^4}+\frac{32 \xi _2^4 a'(t)^2}{a(t)^6 \left(\xi _1^2+\frac{\xi _2^2}{a(t)^2}+\frac{\csc ^2(\eta ) \xi _3^2}{a(t)^2}+\frac{\sec ^2(\eta ) \xi _4^2}{a(t)^2}\right){}^4}+\frac{32 \sec ^4(\eta ) \xi _4^4 a'(t)^2}{a(t)^6 \left(\xi _1^2+\frac{\xi _2^2}{a(t)^2}+\frac{\csc ^2(\eta ) \xi _3^2}{a(t)^2}+\frac{\sec ^2(\eta ) \xi _4^2}{a(t)^2}\right){}^4}+\frac{4 \xi _2^2 a'(t)^2}{a(t)^4 \left(\xi _1^2+\frac{\xi _2^2}{a(t)^2}+\frac{\csc ^2(\eta ) \xi _3^2}{a(t)^2}+\frac{\sec ^2(\eta ) \xi _4^2}{a(t)^2}\right){}^3}+\frac{4 \sec ^2(\eta ) \xi _4^2 a'(t)^2}{a(t)^4 \left(\xi _1^2+\frac{\xi _2^2}{a(t)^2}+\frac{\csc ^2(\eta ) \xi _3^2}{a(t)^2}+\frac{\sec ^2(\eta ) \xi _4^2}{a(t)^2}\right){}^3}+\frac{64 \sec ^2(\eta ) \xi _2^2 \xi _4^2 a'(t)^2}{a(t)^6 \left(\xi _1^2+\frac{\xi _2^2}{a(t)^2}+\frac{\csc ^2(\eta ) \xi _3^2}{a(t)^2}+\frac{\sec ^2(\eta ) \xi _4^2}{a(t)^2}\right){}^4}+\frac{3 a'(t)^2}{a(t)^2 \left(\xi _1^2+\frac{\xi _2^2}{a(t)^2}+\frac{\csc ^2(\eta ) \xi _3^2}{a(t)^2}+\frac{\sec ^2(\eta ) \xi _4^2}{a(t)^2}\right){}^2}+\frac{96 \cot (2 \eta ) \sec ^2(\eta ) \xi _2^2 \xi _4^2 \tan (\eta )}{a(t)^6 \left(\xi _1^2+\frac{\xi _2^2}{a(t)^2}+\frac{\csc ^2(\eta ) \xi _3^2}{a(t)^2}+\frac{\sec ^2(\eta ) \xi _4^2}{a(t)^2}\right){}^4}+\frac{8 \xi _2^2 a''(t)}{a(t)^3 \left(\xi _1^2+\frac{\xi _2^2}{a(t)^2}+\frac{\csc ^2(\eta ) \xi _3^2}{a(t)^2}+\frac{\sec ^2(\eta ) \xi _4^2}{a(t)^2}\right){}^3}+\frac{8 \sec ^2(\eta ) \xi _4^2 a''(t)}{a(t)^3 \left(\xi _1^2+\frac{\xi _2^2}{a(t)^2}+\frac{\csc ^2(\eta ) \xi _3^2}{a(t)^2}+\frac{\sec ^2(\eta ) \xi _4^2}{a(t)^2}\right){}^3}+\frac{6 a''(t)}{a(t) \left(\xi _1^2+\frac{\xi _2^2}{a(t)^2}+\frac{\csc ^2(\eta ) \xi _3^2}{a(t)^2}+\frac{\sec ^2(\eta ) \xi _4^2}{a(t)^2}\right){}^2}-\frac{\sec ^2(\eta )}{a(t)^2 \left(\xi _1^2+\frac{\xi _2^2}{a(t)^2}+\frac{\csc ^2(\eta ) \xi _3^2}{a(t)^2}+\frac{\sec ^2(\eta ) \xi _4^2}{a(t)^2}\right){}^2}-\frac{4}{a(t)^2 \left(\xi _1^2+\frac{\xi _2^2}{a(t)^2}+\frac{\csc ^2(\eta ) \xi _3^2}{a(t)^2}+\frac{\sec ^2(\eta ) \xi _4^2}{a(t)^2}\right){}^2}-\frac{24 \xi _1^2 a''(t)}{a(t) \left(\xi _1^2+\frac{\xi _2^2}{a(t)^2}+\frac{\csc ^2(\eta ) \xi _3^2}{a(t)^2}+\frac{\sec ^2(\eta ) \xi _4^2}{a(t)^2}\right){}^3}-\frac{12 \xi _1^2 a'(t)^2}{a(t)^2 \left(\xi _1^2+\frac{\xi _2^2}{a(t)^2}+\frac{\csc ^2(\eta ) \xi _3^2}{a(t)^2}+\frac{\sec ^2(\eta ) \xi _4^2}{a(t)^2}\right){}^3}-\frac{16 \cot ^2(2 \eta ) \xi _2^2}{a(t)^4 \left(\xi _1^2+\frac{\xi _2^2}{a(t)^2}+\frac{\csc ^2(\eta ) \xi _3^2}{a(t)^2}+\frac{\sec ^2(\eta ) \xi _4^2}{a(t)^2}\right){}^3}-\frac{8 \sec ^4(\eta ) \xi _4^2}{a(t)^4 \left(\xi _1^2+\frac{\xi _2^2}{a(t)^2}+\frac{\csc ^2(\eta ) \xi _3^2}{a(t)^2}+\frac{\sec ^2(\eta ) \xi _4^2}{a(t)^2}\right){}^3}-\frac{12 \sec ^2(\eta ) \xi _4^2 \tan ^2(\eta )}{a(t)^4 \left(\xi _1^2+\frac{\xi _2^2}{a(t)^2}+\frac{\csc ^2(\eta ) \xi _3^2}{a(t)^2}+\frac{\sec ^2(\eta ) \xi _4^2}{a(t)^2}\right){}^3}-\frac{16 \cot (2 \eta ) \sec ^2(\eta ) \xi _4^2 \tan (\eta )}{a(t)^4 \left(\xi _1^2+\frac{\xi _2^2}{a(t)^2}+\frac{\csc ^2(\eta ) \xi _3^2}{a(t)^2}+\frac{\sec ^2(\eta ) \xi _4^2}{a(t)^2}\right){}^3}-\frac{32 \xi _1^2 \xi _2^2 a''(t)}{a(t)^3 \left(\xi _1^2+\frac{\xi _2^2}{a(t)^2}+\frac{\csc ^2(\eta ) \xi _3^2}{a(t)^2}+\frac{\sec ^2(\eta ) \xi _4^2}{a(t)^2}\right){}^4}-\frac{32 \sec ^2(\eta ) \xi _1^2 \xi _4^2 a''(t)}{a(t)^3 \left(\xi _1^2+\frac{\xi _2^2}{a(t)^2}+\frac{\csc ^2(\eta ) \xi _3^2}{a(t)^2}+\frac{\sec ^2(\eta ) \xi _4^2}{a(t)^2}\right){}^4}-\frac{48 \xi _1^2 \xi _2^2 a'(t)^2}{a(t)^4 \left(\xi _1^2+\frac{\xi _2^2}{a(t)^2}+\frac{\csc ^2(\eta ) \xi _3^2}{a(t)^2}+\frac{\sec ^2(\eta ) \xi _4^2}{a(t)^2}\right){}^4}-\frac{48 \sec ^2(\eta ) \xi _1^2 \xi _4^2 a'(t)^2}{a(t)^4 \left(\xi _1^2+\frac{\xi _2^2}{a(t)^2}+\frac{\csc ^2(\eta ) \xi _3^2}{a(t)^2}+\frac{\sec ^2(\eta ) \xi _4^2}{a(t)^2}\right){}^4}-\frac{192 \xi _1^2 \xi _2^4 a'(t)^2}{a(t)^6 \left(\xi _1^2+\frac{\xi _2^2}{a(t)^2}+\frac{\csc ^2(\eta ) \xi _3^2}{a(t)^2}+\frac{\sec ^2(\eta ) \xi _4^2}{a(t)^2}\right){}^5}-\frac{192 \sec ^4(\eta ) \xi _1^2 \xi _4^4 a'(t)^2}{a(t)^6 \left(\xi _1^2+\frac{\xi _2^2}{a(t)^2}+\frac{\csc ^2(\eta ) \xi _3^2}{a(t)^2}+\frac{\sec ^2(\eta ) \xi _4^2}{a(t)^2}\right){}^5}-\frac{384 \sec ^2(\eta ) \xi _1^2 \xi _2^2 \xi _4^2 a'(t)^2}{a(t)^6 \left(\xi _1^2+\frac{\xi _2^2}{a(t)^2}+\frac{\csc ^2(\eta ) \xi _3^2}{a(t)^2}+\frac{\sec ^2(\eta ) \xi _4^2}{a(t)^2}\right){}^5}-\frac{192 \sec ^4(\eta ) \xi _2^2 \xi _4^4 \tan ^2(\eta )}{a(t)^8 \left(\xi _1^2+\frac{\xi _2^2}{a(t)^2}+\frac{\csc ^2(\eta ) \xi _3^2}{a(t)^2}+\frac{\sec ^2(\eta ) \xi _4^2}{a(t)^2}\right){}^5}. 
\end{math}
\end{center}
Note that $b_{-4}$ as well as $\sigma_{-4}$ has no dependence on $\phi_1$ and $\phi_2$. 

\smallskip

To express the expression in \eqref{bminus4} 
as a rational function, we introduce the change of coordinates
\begin{equation}\label{u-coords}
\begin{array}{ccc}
u_0 = \sin^2(\eta), & u_1=\xi_1, & u_2 =\xi_2,  \\
u_3 = \csc(\eta)\, \xi_3, & \qquad  u_4 = \sec(\eta)\, \xi_4,  & 
\end{array}
\end{equation}
and we write $\alpha$ for the term $a(t)$, considering it as an affine variable,
momentarily forgetting the time dependence, since we are omitting the time integration.
The trigonometric expressions appearing in $b_4$ transform 
to the following expressions in the new coordinates: 
\[
\xi _1^2+\frac{\xi _2^2}{a(t)^2}+ \frac{\xi _3^2 \csc ^2(\eta )}{a(t)^2}+\frac{\xi _4^2 \sec ^2(\eta )}{a(t)^2}
= u_1^2 + \frac{1}{a(t)^2}( u_2^2 + u_3^2+u_4^2 ), 
\]
\[
\cot^2(\eta) = \frac{1-u_0}{u_0}, 
\]
\[
\csc^2(\eta) = \frac{1}{u_0}, 
\]
\[
\sec^2(\eta) = \frac{1}{1-u_0}, 
\]
\[
\cot (\eta ) \cot (2 \eta )  = \frac{\cot ^2(\eta )}{2}-\frac{1}{2}, 
\]
\[
\csc ^2(2 \eta ) = \frac{1}{4} \csc ^2(\eta ) \sec ^2(\eta ), 
\]
\[
\tan^2(\eta) = \sec^2(\eta) -1, 
\]
\[
\tan (\eta ) \cot (2 \eta ) = \frac{1}{2}-\frac{\tan ^2(\eta )}{2}, 
\]
\[
\cot ^2(2 \eta ) = \frac{\tan ^2(\eta )}{8}+\frac{\cot ^2(\eta )}{8}+\frac{1}{8} \csc ^2(\eta ) \sec ^2(\eta )-\frac{3}{4}. 
\]
Moreover, the exponents of the variables $\xi_j$ in the expression \eqref{bminus4} 
for $b_{-4}$ are even positive integers.

\smallskip

Thus, it is clear that with the change of variables \eqref{u-coords} 
the expression above becomes a $\Q$-linear combination
of rational functions with numerators that are rational polynomials in the variables $(u_0,u_1,u_2,u_3,u_4,\alpha)$
and with denominators that are all of the form $\alpha^{2r} u_0^k (1-u_0)^m (u_1^2 + \alpha^{-2} (u_2^2 + u_3^2 + u_4^2))^\ell$,
with $r,k,m,\ell\in \Z_{\geq 0}$. 

\smallskip

We also have
\begin{equation}\label{tildesigma}
\sigma_{\xi, \, 3} =   \sum_{j=1}^4 (-1)^{j-1} \xi_j \, d\xi_1 
\wedge \cdots \wedge {\widehat d \xi_j} \wedge \cdots \wedge d \xi_4.
\end{equation}
With the change of variables \eqref{u-coords} we obtain
\begin{eqnarray*}
d \eta &=& \frac{1}{2 \sin(\eta) \cos(\eta)} \, du_0 = \frac{1}{2} \csc(\eta) \sec(\eta) \, du_0, 
\\
d \xi_1 &=& du_1, \\
d \xi_2 &=& du_2, \\
d \xi_3 &=&   \cos (\eta ) u_3 \, d\eta + \sin(\eta) \, du_3, \\
d \xi_4 &=& - \sin(\eta) u_4 \, d\eta + \cos(\eta) \, du_4,  
\end{eqnarray*}
which yields
\begin{eqnarray*}
\widetilde \sigma_3 &:=& d\eta \wedge \sigma_{\xi, \,3} \\
&=& 
 \sum_{j=1}^4 (-1)^{j-1} \xi_j \, d \eta  \wedge  d\xi_1 
\wedge \cdots \wedge {\widehat d \xi_j} \wedge \cdots \wedge d \xi_4 \\
&=& \sin(\eta) \cos(\eta) \Big ( u_1  \, d\eta \, du_2 \, du_3 \, du_4 - u_2  \, d\eta\, du_1 \, du_3 \, du_4 \\
&& \qquad \qquad \qquad \qquad \qquad\qquad\qquad+ u_3 \, d \eta \, du_1 \, du_2 \, du_4 -  u_4   \, d \eta \, du_1 \, du_2 \, du_3 \Big ) 
\\
&=& \frac{1}{2} \Big ( u_1  \, du_0 \, du_2 \, du_3 \, du_4 - u_2  \, du_0\, du_1 \, du_3 \, du_4
+ u_3 \, d u_0 \, du_1 \, du_2 \, du_4   \\ 
&& \qquad \qquad \qquad \qquad \qquad\qquad\qquad   \qquad\qquad\qquad  -  u_4   \, d u_0 \, du_1 \, du_2 \, du_3 \Big ). 
\end{eqnarray*}
Thus we obtain an algebraic differential form $\Omega$, defined over $\Q$, 
whose singular locus is defined by the vanishing of the quadratic form
$$ u_1^2 + \alpha^{-2} (u_2^2 + u_3^2 + u_4^2), $$
where we assume $\alpha \in \bG_m=\A^1 \smallsetminus \{ 0 \}$. 
This quadratic form defines, for each fixed value of $\alpha$, a quadric surface $Z_\alpha$ 
in $\P^3$. We write $\hat Z_\alpha$ for the corresponding affine hypersurface in $\A^4$,
the affine cone over $Z_\alpha$.  We also denote by
$CZ_\alpha$ the projective cone of $Z_\alpha$ in $\P^4$ and by $\widehat{CZ}_\alpha$ its affine cone in $\A^5$.
The latter is a product of the hypersurface $\hat Z_\alpha$ in $\A^4$ with the line $\A^1$ of the $u_0$-coordinate.

\smallskip

The restriction $\Omega^\alpha$ of the form $\Omega$ to a fixed (rational) value of $\alpha$ determines
a two-parameter family of rational differential forms, by viewing the two variables $(\alpha_1,\alpha_2)$
that appear in the numerators as parameters,
$$ P(u_0,u_1,u_2,u_3,u_4,\alpha, \alpha_1, \alpha_2) = P_{(\alpha_1, \alpha_2)}(u_0,u_1,u_2,u_3,u_4,\alpha), $$
with $P_{(\alpha_1, \alpha_2)} \in \Q[u_0,u_1,u_2,u_3,u_4,\alpha]$ for all $(\alpha_1, \alpha_2)\in \Q^2$.
For any choice of the parameters $(\alpha_1, \alpha_2)\in \Q^2$, the resulting form $\Omega^\alpha_{(\alpha_1,\alpha_2)}$ is an algebraic differential form on the same variety $\A^5 \smallsetminus (\widehat{CZ}_\alpha \cup H_0 \cup H_1)$.
\endproof

\medskip

A $\Q$-semialgebraic set is a subset $S$ of some $\R^n$ that is of the form
$$ S = \{ (x_1,\ldots, x_n)\in \R^n \,:\, P(x_1,\ldots, x_n) \geq 0 \}, $$
for some polynomial $P\in \Q[x_1,\ldots, x_n]$, or obtained from such sets by
taking a finite number of complements, intersections, and unions.

\medskip

\begin{thm}\label{a2intperiod}
When computed without performing the time integration, the $a_2$ term in the heat kernel expansion
is a period integral 
\begin{equation}\label{a2period}
 C \cdot \int_{A} \Omega^\alpha_{(\alpha_1,\alpha_2)}
\end{equation} 
with the algebraic differential form of Proposition~\ref{a2compute}, and with 
domain of integration the $\Q$-semialgebraic set
$$ A_4 = \left\{ (u_0,u_1,u_2,u_3,u_4)\in \A^5(\R)\,:\, \begin{array}{ll} u_1^2 + u_2^2 +u_0 u_3^2 + (1-u_0)u_4^2 =1, \\ 
0< u_i <1, \text{ for } i=0,1,2 \end{array} \right\}. $$
The coefficient $C$ is in $\Q[(2\pi i)^{-1}]$.
This integral is a period of the mixed motive 
\begin{equation}\label{MMa2}
\m(\A^5\smallsetminus (\widehat{CZ}_\alpha \cup H_0 \cup H_1), \Sigma), 
\end{equation}
where $\widehat{CZ}_\alpha$ is the hypersurface in $\A^5$ defined by the vanishing of the quadric 
$Q_{\alpha, 2}$ of \eqref{Qalpha}, and $\Sigma=\cup_{i,a} H_{i,a}$
is the divisor given by the union of the hyperplanes $H_{i,a}=\{ u_i = a \}$, with $i\in \{ 0,1,2 \}$ and $a\in \{ 0, 1 \}$.
\end{thm}

\proof 
With the notation \eqref{tildesigma} as above, 
and leaving out an integration with respect to $t$, we have 
\begin{eqnarray}\label{a2prerationalform}
a_2 &=&
 \frac{1}{2^5 \pi^4} \int_0^{\pi/2} d \eta \int_0^{2 \pi} d\phi_1 \int_0^{2 \pi} d\phi_2 
\int_{\xi_1^2+\xi_2^2+\xi_3^2+\xi_4^2 = 1} d^3\xi \, \cdot b_{-4}(t, \eta, \xi) \,\cdot  \sigma_{\xi, \,3} \nonumber \\
&=&
 \frac{1}{2^3 \pi^2} \int_0^{\pi/2} d \eta  
\int_{\xi_1^2+\xi_2^2+\xi_3^2+\xi_4^2 = 1} d^3\xi  \cdot b_{-4}(t, \eta, \xi) \cdot  \sigma_{\xi, \,3} \nonumber \\
&=& \frac{1}{2^3 \pi^2}  \int_{(0, \frac{\pi}{2}) \times \mathbb{S}^3}  b_{-4}(t, \eta, \xi)  
 \sum_{j=1}^4 (-1)^{j-1} \xi_j \, d \eta  \wedge  d\xi_1 
\wedge \cdots \wedge {\widehat d \xi_j} \wedge \cdots \wedge d \xi_4. 
\end{eqnarray}
We use the change of variables \eqref{u-coords} as in Proposition~\ref{a2compute} to rewrite both
the form and the domain of integration and we obtain, up to a coefficient in $\Q[(2\pi i)^{-1}]$, 
an integral of the form
$$ \int_{A} \Omega^\alpha_{(\alpha_1,\alpha_2)}, $$
with the form $\Omega^\alpha_{(\alpha_1,\alpha_2)}$ as obtained in Proposition~\ref{a2compute}.
The same change of variables \eqref{u-coords} transforms the domain of
integration given by the set $(0, \frac{\pi}{2}) \times \mathbb{S}^3$ appearing in the integral 
\eqref{a2prerationalform}, into the set
\begin{equation} \label{Aset}
A_4 := \{ (u_0, u_1, u_2, u_3, u_4) \in (0, 1)^3 \times \mathbb{R}^2:  \qquad u_1^2 + u_2^2 + u_0 u_3^2+(1-u_0) u_4^2 = 1\},
\end{equation} 
which is a $\Q$-semialgebraic set. The form $\Omega^\alpha$ is defined on the complement in $\A^5$
of the union of the hyperplanes $H_0$ and $H_1$ and the hypersurface $\widehat{CZ}_\alpha$ 
given by the vanishing of the quadric $Q_\alpha$ of \eqref{Qalpha}.
Thus, it is an algebraic differential form on the algebraic variety $\A^5 \smallsetminus (\widehat{CZ}_\alpha\cup H_0 \cup H_1)$.
The domain of integration is not a closed cycle: it has a boundary $\partial A$ which is 
contained in the union of the hyperplanes $H_{i,a}=\{ u_i = a \}$, with $i\in \{ 0,1,2 \}$ and $a\in \{ 0, 1 \}$.
Thus, the period corresponds to the relative motive 
$\m(\A^5 \smallsetminus (\widehat{CZ}_\alpha \cup H_0 \cup H_1),\Sigma)$, 
where the divisor is the union of these hyperplanes, $\Sigma=\cup_{i,a} H_{i,a}$.
\endproof

\medskip

\begin{rem}\label{nodivergence} {\rm 
The singular locus $\widehat{CZ}_\alpha \cup H_0 \cup H_1$ of the algebraic 
differential form and the divisor $\Sigma$ containing the boundary of the domain of
integration $A_4$ have nonempty intersection along $H_0 \cup H_1$. 
However, unlike the case of quantum
field theory where the intersection of the boundary of the domain of integration with the
graph hypersurface is the source of infrared divergences, here we know a priori that the 
integral \eqref{a2period} is convergent, 
and so are all the other analogous integrals for the higher order $a_{2n}$ terms,
as one can see by computing them in the original spherical coordinates. 
Thus, we do not have a renormalization problem for these integrals.  }
\end{rem}

\section{The $a_4$ term and quadric hypersurfaces in $\P^5$}\label{a4Sec}

In order to compute the term $a_4$ appearing in the asymptotic expansion 
\eqref{HeatExp}, we use formula \eqref{HeatCoefResidue} in the special 
case when $r$ is set equal to $2$. That is, we need to consider the 
operator 
\[
\Delta_4 = D^2 \otimes 1 + 1 \otimes \Delta_{\mathbb{T}^2},
\]
in which $\Delta_{\mathbb{T}^2}$ is the flat Laplacian on 
the $2$-dimensional torus $\mathbb{T}^2 = \left ( \mathbb{R}/\mathbb{Z} \right )^2$. 
The formula  \eqref{HeatCoefResidue} allows us to write 
\begin{equation*} 
a_{4} = \frac{1}{2^5\, \pi^{5}} \textnormal{Res}(\Delta_4^{-1}). 
\end{equation*}
Since the operator $\Delta_4$ and its parametrix $\Delta_4^{-1}$ act on 
the smooth sections of a vector bundle on a 6-dimensional manifold, in 
order to compute $\textnormal{Res}(\Delta_4^{-1})$, 
we first need to calculate $\textnormal{tr}(\sigma_{-6}(\Delta_4^{-1}))$, where $\sigma_{-6}(\Delta_4^{-1})$ is 
the term of order $-6$ in the expansion of the symbol of $\Delta_4^{-1}$. 

\smallskip

This can be done by preforming symbolic calculations as explained in 
\cite{FFMRationality}, which works for a general 
positive even integer $r$ and the operator 
\[
\Delta_{r+2} = D^2 \otimes 1 + 1 \otimes \Delta_{\mathbb{T}^r},
\]
in which $\Delta_{\mathbb{T}^r}$ is the flat Laplacian on 
the $r$-dimensional torus $\mathbb{T}^r = \left ( \mathbb{R}/\mathbb{Z} \right )^r$. 
In order to calculate the homogeneous terms in the expansion of the symbol of 
$\Delta_{r+2}^{-1}$, one can start by writing 
\begin{equation} \label{order-2}
\sigma_{-2}(\Delta_{r+2}^{-1}) =  \left  (p_2(x, \xi_1, \xi_2, \xi_3, \xi_4) + (\xi_5^2 + \cdots + \xi_{4+r}^2) I_{4\times 4} \right )^{-1}. 
\end{equation}
Then, the following formula, for $n>0$, can be used to calculate the next terms recursively:  
\begin{equation}
 \sigma_{-2-n}(\Delta_{r+2}^{-1})  = 
\end{equation}

\begin{eqnarray*} \label{order-n}
- \left ( \sum_{\substack{ 0 \leq j < n,\, 0 \leq k \leq 2 \\ \alpha \in \mathbb{Z}_{\geq 0}^4 
 \\-2 - j - |\alpha| + k = -n }} 
 \frac{(-i)^{|\alpha|}}{\alpha! } 
\left (\partial_{\xi}^{\alpha }  \sigma_{-2-j}(\Delta_{r+2}^{-1}) \right ) \left ( \partial_x^\alpha  p_k \right ) \right ) \sigma_{-2}(\Delta_{r+2}^{-1}). 
\end{eqnarray*}

\smallskip

\begin{thm}\label{a4compute}
The form $b_{-6}=b_{-6}(t, \eta, \xi_1, \dots, \xi_6)$ computing the 
$a_4$ term of the heat kernel expansion of $D^2$ is a rational differential form
$$ \Omega = f \, \widetilde\sigma_5, $$
in affine coordinates $(u_0,u_1,u_2,u_3,u_4,u_5,u_6)\in \A^7$, $\alpha\in \bG_m$,
and $(\alpha_1,\alpha_2,\alpha_3,\alpha_4)\in \A^4$,
where the functions $$ f (u_0,u_1,u_2,u_3,u_4,\alpha, \alpha_1, \alpha_2, \alpha_3,\alpha_4)= 
f_{(\alpha_1,\alpha_2, \alpha_3,\alpha_4)}(u_0,u_1,u_2,u_3,u_4,u_5,u_6, \alpha) $$ are 
$\Q$-linear combinations of rational functions of the form
$$ \frac{P(u_0,u_1,u_2,u_3,u_4,u_5,u_6,\alpha, \alpha_1, \alpha_2,\alpha_3,\alpha_4)}
{\alpha^{2r} u_0^k (1-u_0)^m (u_1^2 + \alpha^{-2} (u_2^2 + u_3^2 + u_4^2)+u_5^2 +u_6^2)^\ell}, $$
where
$$ P(u_0,u_1,u_2,u_3,u_4,u_5,u_6,\alpha, \alpha_1, \alpha_2,\alpha_3,\alpha_4) = P_{(\alpha_1, \alpha_2,\alpha_3,\alpha_4)}(u_0,u_1,u_2,u_3,u_4,u_5,u_6,\alpha) $$
are polynomials in $\Q[u_0,u_1,u_2,u_3,u_4,u_5,u_6\alpha, \alpha_1, \alpha_2,\alpha_3,\alpha_4]$ and 
where $r$, $k$, $m$ and $\ell$ are non-negative integers, and with 
$\widetilde\sigma_5=\widetilde\sigma_5(u_0,u_1,u_2,u_3,u_4,u_5,u_6)$ the form
\begin{eqnarray*}
\widetilde \sigma_5 & =&  \frac{1}{2} \Big ( u_1\, du_0 \, du_2 \, du_3 \, du_4 \, du_5 \, du_6 - 
u_2 \, du_0 \, du_1 \, du_3 \, du_4 \, du_5 \, du_6 \\
&&+ u_3\, du_0 \, du_1 \, du_2 \, du_4 \, du_5 \, du_6 
-  u_4 \,du_0 \, du_1 \, du_2 \, du_3 \, du_5 \, du_6 \\
&& +  u_5\, du_0 \, du_1 \, du_2 \, du_3 \, du_4 \, du_6 
-  u_6 \,du_0 \, du_1 \, du_2 \, du_3 \, du_4 \, du_5 \Big ).
\end{eqnarray*}
The forms $\Omega^\alpha=\Omega^\alpha_{(\alpha_1,\alpha_2,\alpha_3,\alpha_4)}$ 
obtained by restricting the above to a fixed value of 
$\alpha\in \A^1\smallsetminus \{ 0 \}$ are a four-parameter family 
of algebraic differential forms on the algebraic variety over $\Q$
given by the complement in $\A^7$ of the union of the affine hyperplanes 
$H_0=\{ u_0=0 \}$ and $H_1=\{ u_0=1 \}$ and the 
hypersurface $\widehat{CZ}_\alpha$ defined by 
the vanishing of the quadratic form
\begin{equation}\label{a4Qalpha}
 Q_{\alpha, 4} = u_1^2 + \alpha^{-2} (u_2^2 + u_3^2 + u_4^2) + u_5^2 + u_6^2. 
\end{equation} 
\end{thm}

\proof
As we explained earlier, for the calculation of the term $a_4$, we need to set 
$r=2$. In this case, after performing the algebraic calculations we find a lengthy 
expression for $\textnormal{tr}(\sigma_{-6}(\Delta_4^{-1}))$. Like the case of 
$a_2$,  the expression for  $\textnormal{tr}(\sigma_{-6}(\Delta_4^{-1}))$ has 
terms that have odd powers of $\xi_j$ in their numerators. Since these terms, 
following formulas \eqref{WodResDensity} and \eqref{WodResFormula},  
will vanish under the necessary integrations for calculating $a_4$, 
we eliminate them from $\textnormal{tr}(\sigma_{-6}(\Delta_4^{-1}))$ and 
denote the reduced expression by $b_{-6}$. Using the notation  
\[
\alpha = a(t), 
\]
\[ Q_{\alpha, 4} = \xi _1^2+\frac{\xi _2^2}{\alpha^2}+
\frac{\xi _3^2 \csc ^2(\eta )}{\alpha^2}+\frac{\xi _4^2 \sec ^2(\eta )}{\alpha^2}+\xi _5^2+\xi _6^2, 
\]
we find that 
\begin{equation} \label{b-6shortexpression}
b_{-6}=b_{-6}(t, \eta, \xi_1, \dots, \xi_6) = 
\end{equation}
\begin{center}
\begin{math}
\frac{2560 \xi _3^8 a'(t)^4 \csc ^8(\eta )}{\alpha ^{12} Q_{\alpha ,4}^7}-\frac{53760 \xi
   _1^2 \xi _3^8 a'(t)^4 \csc ^8(\eta )}{\alpha ^{12} Q_{\alpha ,4}^8}+\frac{107520 \xi
   _1^4 \xi _3^8 a'(t)^4 \csc ^8(\eta )}{\alpha ^{12} Q_{\alpha ,4}^9}-\frac{640 \xi _3^6
   a'(t)^2 \csc ^8(\eta )}{\alpha ^{10} Q_{\alpha ,4}^6}+\frac{5120 \cot ^2(\eta ) \xi
   _3^8 a'(t)^2 \csc ^8(\eta )}{\alpha ^{12} Q_{\alpha ,4}^7}+\frac{5760 \xi _1^2 \xi
   _3^6 a'(t)^2 \csc ^8(\eta )}{\alpha ^{10} Q_{\alpha ,4}^7}+\frac{5120 \xi _2^2 \xi
   _3^6 a'(t)^2 \csc ^8(\eta )}{\alpha ^{12} Q_{\alpha ,4}^7}+\frac{10240 \xi _3^6 \xi
   _4^2 a'(t)^2 \csc ^8(\eta )}{\alpha ^{12} Q_{\alpha ,4}^7}-\frac{53760 \cot ^2(\eta )
   \xi _1^2 \xi _3^8 a'(t)^2 \csc ^8(\eta )}{\alpha ^{12} Q_{\alpha ,4}^8}-\frac{53760
   \cot ^2(\eta ) \xi _2^2 \xi _3^8 a'(t)^2 \csc ^8(\eta )}{\alpha ^{14} Q_{\alpha
   ,4}^8}-\frac{53760 \xi _1^2 \xi _2^2 \xi _3^6 a'(t)^2 \csc ^8(\eta )}{\alpha ^{12}
   Q_{\alpha ,4}^8}-\frac{107520 \xi _1^2 \xi _3^6 \xi _4^2 a'(t)^2 \csc ^8(\eta
   )}{\alpha ^{12} Q_{\alpha ,4}^8}-\frac{107520 \xi _2^2 \xi _3^6 \xi _4^2 a'(t)^2 \csc
   ^8(\eta )}{\alpha ^{14} Q_{\alpha ,4}^8}+\frac{645120 \cot ^2(\eta ) \xi _1^2 \xi _2^2
   \xi _3^8 a'(t)^2 \csc ^8(\eta )}{\alpha ^{14} Q_{\alpha ,4}^9}+\frac{1290240 \xi _1^2
   \xi _2^2 \xi _3^6 \xi _4^2 a'(t)^2 \csc ^8(\eta )}{\alpha ^{14} Q_{\alpha
   ,4}^9}+\frac{112 \xi _3^4 \csc ^8(\eta )}{\alpha ^8 Q_{\alpha ,4}^5}-\frac{1664 \cot
   ^2(\eta ) \xi _3^6 \csc ^8(\eta )}{\alpha ^{10} Q_{\alpha ,4}^6}-\frac{1664 \xi _2^2
   \xi _3^4 \csc ^8(\eta )}{\alpha ^{10} Q_{\alpha ,4}^6}+\frac{2560 \cot ^4(\eta ) \xi
   _3^8 \csc ^8(\eta )}{\alpha ^{12} Q_{\alpha ,4}^7}+\frac{30080 \cot ^2(\eta ) \xi _2^2
   \xi _3^6 \csc ^8(\eta )}{\alpha ^{12} Q_{\alpha ,4}^7}+\frac{2560 \xi _2^4 \xi _3^4
   \csc ^8(\eta )}{\alpha ^{12} Q_{\alpha ,4}^7}-\frac{10240 \xi _3^6 \xi _4^2 \csc
   ^8(\eta )}{\alpha ^{12} Q_{\alpha ,4}^7}-\frac{53760 \cot ^4(\eta ) \xi _2^2 \xi _3^8
   \csc ^8(\eta )}{\alpha ^{14} Q_{\alpha ,4}^8}-\frac{53760 \cot ^2(\eta ) \xi _2^4 \xi
   _3^6 \csc ^8(\eta )}{\alpha ^{14} Q_{\alpha ,4}^8}+\frac{215040 \xi _2^2 \xi _3^6 \xi
   _4^2 \csc ^8(\eta )}{\alpha ^{14} Q_{\alpha ,4}^8}+\frac{107520 \cot ^4(\eta ) \xi
   _2^4 \xi _3^8 \csc ^8(\eta )}{\alpha ^{16} Q_{\alpha ,4}^9}-\frac{430080 \xi _2^4 \xi
   _3^6 \xi _4^2 \csc ^8(\eta )}{\alpha ^{16} Q_{\alpha ,4}^9}-\frac{2752 \xi _3^6
   a'(t)^4 \csc ^6(\eta )}{\alpha ^{10} Q_{\alpha ,4}^6}+\frac{47040 \xi _1^2 \xi _3^6
   a'(t)^4 \csc ^6(\eta )}{\alpha ^{10} Q_{\alpha ,4}^7} + \cdots .
\end{math}
\end{center}
Since the full expression for $b_{-6}$ is quite lengthy, we have recorded it 
in Appendix A.  

\smallskip

In order to express $b_{-6}$ as a rational function, we introduce the following 
coordinates: 
\begin{equation} \label{u-coords6}
u_0= \sin^2(\eta), \qquad u_1 = \xi_1, \qquad u_2 = \xi_2, 
\end{equation}
\[
u_3 = \csc(\eta) \, \xi_3, \qquad u_4 = \sec(\eta) \, \xi_4, \qquad u_5 = \xi_5, \qquad u_6 = \xi_6.
\]
First, let us note that in the new coordinates we have
\begin{eqnarray*}
\widetilde \sigma_5 &:=& d\eta \wedge \sigma_{\xi, \,5} \\
&=& \sum_{j=1}^6 (-1)^{j-1} \xi_j \,d\eta \wedge d\xi_1 
\wedge \cdots \wedge {\widehat d \xi_j} \wedge \cdots \wedge d \xi_6 \\
&=& \frac{1}{2} \Big ( u_1\, du_0 \, du_2 \, du_3 \, du_4 \, du_5 \, du_6 - 
u_2 \, du_0 \, du_1 \, du_3 \, du_4 \, du_5 \, du_6 \\
&&+ u_3\, du_0 \, du_1 \, du_2 \, du_4 \, du_5 \, du_6 
-  u_4 \,du_0 \, du_1 \, du_2 \, du_3 \, du_5 \, du_6 \\
&& +  u_5\, du_0 \, du_1 \, du_2 \, du_3 \, du_4 \, du_6 
-  u_6 \,du_0 \, du_1 \, du_2 \, du_3 \, du_4 \, du_5 \Big ).
\end{eqnarray*}
Finally, the rationality of $b_{-6}$ in the new coordinates \eqref{u-coords6} follows from the fact 
that the exponents of the $\xi_j$ appearing in the numerators in the expression of $b_{-6}$ are 
non-negative even integers, and all trigonometric terms in the expression, which are listed below, 
transform to the following rational expressions in the $u_j$: 
\[
\xi _1^2+\frac{\xi _2^2}{a(t)^2}+
\frac{\xi _3^2 \csc ^2(\eta )}{a(t)^2}+\frac{\xi _4^2 \sec ^2(\eta )}{a(t)^2}+\xi _5^2+\xi _6^2
= u_1^2 + \frac{1}{a(t)^2}( u_2^2 + u_3^2+u_4^2  ) + u_5^2 + u_6^2, 
\]
\[
\cot^2(\eta) = \frac{1-u_0}{u_0}, 
\]
\[
\csc^2(\eta) = \frac{1}{u_0}, 
\]
\[
\sec^2(\eta) = \frac{1}{1-u_0}, 
\]
\[
\cot ^3(\eta ) \cot (2 \eta ) = \frac{\cot ^4(\eta )}{8}-\frac{3 \cot ^2(\eta )}{4}+\frac{\csc ^4(\eta )}{8}-\frac{\csc ^2(\eta )}{4}+\frac{1}{4} \cot ^2(\eta ) \csc ^2(\eta )+\frac{1}{8}, 
\]
\[
\cot (2 \eta ) \csc ^5(\eta ) \sec (\eta ) = \frac{\csc ^6(\eta )}{2}-\frac{1}{2} \csc ^4(\eta ) \sec ^2(\eta ), 
\]
\[
\cot (2 \eta ) \csc (\eta ) \sec ^5(\eta ) =  \frac{1}{2} \csc ^2(\eta ) \sec ^4(\eta )-\frac{\sec ^6(\eta )}{2}, 
\]
\[
\cot ^2(2 \eta )= \frac{\tan ^2(\eta )}{8}+\frac{\cot ^2(\eta )}{8}+\frac{1}{8} \csc ^2(\eta ) \sec ^2(\eta )-\frac{3}{4}, 
\]
\[
\cot (\eta ) \cot (2 \eta )= \frac{\cot ^2(\eta )}{2}-\frac{1}{2}, 
\]
\[
\cot (2 \eta ) \csc ^3(\eta ) \sec (\eta ) = 
\frac{\csc ^4(\eta )}{2}-\frac{1}{2} \csc ^2(\eta ) \sec ^2(\eta ), 
\]
\[
\csc ^2(2 \eta ) = \frac{1}{4} \csc ^2(\eta ) \sec ^2(\eta ), 
\]
\[
\cot (2 \eta ) \csc (\eta ) \sec ^3(\eta ) = \frac{1}{2} \csc ^2(\eta ) \sec ^2(\eta )-\frac{\sec ^4(\eta )}{2}, 
\]
\[
\tan (\eta ) \cot (2 \eta ) = \frac{1}{2}-\frac{\tan ^2(\eta )}{2},
\]
\[
\tan (\eta ) \cot ^3(2 \eta ) = 
\]
\[
-\frac{\tan ^4(\eta )}{32}+\frac{15 \tan ^2(\eta )}{32}+\frac{\cot ^2(\eta )}{32}-\frac{1}{32} 3 \sec ^4(\eta )+\frac{3}{32} \csc ^2(\eta ) \sec ^2(\eta )-\frac{15}{32}. 
\]

\smallskip
We note that, in the case of $a_2$, the rational expression has $a(t), a'(t), a''(t)$ 
appearing in the numerators, while in the case of $a_4$ we have $a(t), a'(t), a''(t), a^{(3)}(t), 
a^{(4)}(t)$ appearing in the numerators.  As in the $a_2$ case, we treat these as variables
$(\alpha_1,\ldots, \alpha_4) $ \,\,$\in \A^4$. Since the denominators do not depend on the
variables $(\alpha_1,\ldots, \alpha_4)$, we can regard the forms $\Omega^\alpha$ as
a four-parameter family of algebraic differential forms defined over the same algebraic
variety, given by the complement in the affine space $\A^7$ of the union of the hyperplanes $H_0$ and $H_1$ and the variety $\widehat{CZ}_\alpha$ determined by the vanishing of the quadratic form 
$$  Q_{\alpha, 4} = u_1^2 + \alpha^{-2} (u_2^2 + u_3^2 + u_4^2) + u_5^2 + u_6^2 . $$
We view $\widehat{CZ}_\alpha$ as the affine cone in $\A^7$ over the projective cone $CZ_\alpha$
in $\P^6$ of the quadric $Z_\alpha$ in $\P^5$ defined by the vanishing of $Q_\alpha$.
\endproof

\medskip

\begin{thm}\label{a4intperiod}
When computed without performing the time integration, the $a_4$ term in the heat 
kernel expansion of $D^2$ is a period integral 
\begin{equation}\label{a4period}
 C \cdot \int_{A} \Omega^\alpha_{(\alpha_1,\alpha_2, \alpha_3, \alpha_4)}
\end{equation} 
with the algebraic differential forms $\Omega^\alpha_{(\alpha_1,\alpha_2, \alpha_3, \alpha_4)}$ as in 
Proposition~\ref{a4compute}, and with 
domain of integration the $\Q$-semialgebraic set
\begin{equation}\label{a4A}
  A_6 = \left\{ (u_0,\ldots, u_6)\in \A^7(\R)\,:\, \begin{array}{l} 
u_1^2 + u_2^2 + u_0 u_3^2 + (1-u_0) u_4^2 + u_5^2 + u_6^2 =1  \\ 
0< u_i <1, \ \ \  i=0,1,2,5,6 \end{array} \right\}. 
\end{equation}
The coefficient $C$ is in $\Q[(2\pi i)^{-1}]$.
This integral is a period of the mixed motive 
\begin{equation}\label{MMa4}
\m(\A^7\smallsetminus (\widehat{CZ}_\alpha \cup H_0 \cup H_1), \Sigma), 
\end{equation}
where $\widehat{CZ}_\alpha$ is the hypersurface in $\A^7$ defined by the vanishing of the quadric 
$Q_\alpha$ of \eqref{a4Qalpha}, and $\Sigma=\cup_{i,a} H_{i,a}$
is the divisor given by the union of the hyperplanes $H_{i,a}=\{ u_i = a \}$, with $i\in \{ 0,1,2,5,6 \}$ and 
$a\in \{ 0, 1 \}$.
\end{thm}

\proof The term $a_4$ is computed as
$$ a_{4} = \frac{1}{2^5\, \pi^{5}} \textnormal{Res}(\Delta_4^{-1}) . $$
Therefore, before performing the integration in the time variable $t$, we can write
explicitly the expression for the term $a_4$ as 
\begin{eqnarray} \label{a4prerationalform}
a_{4} &=&  \frac{1}{2^5 \pi^5} \int_0^{\pi/2} d \eta \int_0^{2 \pi} d\phi_1 \int_0^{2 \pi} d\phi_2 
 \int_{\mathbb{T}^2} \, d^2x'
\int_{\xi_1^2+\xi_2^2+\xi_3^2+\xi_4^2+\xi_5^2+\xi_6^2 = 1} d^5\xi \, \cdot b_{-6} \,\cdot  \sigma_{\xi, \, 5} \nonumber \\
 &=&  \frac{1}{2^3 \pi^3} \int_0^{\pi/2} d \eta  
\int_{\xi_1^2+\xi_2^2+\xi_3^2+\xi_4^2+\xi_5^2+\xi_6^2 = 1} d^5\xi \, \cdot b_{-6}(t, \eta, \xi_1, \dots, \xi_6) \,\cdot  \sigma_{\xi, \,5} \nonumber \\
&=& 
 \frac{1}{2^3 \pi^3} \int_{(0, \frac{\pi}{2}) \times \mathbb{S}^5}  
  b_{-6}(t, \eta, \xi_1, \dots, \xi_6) 
\sum_{j=1}^6 (-1)^{j-1} \xi_j \,d\eta \wedge d\xi_1 
\wedge \cdots \wedge {\widehat d \xi_j} \wedge \cdots \wedge d \xi_6. \nonumber \\
\end{eqnarray}
As in Proposition~\ref{a4compute}, we perform the change of variables \eqref{u-coords6}.
In these coordinates the domain of integration $(0, \frac{\pi}{2}) \times \mathbb{S}^5$ 
in \eqref{a4prerationalform} transforms to the set 
\begin{equation} \label{A6set}
A_6 := 
\{ (u_0, \ldots, u_6) \in (0, 1)^3 \times \mathbb{R}^2 \times (0, 1)^2;   u_1^2 + u_2^2 + u_0 u_3^2+(1-u_0) u_4^2 + u_5^2 + u_6^2= 1\},
\end{equation}
which is a $\Q$-semialgebraic set as in \eqref{a4A}. The boundary $\partial A$ of the domain
of integration is contained in the divisor $\Sigma=\cup_{i,a} H_{i,a}$, with
$i\in \{ 0,1,2,5,6 \}$ and $a\in \{ 0, 1 \}$. For any fixed $\alpha \in \bG_m$, and 
for any choice of $(\alpha_1,\ldots,\alpha_4)\in \A^4$, 
the forms $\Omega^\alpha_{(\alpha_1,\alpha_2, \alpha_3, \alpha_4)}$ are algebraic
differential forms defined on the complement in $\A^7$ of  the union of $\widehat{CZ}_\alpha$
and $H_0 \cup H_1$,
hence the integrals in \eqref{a4period} are periods of the motive 
$\m(\A^7\smallsetminus (\widehat{CZ}_\alpha \cup H_0 \cup H_1), \Sigma)$.
\endproof

\medskip

Again, as mentioned in Remark~\ref{nodivergence}, although the 
singular set of the algebraic differential form intersects the divisor $\Sigma$, 
the integral is convergent, hence there
is no renormalization issue involved. 

\medskip

\section{The $a_{2n}$ term}\label{a2nSec}

We start by using \eqref{HeatCoefResidue} to express the term $a_{2n}$ as the 
noncommutative residue of an operator, namely 
\begin{equation} \label{a_2nncresidue}
a_{2n} = \frac{1}{2^5\, \pi^{3+n}} \textnormal{Res}(\Delta_{2n}^{-1}),
\end{equation}
where 
\[
\Delta_{2n} = D^2 \otimes 1 + 1 \otimes \Delta_{\mathbb{T}^{2n-2}}. 
\]
Here $\Delta_{\mathbb{T}^{2n-2}}$ is the Laplacian of the flat metric on the 
torus of dimension $2n-2$. Since $\Delta_{2n}$ acts on the smooth sections 
of a vector bundle on a manifold of dimension $2n+2$, in order to compute 
$\textnormal{Res}(\Delta_{2n}^{-1})$ one needs to calculate 
the term $\sigma_{-2n-2}$ which is homogeneous of order $-2n-2$ in the 
expansion of the pseudodifferential symbol of $\Delta_{2n}^{-1}$. 

%%%%%
\begin{thm}\label{a2nsigma}
The term $\sigma_{-2n-2}$ satisfies
\begin{equation} \label{trsigma-2n-2} 
\textnormal{tr}(\sigma_{-2n-2} ) 
= 
\end{equation}
\[
\sum_{j=1}^{M_n} c_{j, 2n }\,
 u_0^{\beta_{0,1, j}/2} \, (1-u_0)^{\beta_{0,2, j}/2}\, 
 \frac{u_1^{\beta_{1, j}} \, u_2^{\beta_{2,j}} \, \cdots \, u_{2n+2}^{\beta_{2n+2, j}}}{  Q_{\alpha, 2n} ^{\rho_{j,2n}}} \,
 \alpha^{k_{0,j}} \, \alpha_1^{k_{1,j}} \, \cdots \, \alpha_{2n}^{k_{2n,j}}, 
\]
where 
\[
\alpha = a(t), \qquad \alpha_1 = a'(t), \qquad \alpha_2= a''(t), \qquad  \dots, \qquad \alpha_{2n} = a^{2n}(t), 
\]

\[
Q_{\alpha,2n} = u_1^2 + \frac{1}{\alpha^2} (u_2^2 + u_3^2 +u_4^2) + u_5^2 + \cdots + u_{2n+2}^2, 
\]

\[
c_{j, 2n} \in \Q, \qquad \beta_{0,1, j}, \beta_{0,2, j}, k_{0,j} \in \Z, \qquad \beta_{1,j}, \dots, \beta_{2n+2, j}, 
\rho_{j, 2n}, k_{1,j}, \dots, k_{2n,j}  \in \Z_{\geq 0}. 
\]
\end{thm}

\proof The argument is similar to our 
treatment of the terms $a_2$ and $a_4$. In order to finally see rational differential forms  
that integrate to the term $a_{2n}$, 
it will be useful to perform calculations in the following coordinates: 
\begin{equation} \label{u-coords2nplus2}
u_0= \sin^2(\eta), \qquad u_3 = \csc(\eta) \, \xi_3, \qquad u_4 = \sec(\eta) \, \xi_4,   
\end{equation}
\[
u_j = \xi_j,  \qquad j=1, 2, 5, 6, \dots, 2n+2. 
\]

\smallskip
In this regard, we use the new coordinates for writing the pseudodifferential symbol of the Dirac operator $D$ 
of the Robertson-Walker metric, namely that we rewrite 
\[
\sigma(D) = q_1 + q_0, 
\]
where $q_1$ and $q_0$ are now expressed as 
\[
q_1 =
\left(
\begin{array}{cccc}
 0 & 0 & \frac{i u_4}{a(t)}-u_1 & \frac{i u_2}{a(t)}+\frac{u_3}{a(t)} \\
 0 & 0 & \frac{i u_2}{a(t)}-\frac{u_3}{a(t)} & -u_1-\frac{i u_4}{a(t)} \\
 -u_1-\frac{i u_4}{a(t)} & -\frac{i u_2}{a(t)}-\frac{u_3}{a(t)} & 0 & 0 \\
 \frac{u_3}{a(t)}-\frac{i u_2}{a(t)} & \frac{i u_4}{a(t)}-u_1 & 0 & 0 \\
\end{array}
\right), 
\]

\[
q_0 = \left(
\begin{array}{cccc}
 0 & 0 & \frac{3 i a'(t)}{2 a(t)} & \frac{1-2 u_0}{2 a(t) \sqrt{\left(1-u_0\right) u_0}} \\
 0 & 0 & \frac{1-2 u_0}{2 a(t) \sqrt{\left(1-u_0\right) u_0}} & \frac{3 i a'(t)}{2 a(t)} \\
 \frac{3 i a'(t)}{2 a(t)} & -\frac{1-2 u_0}{2 a(t) \sqrt{\left(1-u_0\right) u_0}} & 0 & 0 \\
 -\frac{1-2 u_0}{2 a(t) \sqrt{\left(1-u_0\right) u_0}} & \frac{3 i a'(t)}{2 a(t)} & 0 & 0 \\
\end{array}
\right).
\]
Since $q_1$ and $q_0$ depend only on $t$ and $u_0$, or equivalently only on $t$ and $\eta$, 
for the symbol of $D^2$, we have
\[
\sigma(D^2) = p_2 + p_1 +p_0, 
\]
where 
\begin{eqnarray*}
p_2 &=& q_1^2 = \left (u_1^2 + \frac{1}{a(t)^2}(u_2^2 + u_3^2 + u_4^2) \right) I_{4\times4}, 
\\
p_1
&=& 
q_0 \, q_1 + q_1 \, q_0 + 
\left( -i \frac{\partial q_1}{\partial \xi_1} \, \frac{\partial q_1}{\partial t}  
-i \frac{\partial q_1}{\partial \xi _2} \, \frac{\partial q_1}{\partial \eta} 
\right ), \\
p_0 &=& 
q_0^2 + \left ( -i \frac{\partial q_1}{\partial \xi _1} \, \frac{\partial q_0}{\partial t}
-i \frac{\partial q_1}{\partial \xi _2} \, \frac{\partial q_0}{\partial \eta} 
\right ). 
\end{eqnarray*}

\smallskip
It is possible to keep track of the general form of the expressions that will appear in the  
calculations, when written in the new coordinates given by \eqref{u-coords2nplus2}, as 
follows. We achieve this by using the fact that  if, 
for a smooth function $f$ of our variables,  we use the notation 
\[
f(t, \eta, \xi_1, \xi_2, \dots, \xi_{2n+2}) 
=
\tilde f(t, u_0, u_1, u_2, \dots, u_{2n+2}), 
\]
then 
\begin{eqnarray} \label{chainrule}
&& \partial_t f = \partial_t \tilde f, \qquad \partial_{\xi_j} f = \partial_{u_j} \tilde f, \qquad j=1,2, \nonumber \\
&& 
\partial_\eta f =  2 \sqrt{u_0(1-u_0)} \,  \partial_{u_0} \tilde f 
- u_3 \sqrt{\frac{1-u_0}{u_0}} \,  \partial_{u_3} \tilde f 
+ u_4 \sqrt{\frac{u_0}{1-u_0}} \,  \partial_{u_4} \tilde f.  
\end{eqnarray}

We can now focus on the homogeneous terms $\sigma_{-2-j}(\Delta_{2n}^{-1})$ of order $-2-j$ 
in the expansion of the pseudodifferential symbol of the parametrix of $\Delta_{2n}$. As mentioned 
earlier, the relevant term for the calculation of $a_{2n}$ is the term $\sigma_{-2n-2}(\Delta_{2n}^{-1})$.  
By using \eqref{order-n} and considering the independence of the symbols from the variables 
$\phi_1$ and $\phi_2$, we have  
\begin{equation} 
\sigma_{-2}(\Delta_{2n}^{-1}) 
=  
\left  (p_2 + (u_5^2 + \cdots + u_{2n+2}^2) I_{4\times 4} \right )^{-1} = \frac{1}{Q_{\alpha, 2n}} I_{4\times4}, 
\end{equation}
where 
\begin{equation}\label{1Qalpha2n}
Q_{\alpha,2n} = u_1^2 + \frac{1}{\alpha^2} (u_2^2 + u_3^2 +u_4^2) + u_5^2 + \cdots + u_{2n+2}^2,  
\end{equation}
and the desired $\sigma_{-2n-2}(\Delta_{2n}^{-1})$ can be calculated recursively:  
\begin{equation}
 \sigma_{-2n-2}(\Delta_{2n}^{-1})  = 
\end{equation}
\begin{eqnarray*} \label{order-n}
- \left ( \sum_{\substack{ 0 \leq j < 2n,\, 0 \leq k \leq 2 \\ \ell_1, \ell_2 \in \mathbb{Z}_{\geq 0} 
 \\-2 - j - |\alpha| + k = -2n }} 
 \frac{(-i)^{\ell_1+\ell_2}}{\ell_1! \, \ell_2! } 
\left (\partial_{\xi_1}^{\ell_1 } \partial_{\xi_2}^{\ell_2 }  \sigma_{-2-j}(\Delta_{2n}^{-1}) \right ) 
\left ( \partial_t^{\ell_1} \partial_\eta^{\ell_2}  p_k \right ) \right ) \sigma_{-2}(\Delta_{2n}^{-1}). 
\end{eqnarray*} 
Combining these formulas with the equations given by \eqref{chainrule}, one can see by induction that 
\eqref{trsigma-2n-2} holds as stated.
\endproof

\medskip

\begin{thm}\label{a2nalgformset}
When computed without performing the time integration, the coefficient $a_{2n}$ is a period integral
$$ C \cdot \int_{A_{2n}} \Omega^\alpha_{\alpha_1, \ldots, \alpha_{2n}} $$
of an algebraic differential form $\Omega^\alpha_{\alpha_1, \ldots, \alpha_{2n}} (u_0,u_1,\ldots,u_{2n+2})$
defined on the complement $\A^{2n+3}\smallsetminus (\widehat{CZ}_{\alpha,2n} \cup H_0 \cup H_1)$, with 
$\widehat{CZ}_{\alpha,2n}$ the hypersurface defined by the vanishing of the quadric $Q_{\alpha,2n}$ of
\eqref{1Qalpha2n}, and hyperplanes $H_0=\{ u_0=0 \}$ and $H_1=\{ u_0=1 \}$. 
The period integral is performed over the $\Q$-semialgebraic set 
\begin{equation}\label{a2nA}
  A_{2n+2} =   
\end{equation}
\[
\left\{ (u_0,  \ldots, u_{2n+2})\in \A^{2n+3}(\R)\,:\, \begin{array}{l} 
u_1^2 + u_2^2 + u_0 u_3^2 + (1-u_0) u_4^2 + \sum_{i=5}^{2n+2} u_i^2 =1  \\ 
0< u_i <1, \ \ \  i=0,1,2,5,6, \dots , 2n+2 \end{array} \right\}.
\]
\end{thm}

\proof
Having obtained in Proposition~\ref{a2nsigma} a general form for 
$\textnormal{tr}(\sigma_{-2n-2} )$, we use the explicit definition of the 
Wodzicki residue, while leaving out an integration with respect to $t$, 
to expand the formula \eqref{a_2nncresidue} 
for the term $a_{2n}$ and we write: 
\begin{equation} \label{a2nprerationalform}
a_{2n}
\end{equation}

\begin{eqnarray} 
&=& \frac{1}{2^5 \pi^{3+n}} \int_0^{\pi/2} d \eta \int_0^{2 \pi} d\phi_1 \int_0^{2 \pi} d\phi_2 
 \int_{\mathbb{T}^{2n-2}} \, dx'
\int_{\sum_{j=1}^{2n+2} \xi_j^2 = 1} d \xi \, \cdot \textnormal{tr}( \sigma_{-2n-2} ) \,\cdot  \sigma_{\xi, \, 2n+1} \nonumber \\
 &=&  \frac{1}{2^3 \pi^{1+n}} \int_0^{\pi/2} d \eta  
\int_{\sum_{j=1}^{2n+2} \xi_j^2 = 1} d\xi \, \cdot \textnormal{tr} (\sigma_{-2n-2}) \,\cdot  \sigma_{\xi, \,2n+1} \nonumber 
\end{eqnarray}
\begin{eqnarray}
&=& 
 \frac{1}{2^3 \pi^{1+n}} \int_{(0, \frac{\pi}{2}) \times \mathbb{S}^{2n+1}}  
 \textnormal{tr}(\sigma_{-2n-2}) 
\sum_{j=1}^{2n+2} (-1)^{j-1} \xi_j \,d\eta \wedge d\xi_1 
\wedge \cdots \wedge {\widehat d \xi_j} \wedge \cdots \wedge d \xi_{2n+2} \nonumber \\
&=& 
 \frac{1}{2^3 \pi^{1+n}} \int_{(\eta, \, \xi) \in (0, \frac{\pi}{2}) \times \mathbb{S}^{2n+1}}  
 \textnormal{tr}(\sigma_{-2n-2})  \, \widetilde \sigma_{2n+1}, \nonumber 
\end{eqnarray}
where 
\begin{eqnarray*}
\widetilde \sigma_{2n+1} &:=& d\eta \wedge \sigma_{\xi, \,2n+1} \\
&=& \sum_{j=1}^{2n+2} (-1)^{j-1} \xi_j \,d\eta \wedge d\xi_1 
\wedge \cdots \wedge {\widehat d \xi_j} \wedge \cdots \wedge d \xi_{2n+2}\\
&=& \frac{1}{2} \sum_{j=1}^{2n+2} (-1)^{j-1} u_j \, du_0 \, du_1 \wedge \, \cdots \,  \wedge\widehat du_j 
\wedge \, \cdots \, \wedge du_{2n+2} .
\end{eqnarray*}

\smallskip
Considering the general form of $\textnormal{tr}(\sigma_{-2n-2})$ given by \eqref{trsigma-2n-2}, 
which is a rational expression in $\sqrt{u_0}$, $\sqrt{1-u_0}$, $u_1, \dots, u_{2n+2}$, $\alpha, \alpha_1, 
\dots, \alpha_{2n}$, in order to prove that $a_{2n}$ is the integral of a rational differential form, our next 
task is apparently to argue that only the terms that have even powers of $\sqrt{u_0}$ and $\sqrt{1-u_0}$ 
involved contribute to the calculation of the term $a_{2n}$. This will in particular show that 
the last integral expression given for $a_{2n}$ by \eqref{a2nprerationalform} is equal to 
the integral of a rational differential form in $u_0$, $u_1$, $\dots,$ $u_{2n+2}$, $\alpha, \alpha_1, 
\dots, \alpha_{2n}$ over the $\Q$-semialgebraic set \eqref{a2nA}.

\smallskip 

The claim that only the terms with $\beta_{0,1,j}, \beta_{0,2,j} \in 2 \Z$ in the summation given 
by \eqref{trsigma-2n-2}  contribute to the calculation of the term 
$a_{2n}$ can be proved as follows. Fixing a point $(x, x')=(t, \eta, \phi_1, \phi_2, x') \in M \times \mathbb{T}^{2n-2}$, 
the differential form  $\textnormal{tr}(\sigma_{-2n-2}) \, \sigma_{\xi,\, 2n+1}$ on the Euclidean space $\R^{2n+2} \simeq T^*_{(x, x')} (M \times \mathbb{T}^{2n-2})$ is a closed differential form of degree $2n+1$. The reason is that 
$\textnormal{tr}(\sigma_{-2n-2})$ is homogeneous of order $-2n-2$ in $\xi \in \R^{2n+2}$, cf. Proposition 
7.3  on page 265 of \cite{GVFbook}. Therefore, using the Stokes theorem, the integral of this differential form over the 
unit sphere $|\xi| =1$ is the same as its integral  over the cosphere of the metric in the cotangent 
fibre given by 
\begin{eqnarray*}
|\xi|_g^2 
&=&  
\xi _1^2+\frac{\xi _2^2}{a(t)^2}+\frac{\csc ^2(\eta ) \xi _3^2}{a(t)^2}+\frac{\sec ^2(\eta ) \xi _4^2}{a(t)^2} 
+ \xi_5^2 + \cdots + \xi_{2n+2}^2 \\
&=& u_1^2 + \frac{1}{\alpha^2}(u_2^2 + u_3^2 + u_4^2) + u_5^2 + \cdots + u_{2n+2}^2 \\
&=& 1. 
\end{eqnarray*}
We parametrize the cosphere $|\xi|_g=1$ by writing 
\begin{eqnarray} \label{metriccospherepara}
\xi_1&=& \sin (\psi_{2n+1}) \sin (\psi_{2n}) \, \cdots \, \sin (\psi_{2})  \cos (\psi_1),  \nonumber \\
\xi_2 &=& \alpha \sin (\psi_{2n+1}) \sin (\psi_{2n}) \, \cdots \, \sin (\psi_{2})  \sin (\psi_1), \nonumber \\
\xi_3 &=& \frac{\alpha}{\csc(\eta)} \sin (\psi_{2n+1}) \sin (\psi_{2n}) \, \cdots \, \sin (\psi_{3}) \cos (\psi_{2}),  \nonumber \\
\xi_4 &=&  \frac{\alpha}{\sec(\eta)} \sin (\psi_{2n+1}) \sin (\psi_{2n})\, \cdots \, \sin (\psi_{4}) \cos (\psi_{3}),
\end{eqnarray}
\begin{eqnarray}
\xi_5 &=& \sin (\psi_{2n+1}) \sin (\psi_{2n})\, \cdots \, \sin (\psi_{5}) \cos (\psi_{4}), \nonumber  \\
\xi_6 &=& \sin (\psi_{2n+1}) \sin (\psi_{2n})\, \cdots \, \sin (\psi_{6}) \cos (\psi_{5}), \nonumber \\
&\cdots&  \nonumber \\
\xi_{2n+1} &=& \sin (\psi_{2n+1}) \cos (\psi_{2n}), \nonumber \\
\xi_{2n+2} &=& \cos (\psi_{2n+1}),  \nonumber
\end{eqnarray}
with the variables $\psi_1, \dots, \psi_{2n+1}$ having the following ranges: 
\[
0 < \psi_1 < 2 \pi, \qquad 0 < \psi_2 <  \pi, \qquad 0 < \psi_3 <  \pi,  \qquad \dots, \qquad 0 < \psi_{2n+1} <  \pi. 
\]
Using this parametrization, over the cosphere $|\xi|_g=1$, we have 
\begin{eqnarray*}
\sigma_{\xi,\,2n+1} &=& \sum_{j=1}^{2n+2} (-1)^{j-1} \xi_j \, d\xi_1 
\wedge \cdots \wedge {\widehat d \xi_j} \wedge \cdots \wedge d \xi_{2n+2} \\
&=& \alpha^3 \sin(\eta) \cos (\eta) \sin (\psi_2) \sin^2(\psi_3) \, \cdots \, \sin^{2n} (\psi_{2n+1}) 
\, d \psi_1 \, d \psi_2 \, \cdots \, d\psi_{2n+1}. 
\end{eqnarray*}
Combining this form with the expression given by \eqref{trsigma-2n-2}, we conclude that  over 
$|\xi|_g=1$ we have: 
\[
\textnormal{tr}(\sigma_{-2n-2} ) \, \sigma_{\xi, \, 2n+1} =
\]
\[ 
\sin (\eta) \cos (\eta) \sum_{j=1}^{M_n}  \Big \{ c_{j, 2n }\,
  \alpha^{\beta_{2,j}+\beta_{3,j}+\beta_{4,j}+k_{0,j}}   \ \, \alpha_1^{k_{1,j}} \, \cdots \, \alpha_{2n}^{k_{2n,j}} 
  \sin^{\beta_{0,1, j}}(\eta) \, \cos^{\beta_{0,2, j}}(\eta) 
  \]
  \[
\cos^{\beta_{1,j}} (\psi_1) \sin^{\beta_{2,j}}(\psi_1)
\prod_{\ell=2}^{2n+1} \left ( \sin \psi_\ell \right )^{\ell - 1+ \sum_{i=1}^\ell \beta_{i, j} } \left ( \cos \psi_\ell \right )^{\beta_{\ell+1,j}} 
\Big \} \, d \psi_1 \, d \psi_2 \, \cdots \, d\psi_{2n+1}. 
\]

\smallskip 

By exploiting symmetries of the Robertson-Walker metric and its consequent rich isometry group, it is shown 
in Lemma 1 of \cite{FGK} that the local invariant that integrates to the term $a_{2n}$ has a spatial 
independence.  
This in particular means that the following expression is independent of the variable $\eta$:
\begin{eqnarray} \label{sumindepofeta}
&& \frac{1}{\sin (\eta) \cos (\eta)}
\int_{|\xi|_g=1}
\textnormal{tr}(\sigma_{-2n-2} ) \, \sigma_{\xi, \,2n+1} \nonumber \\
&=& 
 \sum_{j=1}^{M_n}   c_{j, 2n }\, d_{j, 2n} \, 
  \alpha^{\beta_{2,j}+\beta_{3,j}+\beta_{4,j}+k_{0,j}}   \, \alpha_1^{k_{1,j}} \, \cdots \, \alpha_{2n}^{k_{2n,j}} 
  \sin^{\beta_{0,1, j}}(\eta) \, 
  \cos^{\beta_{0,2, j}}(\eta),  
  \end{eqnarray}
  where 
  \begin{eqnarray*}
  d_{j, 2n} &=& 
  \int_{0}^{2 \pi} \cos^{\beta_{1,j}} (\psi_1) 
  \sin^{\beta_{2,j}}(\psi_1) \, d \psi_1 \times \\
&& \qquad \int_{0}^{\pi} d \psi_2 \cdots \int_{0}^{\pi} d \psi_{2n+1} 
\prod_{\ell=2}^{2n+1} \left ( \sin \psi_\ell \right )^{\ell - 1+ \sum_{i=1}^\ell \beta_{i, j} } \left ( \cos \psi_\ell \right )^{\beta_{\ell+1,j}}. 
\end{eqnarray*}

\smallskip 
We now exploit the independence from $\eta$ of the sum given by \eqref{sumindepofeta}  to 
show that only the terms in \eqref{trsigma-2n-2} for which $\beta_{0, 1, j}$ 
and $\beta_{0,2, j}$ are both even integers contribute to the calculation of the heat coefficient $a_{2n}$. 
This follows easily from the fact that if for some coefficients $c_j$ and some integers 
$\gamma_j$ and $\nu_j$, a finite  summation of the form $\sum_j c_j \sin^{\gamma_j}(\eta) \cos^{\nu_j}(\eta) $
is identically equal to a non-zero constant, or without loss in generality equal to $1$,   then all the 
exponents $\gamma_j$ and $\nu_j$ are even integers in the sense that possible terms 
with odd exponents involved have to inevitably cancel each other out.  Since $\eta$ varies between $0$ and $\pi/2$, this is 
equivalent to saying that if 
\[
\sum_j c_j \,s^{\gamma_j} \,(1-s^2)^{\nu_j/2} =1, \qquad s \in (0, 1), 
\]
then all $\gamma_j$ and $\nu_j$ are even integers in the mentioned sense. It is useful to point out that a 
simple replacement of 
$s$ in the above equation by $s_1= (1-s^2)^{1/2}$ shows that our claim is symmetric with respect to the $\gamma_j$ and $\nu_j$ (therefore it suffices to argue that all $\gamma_j$ are even integers).  
We decompose the summation on the left side of the above identity and write 
\[
\sum_{j_o} c_{j_o} \,s^{\gamma_{j_o} }\,(1-s^2)^{\nu_{j_o}/2}+
\sum_{j_e} c_{j_e} \,s^{\gamma_{j_e} }\,(1-s^2)^{\nu_{j_e}/2} =1, \qquad s \in (0, 1), 
\]
where for each term in the first summation either $\gamma_{j_o}$ or $\nu_{j_o}$ is odd, 
and in the second summation the $\gamma_{j_e}$ and $\nu_{j_e}$ are even integers. Therefore we 
have 
\[
\sum_{j_o} c_{j_o} \,s^{\gamma_{j_o} }\,(1-s^2)^{\nu_{j_o}/2} =1 - \sum_{j_e} c_{j_e} \,s^{\gamma_{j_e} }\,(1-s^2)^{\nu_{j_e}/2} , \qquad s \in (0, 1), 
\]
and we proceed by  considering the binomial series of the two sides of this equation. Since 
the series of the right hand side has only even powers of the variable $s \in (0, 1)$, 
it follows that the terms on left side whose $\gamma_{j_o}$ are odd cancel each other out, 
therefore with no loss in generality we can assume all the $\gamma_{j_o}$ are even, 
which implies that all the  $\nu_{j_o}$ have to be 
odd integers. Now by making the replacement $s_1= (1-s^2)^{1/2}$ we are led to 
\[
\sum_{j_o} c_{j_o} \,(1-s_1^2)^{\gamma_{j_o}/2 }\,s_1^{\nu_{j_o}} =1 - \sum_{j_e} c_{j_e} \, 
(1-s_1^2)^{\gamma_{j_e}/2 }\,s_1^{\nu_{j_e}} , \qquad s_1 \in (0, 1). 
\] 
Finally we compare the binomial series in $s_1$ of the two sides of this equation: since 
the series of the right hand side has only even exponents and all the $\nu_{j_o}$ 
on the left side are odd integers, we conclude that 
\[
\sum_{j_o} c_{j_o} \,(1-s_1^2)^{\gamma_{j_o}/2 }\,s_1^{\nu_{j_o}} = 0, \qquad s_1 \in (0, 1). 
\]
\endproof

Again, as mentioned in Remark~\ref{nodivergence}, 
the integrals are all convergent, hence there is no
renormalization problem caused by the intersection of
the boundary of the domain of integration with the
singular set of the algebraic differential form.

\medskip
\section{The motives}\label{MotSec}

In this section we analyze the motives associated to the periods obtained
from the coefficients $a_{2n}$ of the spectral action.
We are considering a family of quadrics
\begin{equation}\label{quadQalpha}
Q_{\alpha,2n} = u_1^2 + \frac{1}{\alpha^2} (u_2^2 + u_3^2 +u_4^2) + u_5^2 + \cdots + u_{2n+2}^2,
\end{equation}
where $\alpha$ is a (rational) parameter. 
These define quadric hypersurfaces $Z_{\alpha,2n}$ in $\P^{2n+1}$. We will also be considering
the projective cone $CZ_{\alpha,2n}$ in $\P^{2n+2}$ and the affine cone $\widehat{CZ}_{\alpha,2n}$
in the affine space $\A^{2n+3}$. 

\smallskip
\subsection{Pencils of quadrics}
A quadratic form $Q$ on a vector space $V$ determines a quadric $Z_Q \subset \P(V)$.
Given two quadratic forms $Q_1$ and $Q_2$ on $V$, a pencil $\cZ_\cQ$ of quadrics in $\P(V)$
is obtained by considering, for each $z=(\lambda : \mu)\in \P^1$, the quadric $Z_{Q_z}$ 
defined by the quadratic form $\lambda Q_1 + \mu Q_2$. 
Let $\cZ_\cQ =\{ (z,u)\in \P^1\times \P(V)\,:\, u \in Z_{Q_z}\} \subset \P^1\times \P(V)$.

\smallskip

In particular, we can view the quadrics $Z_{\alpha,2n}$ defined by the quadratic forms $Q_{\alpha,2n}$ of
\eqref{quadQalpha} as defining a pencil of quadrics in $\P^1\times \P^{2n+1}$, with $\lambda/\mu=\alpha^2$.
Namely, we regard the quadric $Z_{\alpha,2n}$ as part of the pencil of quadrics $\cZ_{2n}=\{ Z_{z,2n} \}_{z\in \P^1}$,
defined by
\begin{equation}\label{pencilQn}
Q_{z,2n} = \lambda (u_1^2 + u_5^2 + \cdots + u_{2n+2}^2) + \mu (u_2^2 + u_3^2 +u_4^2),
\end{equation}
for $z=(\lambda:\mu)\in \P^1$. The quadric $Z_{z,2n}$ becomes degenerate over the set
$X=\{0,1\} \subset \P^1$, where it reduces, in the case $\lambda=0$ to a projective cone 
$Z_{Q_1,2n}=C^{2n-1}B_1$ over the conic $B_1=\{ u_2^2+u_3^2+u_4^2 =0\}$ in $\P^2$, and in the case
$\mu=0$ to a projective cone $Z_{Q_2,2n}=C^3 B_2$ over the quadric $B_2=\{ u_1^2 + 
u_5^2 + \cdots + u_{2n+2}^2 =0 \}$ in $\P^{2n-2}$. There is a correspondence, as in
\S 10 of \cite{BEK},
$$ \xymatrix{ (\P^1 \times \P^{2n+1}) \smallsetminus \cZ_n \ar[r] \ar[d] & \P^{2n+1} \smallsetminus 
(Z_{Q_1,2n}\cap Z_{Q_2,2n}) \\ \P^1 &  } $$
where the horizontal map is an $\A^1$-fibration and the vertical map is the projection to
$z=(\lambda:\mu)\in \P^1$. By homotopy invariance, we can identify 
$H^{2n+2}_c( (\P^1 \times \P^{2n+1}) \smallsetminus \cZ_{2n})$ with the Tate twisted 
$H^{2n+1}_c(\P^{2n+1}\smallsetminus (Z_{Q_1,2n}\cap Z_{Q_2,2n}))(-1)$.

\smallskip
\subsection{Motives of quadrics}

The theory of motives of quadrics is a very rich and interesting topic, see \cite{Rost},
\cite{Vishik1}, \cite{Vishik2}. We recall here only a few essential facts that we need in
our specific case.  Suppose given a quadratic form $Q$ on an $n$-dimensional vector space $V$ over
a field $\K$ of characteristic not equal to $2$. For our purposes, we will focus on the
case where $\K=\Q$. We write $\langle a_1, \ldots, a_n \rangle$ for the matrix of $Q$
in diagonal form. The quadratic form $\H:= \langle 1, -1 \rangle$ is the elementary hyperbolic form.
A quadratic form $Q$ is isotropic if $\H$ is a direct summand, hence $Q=\H \perp Q'$. It is
anisotropic otherwise. Any quadratic form can be written in the form $Q= d\cdot \H \perp Q'$, 
where $Q'$ is a uniquely determined anisotropic quadratic form. The integer $d$ is the Witt
isotropy index of $Q$. Given an anisotropic quadratic form $Q$ over the field $\K$, there is
a tower of field extensions $\K_1=\K(Q)$, $\K_2=\K_1(Q_1)$, $\ldots$, $\K_s=\K_{s-1}(Q_{s-1})$,
such that over $\K_1$ the quadric $Q|_{\K_1}=d_1 \cdot \H \perp Q_1$, with $Q_1$
anisotropic; over $\K_2$ the quadric $Q_1|_{\K_2} d_2 \cdot \H \perp Q_2$, with $Q_2$
anisotropic, and so on, until $Q_{s}=0$. The tower of extensions $\K_1, \ldots, \K_s$ is
the Knebusch universal splitting tower, and $d_1, \ldots, d_s$ are the Witt numbers of $Q$.

\smallskip

Let $Z_Q$ be the quadric defined by the quadratic form $Q$ over $\K$. For a hyperbolic quadratic form
$Q= d\cdot \H$ of dimension $2d$, the motive of $Z_Q$ is given by (see \cite{Vishik2})
\begin{equation}\label{motH2d}
\m( Z_{d\H} ) = \Z(d-1) [2d-2] \oplus \Z(d-1) [2d-2] 
\oplus \bigoplus_{i=0,\ldots, d-2, d, \ldots, 2d-2} \Z(i)[2i] ,
\end{equation}
where $\Z=\m(\Spec(\K))$. In the case where $Q= d\cdot \H \perp \langle 1 \rangle$ in dimension
$2d+1$, the motive of $Z_Q$ is given by (see \cite{Vishik2})
\begin{equation}\label{motH2d1}
\m(Z_{d\H \perp \langle 1 \rangle}) = \bigoplus_{i=0,\ldots, 2d-1} \Z(i) [2i].
\end{equation}

\smallskip

Given a quadric $Z_Q$, we denote by $Z_{Q^i}$ the variety of $i$-dimensional
planes on the quadric $Z_Q$. As in \cite{Vishik2}, we write $\cX_{Q^i}$ for the
associated simplicial scheme (Definition 2.3.1 of \cite{Vishik2})
and $\m(\cX_{Q^i})$ for the corresponding object in the category $\cD\cM^{{\rm eff}}(\K)$
of motives. 

\smallskip

We also recall the following result (see Proposition 4.2 of \cite{Vishik2}) that will be useful in our case.
Let $Z_Q \subset \P^{m+1}$ be a quadratic form of dimension $m=2n$ over $\K$, such that there 
exists a quadratic extension $\K( \sqrt{a} )$ of $\K$ over which $Q$ is hyperbolic. Then the motive
$\m(Z_Q)$ decomposes as a direct sum 
\begin{equation}\label{motZQqext}
\m(Z_Q) =\left\{ \begin{array}{ll}  \m_1 \oplus \m_1(1)[2] & m =2 \mod 4 \\
\m_1 \oplus \cR_{Q,\K} \oplus \m_1(1)[2] & m = 0 \mod 4
\end{array} \right. 
\end{equation}
where the motive $\m_1$ is an extension of the motives $\m(\cX_{Q^i})(i)[2i]$ and 
$\m(\cX_{Q^\ell})(\dim(Q)-\ell)[2\dim(Q)-2\ell]$, for $i$ (respectively, $\ell$) ranging over
all even (respectively, odd) numbers less than or equal to $2 [\dim(Q)/4]$.
The motive we denote by $\cR_{Q,\K}$ is a form of a Tate motive, which is denoted by
$\cR_{Q,\K} =\K(\sqrt{\det(Q)}) (\frac{\dim(Q)}{2})[\dim(Q)] $ in \cite{Vishik2}.

\smallskip

If $Q$ is $d$-times isotropic, $Q= d\cdot \H \perp Q'$, then $\m(\cX_{Q^j})=\Z$ for all
$0\leq j < i$. Thus, the motives $\m(\cX_{Q^j})$ become Tate motives in a field extension
in which the quadric becomes isotropic, and one recovers the motivic decomposition
into a sum of Tate motives mentioned above. The motives $\m(\cX_{Q^j})$ are therefore
{\em forms} of the Tate motive, which means that over the algebraic closure 
$\m(\cX_{Q^j}|_{\bar\K})=\Z$.

\medskip
\subsection{Grothendieck classes}

It if often convenient, instead of working with objects in the category of mixed motives,
to consider a simpler invariant given by the class in the Grothendieck ring of varieties,
which can be regarded as a universal Euler characteristics. The Grothendieck ring
$K_0(\cV_\K)$ of varieties over a field $\K$ is generated by the isomorphism classes
$[X]$ of smooth quasi-projective varieties $X \in \cV_\K$ with the inclusion-exclusion
relations $[X]=[Y]+[X\smallsetminus Y]$ for closed embeddings $Y\subset X$ and
the product $[X\times Y]=[X]\cdot [Y]$. The following simple identities will be useful in the
computations of Grothendieck classes of the motives involved in the period computations
described in the previous sections.

\begin{lem}\label{idGr}
Let $Z$ be a projective subvariety $Z\subset \P^{N-1}$, with $\hat Z \subset \A^N$
the affine cone. Let $CZ$ denote the projective cone in $\P^N$ and $\widehat{CZ}$
the corresponding affine cone in $\A^{N+1}$. Let $H$ and $H'$ be two affine
hyperplanes in $\A^{N+1}$ with $H\cap H'=\emptyset$ and such that the intersections
$\widehat{CZ}\cap H$ and $\widehat{CZ}\cap H'$ are sections of the cone, given by copies of
$\hat Z$. The Grothendieck classes of the
projective and affine complements satisfy
\begin{enumerate} 
\item $[\A^N \smallsetminus \hat Z]= (\bL-1) [\P^{N-1}\smallsetminus Z]$
\item $[\A^{N+1} \smallsetminus \widehat{C Z}]= (\bL-1)[\P^N \smallsetminus CZ]$
\item $[CZ] = \bL \, [Z] +1$
\item $[\A^{N+1} \smallsetminus \widehat{C Z}]=\bL^{N+1} - \bL (\bL-1) [Z] -\bL$
\item $[\A^{N+1} \smallsetminus (\widehat{C Z}\cup H\cup H')]=\bL^{N+1} -2\bL^N - (\bL-2) (\bL-1) [Z] -(\bL-2)$.
\end{enumerate}
where $\bL=[\A^1]$ is the Lefschetz motive, the class of the affine line.
\end{lem} 

\proof The first and second identities follow from the fact that the class of the
affine cone is given by $[\hat Z]=(\bL-1) [Z] +1$, so that
$$ [\A^N \smallsetminus \hat Z] = \bL^N - (\bL-1) [Z] -1 = (\bL-1) ( \frac{(\bL^N-1)}{(\bL-1)} - [Z])
= (\bL-1) [\P^{N-1} - Z]. $$
The identity $[CZ] = \bL \, [Z] +1$ follows by viewing the projective cone over $Z$
as the union of a copy of $Z$ and a copy of the affine cone $\hat Z$ over $Z$, and using
the same identity $[\hat Z]=(\bL-1) [Z] +1$ for the affine cone. The fourth identity
follows from the second and the third,
$$ (\bL-1)[\P^N \smallsetminus CZ] = \bL^{N+1}-1 - (\bL-1)[CZ]  $$
$$ = \bL^{N+1}-1 - (\bL-1) (\bL[Z] +1) =\bL^{N+1} - (\bL^2 -\bL) [Z] -\bL .$$
For the last identity, we write
$$ [\A^{N+1} \smallsetminus (\widehat{C Z}\cup H\cup H')] = \bL^{N+1} - [\widehat{C Z}\cup H\cup H']. $$
The class of the union is given by
$$ [\widehat{C Z}\cup H\cup H'] =[\widehat{C Z}] +[H\cup H']-[\widehat{C Z}\cap (H\cup H')]. $$
Since $H\cap H'=\emptyset$, we have $[H \cup H']=2\bL^N$ and 
$[\widehat{C Z}\cap (H\cup H')]=[\widehat{C Z}\cap H]+[\widehat{C Z}\cap H']=2[\hat Z]=2(\bL-1)[Z]+2$.
Thus, we have
$$ [\A^{N+1} \smallsetminus (\widehat{C Z}\cup H\cup H')] = \bL^{N+1} -
2\bL^N - [\widehat{C Z}] + 2(\bL-1)[Z]+ 2 $$
$$ = \bL^{N+1} - 2\bL^N - \bL(\bL-1)[Z] -\bL + 2(\bL-1)[Z]+ 2 = \bL^{N+1} - 2\bL^N - (\bL-2)(\bL-1)[Z] - (\bL-2). $$
\endproof

\medskip
\subsection{Pencils of quadrics in $\P^3$}\label{penP3sec}

We look first at the case of the quadric $Z_\alpha=Z_{\alpha,2}$ in $\P^3$ that arises in the
computation of the $a_2$ term of the heat kernel expansion.

\smallskip

Over $\C$, any quadric surface $Z_Q$ in $\P^3$ can be put in the standard form $X Y = Z W$ by a simple
change of coordinates. Thus, over $\C$ any quadric surface in $\P^3$ is isomorphic to the
Segre embedding $\P^1 \times \P^1 \hookrightarrow \P^3$. When we consider quadrics over $\Q$,
this is no longer necessarily the case.

\begin{thm}\label{fieldext}
For $\alpha\in \Q$, over the quadratic extension $\K =\Q(\sqrt{-1})$, the quadric $Z_\alpha=Z_{\alpha,2}$
in $\P^3$ is isomorphic to the Segre embedding $\P^1 \times \P^1 \hookrightarrow \P^3$. The class
of the complement in the Grothendieck ring is $[\P^3 \smallsetminus Z_\alpha]=\bL^3 -\bL$, while
the class of the affine complement of $\widehat{CZ}_\alpha$ is 
$[\A^5 \smallsetminus \widehat{CZ}_\alpha]=\bL^5 - \bL^4 - \bL^3 +\bL^2$.
The class of the complement $\A^5 \smallsetminus (\widehat{CZ}_\alpha \cup H_0 \cup H_1)$
with the affine hyperplanes $H_0=\{ u_0=0 \}$ and $H_1=\{ u_0=1 \}$ is given by
$$ [ \A^5 \smallsetminus (\widehat{CZ}_\alpha \cup H_0 \cup H_1) ] = \bL^5 - 3 \bL^4 + \bL^3 + 3 \bL^2 - 2 \bL. $$
\end{thm}

\proof Over the quadratic extension $\K=\Q(i)$ we can consider the change of variables
$$ X = u_1 + \frac{i}{\alpha} u_2, \ \  Y= u_1 - \frac{i}{\alpha} u_2, \ \  Z = \frac{i}{\alpha} (u_3 + i u_4), \ \ 
W = \frac{i}{\alpha} (u_3 - i u_4) ,$$
where we assume that $\alpha \in \Q$. This change of coordinates 
determines the identification of $Z_\alpha$ with the Segre quadric $\{ X Y - Z W =0 \} 
\simeq \P^1 \times \P^1$.

\smallskip 
The classes in the Grothendieck ring are then given by $[Z_\alpha]=[\P^1 \times \P^1]=(\bL+1)^2=\bL^2 +2\bL+1$,
so that $[\P^3 \smallsetminus Z_\alpha] =\bL^3 +\bL^2 +\bL + 1 -(\bL^2 +2\bL+1)=
\bL^3 -\bL$. We then use Lemma~\ref{idGr} to compute the class $[\A^5\smallsetminus \widehat{CZ}_\alpha]$.
We have $$ [\A^5\smallsetminus \widehat{CZ}_\alpha]=\bL^5 - \bL(\bL-1)[Z_\alpha] -\bL =
\bL^5 -\bL - \bL(\bL-1) (\bL+1)^2 = \bL^5 - \bL^4 - \bL^3 +\bL^2. $$
We then use the last identity of Lemma~\ref{idGr} to compute
$$ [ \A^5 \smallsetminus (\widehat{CZ}_\alpha \cup H_0 \cup H_1) ] =\bL^5 -2 \bL^4 - (\bL - 2) (\bL - 1) [Z_\alpha] 
- (\bL-2) $$ $$ = \bL^5 -2 \bL^4 - (\bL - 2) (\bL -1) (\bL+1)^2 
- (\bL-2) =\bL^5 - 3 \bL^4 + \bL^3 + 3 \bL^2 - 2 \bL. $$
\endproof

\smallskip

\begin{thm}\label{a2motQi}
Over the quadratic extension $\K =\Q(\sqrt{-1})$, the motive 
$\m(\A^5\smallsetminus (\widehat{CZ}_\alpha \cup H_0 \cup H_1), \Sigma)$ is mixed Tate.
\end{thm}

\proof Over $\K =\Q(\sqrt{-1})$, the quadric $Q_\alpha$, for $\alpha\in \Q$, satisfies
$$ Q_\alpha |_{\Q(\sqrt{-1})}= 2\cdot \H $$
hence the motive is given by \eqref{motH2d} as
$$ \m(Z_\alpha) = \Z\oplus \Z(1)[2] \oplus \Z(1)[2] \oplus \Z(2)[4]
= \m(\P^1\times \P^1) $$
where $\m(\P^1) =\Z \oplus \Z(1)[2]$. This corresponds to the Grothendieck 
class $[Z_\alpha]=1 + 2\bL + \bL^2$. 

\smallskip

The Gysin distinguished triangle of the closed embedding $Z_\alpha \hookrightarrow \P^3$ of codimension one gives
$$ \m(\P^3\smallsetminus Z_\alpha) \to \m(\P^3) \to \m(Z_\alpha)(1)[2] \to \m(\P^3\smallsetminus Z_\alpha)[1], $$
hence if two of the three terms are in the triangulated subcategory of mixed
Tate motives, the third term also is. This implies that $\m(\P^3\smallsetminus Z_\alpha)$ 
is mixed Tate.

\smallskip

When passing to the projective cone $CZ_\alpha$ in $\P^4$, 
since $\P^4\smallsetminus CZ_\alpha \to \P^3\smallsetminus Z_\alpha$
is an $\A^1$-fibration, by homotopy invariance we have
$\m^j_c(\P^4\smallsetminus CZ_\alpha)=\m^{j-2}_c(\P^3\smallsetminus Z_\alpha)(-1)$,
where we consider here the motive $\m^j_c$ with compact support 
that corresponds to the cohomology $H^j_c$.
Thus, if the motive $\m(\P^3\smallsetminus Z_\alpha)$ is mixed Tate, then so is the
motive $\m(\P^4 \smallsetminus CZ_\alpha)$.

\smallskip

In passing from the motive $\m(\P^4 \smallsetminus CZ_\alpha)$  to the motive
$\m(\A^5 \smallsetminus \widehat{CZ}_\alpha)$, consider the $\P^1$-bundle $\cP$
compactification of the $\bG_m$-bundle 
$\cT=\A^5 \smallsetminus \widehat{CZ}_\alpha \to X=\P^4 \smallsetminus CZ_\alpha$ and
the Gysin distinguished triangle
$$ \m(\cT) \to \m(\cP) \to \m_c(\cP\smallsetminus \cT)^*(1)[2] \to \m(\cT)[1], $$
see \cite{Voe}, p.197.
The motive of a projective bundle satisfies $\m(\cP)$
hence $\m(\cP)$ is mixed Tate, since $\m(X)$ is. The motive $\m_c(\cP\smallsetminus \cT)$
is also mixed Tate since $\cP\smallsetminus \cT$ consists of two copies of $X$, hence
the remaining term $\m(\cT)$ is also mixed Tate. 

\smallskip

We then consider the union of $\widehat{CZ}_\alpha$ and the affine hyperplanes $H_0=\{ u_0=0 \}$
and $H_1=\{ u_0 =1 \}$ in the affine space $\A^5$. In order to check that the motive of the union
$\widehat{CZ}_\alpha \cup H_0 \cup H_1$ is mixed Tate suffices to know that the motives $\m(\A^5
\smallsetminus (H_0\cup H_1))$ and $\m(\A^5\smallsetminus \widehat{CZ}_\alpha)$ as well as the 
motive of the intersection $\m(\widehat{CZ}_\alpha \cap (H_0\cup H_1))$ are mixed Tate. This follows 
by applying the Mayer-Vietoris distinguished triangle
$$ \m(U\cap V) \to \m(U)\oplus \m(V) \to \m(U\cup V) \to \m(U\cap V)[1] $$
with $U=\A^5\smallsetminus \widehat{CZ}_\alpha$ and $V=\A^5\smallsetminus (H_0\cup H_1)$.
This shows that it suffices to know two of the three terms are mixed Tate to know the remaining
one also is. The motive $\m(\A^5\smallsetminus \widehat{CZ}_\alpha)$ is mixed Tate by our previous
argument. The motive $\m(\A^5 \smallsetminus (H_0\cup H_1))$ 
is also mixed Tate by a similar argument, since $\m(H_0\cup H_1)$ clearly is. Thus, 
it suffices to show that the motive $\m(\A^5 \smallsetminus (\widehat{CZ}_\alpha \cap (H_0\cup H_1))$
is mixed Tate, which can be shown by showing that the motive $\m(\widehat{CZ}_\alpha \cap (H_0\cup H_1))$
is mixed Tate. The intersection $\widehat{CZ}_\alpha \cap (H_0\cup H_1)$ consists of two
sections of the cone, hence one has two copies of the motive $\m(\hat Z_\alpha)$ that is also a Tate motive.

\smallskip

The divisor $\Sigma$ in $\A^5$ is a union of coordinate hyperplanes and their translates, and
is also mixed Tate. Thus, the motive $\m(\A^5\smallsetminus (\widehat{CZ}_\alpha \cap (H_0\cup H_1), \Sigma)$
sits in a distinguished triangle in the Voevodsky triangulated category of mixed motives over $\Q$, where 
two of the three terms, $\m( \A^5\smallsetminus (\widehat{CZ}_\alpha \cap (H_0\cup H_1))$ and $\m(\Sigma)$, 
are both mixed Tate. This implies that the remaining term $\m(\A^5\smallsetminus (\widehat{CZ}_\alpha \cap (H_0\cup H_1), \Sigma)$ is also mixed Tate. 
\endproof

%%%%%%%%%%%%%%%%%%
\smallskip
\subsection{Pencils of quadrics in $\P^5$}\label{penP5sec}

We consider then the next step, namely the quadric $Z_{\alpha,4}\subset \P^5$ that occurs
in the computation of the $a_4$ term of the heat kernel expansion.

\begin{thm}\label{a4Grclass}
Over the quadratic field extension $\K=\Q(\sqrt{-1})$ the quadric $Z_{\alpha,4}\subset \P^5$
determined by the quadratic form 
$$ Q_{\alpha,4} = u_1^2 + \frac{1}{\alpha^2}(u_2^2+u_3^2+u_4^2) + u_5^2 + u_6^2 $$
has Grothendieck class $[\P^5\smallsetminus Z_{\alpha,4}]=\bL^5 - \bL^2$. The class
of the complement in $\A^7$ of $\widehat{CZ}_{\alpha,4}$ has Grothendieck class
$$ [ \A^7 \smallsetminus \widehat{CZ}_{\alpha,4} ] =  \bL^7 -\bL^6 -\bL^4 +\bL^3 $$
and the class of the complement of the union $\widehat{CZ}_{\alpha,4} \cup H_0 \cup H_1$
is given by
$$ [ \A^7 \smallsetminus (\widehat{CZ}_{\alpha,4} \cup H_0 \cup H_1) ] =  
\bL^7 - 3 \bL^6 +2 \bL^5 - \bL^4 + 3 \bL^3 -2 \bL^2. $$
\end{thm}

\proof Over $\K=\Q(\sqrt{-1})$ we can consider the change of coordinates
$$ X = u_5 + i u_6, \ \ \ \ Y = u_5 - i u_6. $$
With this change of coordinates, we rewrite the quadratic form as
$$  Q_{\alpha,4} = Q_{\alpha,2} + XY, $$
where $Q_{\alpha,2}=u_1^2 + \frac{1}{\alpha^2}(u_2^2+u_3^2+u_4^2)$ is the
quadratic form of the $a_2$-term that we discussed in \S \ref{penP3sec}.
We can compute the Grothendieck class $[\hat Z_{\alpha,4}]$ by considering
the two possible cases $Y\neq 0$ and $Y=0$. In the first case, since $Y\neq 0$
we have solutions of the form 
$$ X = -\frac{Q_{\alpha,2}(u_1,u_2,u_3,u_4)}{Y}, $$
which give a choice of $Y\in \bG_m$ and arbitrary $(u_1,u_2,u_3,u_4)\in \A^4$,
hence a contribution of $(\bL-1)\bL^4$ to the class. In the case where $Y=0$,
we have solutions given by an arbitrary $X\in \A^1$ and $(u_1,u_2,u_3,u_4)\in \hat Z_{\alpha,1}$.
This gives a contribution of $\bL [\hat Z_{\alpha,2}]$ to the class. Thus, we obtain
$$ [\hat Z_{\alpha,4}]= (\bL-1) \bL^4 + \bL [\hat Z_{\alpha,1}] = (\bL-1) \bL^4 + \bL (\bL-1) (\bL+1)^2 +\bL, $$
since $[\hat Z_{\alpha,2}]=(\bL-1)[Z_{\alpha,1}]+1$ and $[Z_{\alpha,1}]=[\P^1\times \P^1]$. This gives
$$ [\A^6 \smallsetminus \hat Z_{\alpha,4}]= \bL^6 -\bL - (\bL-1) (\bL^4 +\bL^3 +2\bL^2 + \bL) =
(\bL-1) (\bL^5 - \bL^2). $$
Since $[\A^6 \smallsetminus \hat Z_{\alpha,4}] = (\bL-1) [\P^5\smallsetminus Z_{\alpha,2}]$ we then
have $[\P^5\smallsetminus Z_{\alpha,4}]=\bL^5 - \bL^2$, as stated. Thus, we also have
$[Z_{\alpha,4}]=[\P^5]-(\bL^5-\bL^2)=\bL^4+\bL^3 +2 \bL^2 +\bL +1$. Using Lemma~\ref{idGr}
we then obtain 
$$ [ \A^7 \smallsetminus \widehat{CZ}_{\alpha,4} ] = \bL^7 - \bL(\bL-1) [Z_{\alpha,4}]-\bL  $$
$$ = \bL^7 - \bL (\bL-1) (\bL^4 +\bL^3 +2 \bL^2 +\bL +1) -\bL = \bL^7 -\bL^6 -\bL^4 +\bL^3. $$
The Grothendieck class of the complement of the union of $\widehat{CZ}_{\alpha,4}$ and the
two hyperplanes is given by
$$ [ \A^7 \smallsetminus (\widehat{CZ}_{\alpha,4} \cup H_0 \cup H_1) ] =\bL^7 -2 \bL^6 
- (\bL-1) (\bL-2) [Z_{\alpha,4}] - (\bL-2) $$
$$ = \bL^7 -2 \bL^6  - (\bL-1) (\bL-2)(\bL^4 +\bL^3 +2 \bL^2 +\bL +1) - (\bL-2) =
\bL^7 - 3 \bL^6 +2 \bL^5 - \bL^4 + 3 \bL^3 -2 \bL^2 .$$
\endproof

We also have the analog of Theorem \ref{a2motQi}, which we state here. The proof is
analogous and we omit it.

\begin{prop}\label{a4motQi}
Over the quadratic extension $\K =\Q(\sqrt{-1})$, the motive 
$\m(\A^7\smallsetminus (\widehat{CZ}_{\alpha,4} \cup H_0 \cup H_1), \Sigma)$ is mixed Tate.
\end{prop}

\proof The argument is completely analogous to Theorem~\ref{a2motQi}, using the
fact that, over $\K =\Q(\sqrt{-1})$ the quadratic form is
$$ Q_{\alpha,4}|_{\Q(\sqrt{-1})}= 3 \cdot \H. $$
The rest of the argument follows as in Theorem~\ref{a2motQi}.
\endproof

\smallskip
\subsection{The Grothendieck class of $\P^{2n-1}\smallsetminus Z_{\alpha,2n}$ over $\K=\Q(\sqrt{-1})$}

The argument of Theorem~\ref{a4Grclass} can be used to obtain an inductive argument
computing the Grothendieck class $[\P^{2n-1}\smallsetminus Z_{\alpha,2n}]$ for all the
quadrics $Z_{\alpha,n}$ determined by the quadratic forms
\begin{equation}\label{quadQalphan}
Q_{\alpha,2n} = u_1^2 +  \frac{1}{\alpha^2}(u_2^2+u_3^2+u_4^2) + u_5^2 + u_6^2 + \cdots 
+ u_{2n+1}^2 + u_{2n+2}^2,
\end{equation}
for all $n\geq 3$.

\begin{thm}\label{GrQalphan}
Over the quadratic field extension $\K=\Q(\sqrt{-1})$ the quadric $Z_{\alpha,2n}$ has Grothendieck
class $[\P^{2n+1}\smallsetminus Z_{\alpha,2n}]=\bL^{2n+1}-\bL^n$. The affine complement
of $\widehat{CZ}_{\alpha,2n}$ has class
$$ [\A^{2n+3}\smallsetminus \widehat{CZ}_{\alpha,2n}] = \bL^{2n+3} - \bL^{2n+2} - \bL^{n+2} +\bL^{n+1} $$
and the affine complement of the union $\widehat{CZ}_{\alpha,2n} \cup H_0 \cup H_1$ has class
$$ [\A^{2n+3}\smallsetminus (\widehat{CZ}_{\alpha,2n} \cup H_0 \cup H_1)] = \bL^{2n+3} - 3 \bL^{2n+2}
+2 \bL^{2n+1} - \bL^{n+2} + 3 \bL^{n+1} - 2 \bL^n.  $$
\end{thm}

\proof We proceed as in Theorem~\ref{a4Grclass}. Over the field $\K=\Q(\sqrt{-1})$ the
change of coordinates 
$$ X = u_{2n+1} + i u_{2n+2}, \ \ \ \ Y = u_{2n+1} - i u_{2n+2} $$
puts $Q_{\alpha,2n}$ in the form
$$  Q_{\alpha,2n} = Q_{\alpha,2n-2}(u_1,\ldots, u_{2n}) + XY . $$
Thus, the Grothendieck class $[\hat Z_{\alpha,2n}]$ is a sum of a contribution
corresponding to $Y\neq 0$, which is of the form $(\bL-1) \bL^{2n}$ and a
contribution from $Y = 0$, which is of the form $\bL \, [\hat Z_{\alpha,n-1}]$. This gives
$$ [\A^{2n+2}\smallsetminus \hat Z_{\alpha,2n}] = \bL^{2n+2} - 2 \bL^{2n+1} + \bL^{2n}
+ \bL [ \A^{2n}\smallsetminus \hat Z_{\alpha,2n-2} ], $$
hence using the relation between the classes of the affine and projective complements,
$$ [\P^{2n+1}\smallsetminus Z_{\alpha,2n}] =\bL^{2n} (\bL-1) + \bL [\P^{2n-1}\smallsetminus Z_{\alpha,2n-2}]. $$
Assuming inductively that $[\P^{2n-1}\smallsetminus Z_{\alpha,2n-2}]=\bL^{2n-1} - \bL^{n-1}$ we indeed
obtain that the class of the complement is
$[\P^{2n+1}\smallsetminus Z_{\alpha,2n}]=\bL^{2n} (\bL-1) + \bL (\bL^{2n-1} - \bL^{n-1}) =\bL^{2n+1}-\bL^n$.
We then have
$$ [Z_{\alpha,2n}] =[\P^{2n+1} ] - [\P^{2n+1}\smallsetminus Z_{\alpha,2n}] = \bL^{2n} +\bL^{2n-1} +\cdots
+\bL^{n+1} + 2 \bL^n +\bL^{n-1} + \cdots + \bL^2 +\bL +1 .$$
Using Lemma~\ref{idGr}, we obtain
$$ [ \A^{2n+3}\smallsetminus \widehat{CZ}_{\alpha,2n}] = \bL^{2n-3} - \bL (\bL-1) [Z_{\alpha,2n}]  -\bL $$
$$ = \bL^{2n+3} - \sum_{j=2}^n \bL^j  - \bL^{n+1} - 2 \bL^{n+2} - \sum_{j=n+3}^{2n+1} \bL^j  -\bL^{2n+2} 
+ \sum_{j=2}^n \bL^j + 2 \bL^{n+1} + \bL^{n+2} + \sum_{j=n+3}^{2n+1} \bL^j $$
$$ = \bL^{2n+3} + \bL^{n+1} - \bL^{n+2} - \bL^{2n+2}. $$
We proceed in the same way for the computation of the class of the affine complement of the
union $\widehat{CZ}_{\alpha,2n} \cup H_0 \cup H_1$, using Lemma~\ref{idGr}. We have
$$ [\A^{2n+3}\smallsetminus (\widehat{CZ}_{\alpha,2n} \cup H_0 \cup H_1)] = \bL^{2n+3} 
- 2 \bL^{2n+2} - (\bL-2) (\bL-1) [Z_{\alpha,2n}] - (\bL-2) $$
and using again the expression 
$$ [Z_{\alpha,2n}] = \bL^{2n} +\bL^{2n-1} +\cdots
+\bL^{n+1} + 2 \bL^n +\bL^{n-1} + \cdots + \bL^2 +\bL +1 $$
we obtain
$$ (\bL-2)(\bL-1)[Z_{\alpha,2n}] = 2 -\bL + 2 \bL^n -3 \bL^{n+1} + \bL^{n+2} -2 \bL^{n+1} + \bL^{2n+2} $$
due to cancellations of terms similar to the previous case. We then have
$$ \bL^{2n+3} - 2 \bL^{2n+2} - (\bL-2) (\bL-1) [Z_{\alpha,2n}] - (\bL-2) = $$
$$ \bL^{2n+3} - 3 \bL^{2n+2} +2 \bL^{2n+1} - \bL^{n+2} + 3 \bL^{n+1} - 2 \bL^n, $$
which agrees with the cases $n=1$ and $n=2$ computed in Theorems~\ref{fieldext}
and \ref{a4Grclass}.
\endproof

\smallskip

We also have the analog of Theorem~\ref{a2motQi} and Proposition~\ref{a4motQi}, proved by the same argument.

\begin{prop}\label{a2nmotQi}
Over the field extension $\K =\Q(\sqrt{-1})$, the mixed motive 
$$\m(\A^{2n+3}\smallsetminus (\widehat{CZ}_{\alpha,2n} \cup H_0\cup H_1), \Sigma)$$ is mixed Tate.
\end{prop}

\proof The argument is completely analogous to Theorem~\ref{a2motQi}, using the
fact that, over $\K =\Q(\sqrt{-1})$ the quadratic form is
$$ Q_{\alpha,2n}|_{\Q(\sqrt{-1})}= (n+1) \cdot \H, $$
with \eqref{motH2d} giving the motive $\m(Q_{\alpha,2n}|_{\Q(\sqrt{-1})})$. The motives
of complements, and projective and affine cones and the relative motives
$\m(\A^{2n+3}\smallsetminus \widehat{CZ}_{\alpha,2n}, \Sigma)$ and
$\m(\A^{2n+3}\smallsetminus (\widehat{CZ}_{\alpha,2n} \cup H_0\cup H_1), \Sigma)$ 
are then obtained as in Theorem~\ref{a2motQi}.
\endproof

\smallskip
\subsection{The motive of $Z_{\alpha,2n}$ over $\Q$}

Over the rationals, the quadratic form $Q_{\alpha,2n}$ is anisotropic, although, as we have
seen, it becomes isotropic over the field extension $\K =\Q(\sqrt{-1})$, with
$Q_{\alpha,2n}|_{\Q(\sqrt{-1})}= (n+1) \cdot \H$. The motive of $Z_{\alpha,2n}$ over
$\Q(\sqrt{-1})$ is a sum of Tate motives
$$ \m(Z_{\alpha,2n}|_{\K}) = \Z(n)[2n] \oplus \Z(n)[2n] \oplus \bigoplus_{i=0,\ldots,n-1, n+1, \ldots 2n} \Z(i)[2i], $$
which corresponds to the Grothendieck class $[Z_{\alpha,2n}]=[\P^{2n+1}]-[\P^{2n+1}\smallsetminus Z_{\alpha,2n}]
= 1+ \cdots + \bL^{2n+1} - (\bL^{2n+1} - \bL^n) = 1 + \bL + \cdots +\bL^{n-1} + 2 \bL^n + \bL^{n+1} + \cdots + \bL^{2n}$.
Over the field $\Q$, the motive of $Z_{\alpha,2n}$ is given by \eqref{motZQqext}, with
$$ \m(Z_{\alpha,2n}|_\Q) = \m_1 \oplus \m_1(1)[2] $$
when $n$ is odd and 
$$ \m(Z_{\alpha,2n}|_\Q) = \m_1 \oplus \cR_{Q,\Q,n} \oplus \m_1(1)[2] $$
when $n$ is even, where $\cR_{Q,\Q,n}$ is a form of a Tate motive denoted 
by $\cR_{Q,\Q,n}=\Q(\sqrt{\det(Q_{\alpha,2n})}) (n)[2n]$ in \cite{Vishik2}.
When passing to the quadratic field extension $\Q(\sqrt{-1})$ these motivic
decompositions become the decomposition into Tate motives given above.

%%%%%%%%%%%%%%%%%%%%
%%%%%%%%%%%%%%%%%%%%
%\appendix
\section*{Appendix A: Full expression for $b_{-6}$}
\label{Fullbminus6}

Here we provide the full  expression of the term $b_{-6}$ given by \eqref{b-6shortexpression}, which we used for 
the calculation of the term $a_4$: 
\[
b_{-6}(t, \eta, \xi_1, \dots, \xi_6) = 
\]
{\tiny
\begin{center}
\begin{math}
\frac{2560 \xi _3^8 a'(t)^4 \csc ^8(\eta )}{\alpha ^{12} Q_{\alpha ,4}^7}-\frac{53760 \xi
   _1^2 \xi _3^8 a'(t)^4 \csc ^8(\eta )}{\alpha ^{12} Q_{\alpha ,4}^8}+\frac{107520 \xi
   _1^4 \xi _3^8 a'(t)^4 \csc ^8(\eta )}{\alpha ^{12} Q_{\alpha ,4}^9}-\frac{640 \xi _3^6
   a'(t)^2 \csc ^8(\eta )}{\alpha ^{10} Q_{\alpha ,4}^6}+\frac{5120 \cot ^2(\eta ) \xi
   _3^8 a'(t)^2 \csc ^8(\eta )}{\alpha ^{12} Q_{\alpha ,4}^7}+\frac{5760 \xi _1^2 \xi
   _3^6 a'(t)^2 \csc ^8(\eta )}{\alpha ^{10} Q_{\alpha ,4}^7}+\frac{5120 \xi _2^2 \xi
   _3^6 a'(t)^2 \csc ^8(\eta )}{\alpha ^{12} Q_{\alpha ,4}^7}+\frac{10240 \xi _3^6 \xi
   _4^2 a'(t)^2 \csc ^8(\eta )}{\alpha ^{12} Q_{\alpha ,4}^7}-\frac{53760 \cot ^2(\eta )
   \xi _1^2 \xi _3^8 a'(t)^2 \csc ^8(\eta )}{\alpha ^{12} Q_{\alpha ,4}^8}-\frac{53760
   \cot ^2(\eta ) \xi _2^2 \xi _3^8 a'(t)^2 \csc ^8(\eta )}{\alpha ^{14} Q_{\alpha
   ,4}^8}-\frac{53760 \xi _1^2 \xi _2^2 \xi _3^6 a'(t)^2 \csc ^8(\eta )}{\alpha ^{12}
   Q_{\alpha ,4}^8}-\frac{107520 \xi _1^2 \xi _3^6 \xi _4^2 a'(t)^2 \csc ^8(\eta
   )}{\alpha ^{12} Q_{\alpha ,4}^8}-\frac{107520 \xi _2^2 \xi _3^6 \xi _4^2 a'(t)^2 \csc
   ^8(\eta )}{\alpha ^{14} Q_{\alpha ,4}^8}+\frac{645120 \cot ^2(\eta ) \xi _1^2 \xi _2^2
   \xi _3^8 a'(t)^2 \csc ^8(\eta )}{\alpha ^{14} Q_{\alpha ,4}^9}+\frac{1290240 \xi _1^2
   \xi _2^2 \xi _3^6 \xi _4^2 a'(t)^2 \csc ^8(\eta )}{\alpha ^{14} Q_{\alpha
   ,4}^9}+\frac{112 \xi _3^4 \csc ^8(\eta )}{\alpha ^8 Q_{\alpha ,4}^5}-\frac{1664 \cot
   ^2(\eta ) \xi _3^6 \csc ^8(\eta )}{\alpha ^{10} Q_{\alpha ,4}^6}-\frac{1664 \xi _2^2
   \xi _3^4 \csc ^8(\eta )}{\alpha ^{10} Q_{\alpha ,4}^6}+\frac{2560 \cot ^4(\eta ) \xi
   _3^8 \csc ^8(\eta )}{\alpha ^{12} Q_{\alpha ,4}^7}+\frac{30080 \cot ^2(\eta ) \xi _2^2
   \xi _3^6 \csc ^8(\eta )}{\alpha ^{12} Q_{\alpha ,4}^7}+\frac{2560 \xi _2^4 \xi _3^4
   \csc ^8(\eta )}{\alpha ^{12} Q_{\alpha ,4}^7}-\frac{10240 \xi _3^6 \xi _4^2 \csc
   ^8(\eta )}{\alpha ^{12} Q_{\alpha ,4}^7}-\frac{53760 \cot ^4(\eta ) \xi _2^2 \xi _3^8
   \csc ^8(\eta )}{\alpha ^{14} Q_{\alpha ,4}^8}-\frac{53760 \cot ^2(\eta ) \xi _2^4 \xi
   _3^6 \csc ^8(\eta )}{\alpha ^{14} Q_{\alpha ,4}^8}+\frac{215040 \xi _2^2 \xi _3^6 \xi
   _4^2 \csc ^8(\eta )}{\alpha ^{14} Q_{\alpha ,4}^8}+\frac{107520 \cot ^4(\eta ) \xi
   _2^4 \xi _3^8 \csc ^8(\eta )}{\alpha ^{16} Q_{\alpha ,4}^9}-\frac{430080 \xi _2^4 \xi
   _3^6 \xi _4^2 \csc ^8(\eta )}{\alpha ^{16} Q_{\alpha ,4}^9}-\frac{2752 \xi _3^6
   a'(t)^4 \csc ^6(\eta )}{\alpha ^{10} Q_{\alpha ,4}^6}+\frac{47040 \xi _1^2 \xi _3^6
   a'(t)^4 \csc ^6(\eta )}{\alpha ^{10} Q_{\alpha ,4}^7}+\frac{10240 \xi _2^2 \xi _3^6
   a'(t)^4 \csc ^6(\eta )}{\alpha ^{12} Q_{\alpha ,4}^7}+\frac{10240 \sec ^2(\eta ) \xi
   _3^6 \xi _4^2 a'(t)^4 \csc ^6(\eta )}{\alpha ^{12} Q_{\alpha ,4}^7}-\frac{80640 \xi
   _1^4 \xi _3^6 a'(t)^4 \csc ^6(\eta )}{\alpha ^{10} Q_{\alpha ,4}^8}-\frac{215040 \xi
   _1^2 \xi _2^2 \xi _3^6 a'(t)^4 \csc ^6(\eta )}{\alpha ^{12} Q_{\alpha
   ,4}^8}-\frac{215040 \sec ^2(\eta ) \xi _1^2 \xi _3^6 \xi _4^2 a'(t)^4 \csc ^6(\eta
   )}{\alpha ^{12} Q_{\alpha ,4}^8}+\frac{430080 \xi _1^4 \xi _2^2 \xi _3^6 a'(t)^4 \csc
   ^6(\eta )}{\alpha ^{12} Q_{\alpha ,4}^9}+\frac{430080 \sec ^2(\eta ) \xi _1^4 \xi _3^6
   \xi _4^2 a'(t)^4 \csc ^6(\eta )}{\alpha ^{12} Q_{\alpha ,4}^9}+\frac{432 \xi _3^4
   a'(t)^2 \csc ^6(\eta )}{\alpha ^8 Q_{\alpha ,4}^5}-\frac{5760 \cot ^2(\eta ) \xi _3^6
   a'(t)^2 \csc ^6(\eta )}{\alpha ^{10} Q_{\alpha ,4}^6}+\frac{1280 \cot (\eta ) \cot (2
   \eta ) \xi _3^6 a'(t)^2 \csc ^6(\eta )}{\alpha ^{10} Q_{\alpha ,4}^6}-\frac{2640 \xi
   _1^2 \xi _3^4 a'(t)^2 \csc ^6(\eta )}{\alpha ^8 Q_{\alpha ,4}^6}-\frac{6080 \xi _2^2
   \xi _3^4 a'(t)^2 \csc ^6(\eta )}{\alpha ^{10} Q_{\alpha ,4}^6}-\frac{1280 \sec ^2(\eta
   ) \xi _3^4 \xi _4^2 a'(t)^2 \csc ^6(\eta )}{\alpha ^{10} Q_{\alpha ,4}^6}-\frac{6720
   \xi _3^4 \xi _4^2 a'(t)^2 \csc ^6(\eta )}{\alpha ^{10} Q_{\alpha ,4}^6}+\frac{47040
   \cot ^2(\eta ) \xi _1^2 \xi _3^6 a'(t)^2 \csc ^6(\eta )}{\alpha ^{10} Q_{\alpha
   ,4}^7}-\frac{11520 \cot (\eta ) \cot (2 \eta ) \xi _1^2 \xi _3^6 a'(t)^2 \csc ^6(\eta
   )}{\alpha ^{10} Q_{\alpha ,4}^7}+\frac{79040 \cot ^2(\eta ) \xi _2^2 \xi _3^6 a'(t)^2
   \csc ^6(\eta )}{\alpha ^{12} Q_{\alpha ,4}^7}-\frac{15360 \cot (\eta ) \cot (2 \eta )
   \xi _2^2 \xi _3^6 a'(t)^2 \csc ^6(\eta )}{\alpha ^{12} Q_{\alpha ,4}^7}+\frac{10240
   \xi _2^4 \xi _3^4 a'(t)^2 \csc ^6(\eta )}{\alpha ^{12} Q_{\alpha ,4}^7}+\frac{49920
   \xi _1^2 \xi _2^2 \xi _3^4 a'(t)^2 \csc ^6(\eta )}{\alpha ^{10} Q_{\alpha
   ,4}^7}+\frac{5120 \sec ^2(\eta ) \xi _3^4 \xi _4^4 a'(t)^2 \csc ^6(\eta )}{\alpha
   ^{12} Q_{\alpha ,4}^7}-\frac{10240 \sec ^2(\eta ) \xi _3^6 \xi _4^2 a'(t)^2 \csc
   ^6(\eta )}{\alpha ^{12} Q_{\alpha ,4}^7}+\frac{11520 \sec ^2(\eta ) \xi _1^2 \xi _3^4
   \xi _4^2 a'(t)^2 \csc ^6(\eta )}{\alpha ^{10} Q_{\alpha ,4}^7}+\frac{55680 \xi _1^2
   \xi _3^4 \xi _4^2 a'(t)^2 \csc ^6(\eta )}{\alpha ^{10} Q_{\alpha ,4}^7}+\frac{10240
   \sec ^2(\eta ) \xi _2^2 \xi _3^4 \xi _4^2 a'(t)^2 \csc ^6(\eta )}{\alpha ^{12}
   Q_{\alpha ,4}^7}+\frac{89280 \xi _2^2 \xi _3^4 \xi _4^2 a'(t)^2 \csc ^6(\eta )}{\alpha
   ^{12} Q_{\alpha ,4}^7}-\frac{107520 \cot ^2(\eta ) \xi _2^4 \xi _3^6 a'(t)^2 \csc
   ^6(\eta )}{\alpha ^{14} Q_{\alpha ,4}^8}-\frac{779520 \cot ^2(\eta ) \xi _1^2 \xi _2^2
   \xi _3^6 a'(t)^2 \csc ^6(\eta )}{\alpha ^{12} Q_{\alpha ,4}^8}+\frac{161280 \cot (\eta
   ) \cot (2 \eta ) \xi _1^2 \xi _2^2 \xi _3^6 a'(t)^2 \csc ^6(\eta )}{\alpha ^{12}
   Q_{\alpha ,4}^8}-\frac{107520 \xi _1^2 \xi _2^4 \xi _3^4 a'(t)^2 \csc ^6(\eta
   )}{\alpha ^{12} Q_{\alpha ,4}^8}-\frac{53760 \sec ^2(\eta ) \xi _1^2 \xi _3^4 \xi _4^4
   a'(t)^2 \csc ^6(\eta )}{\alpha ^{12} Q_{\alpha ,4}^8}-\frac{53760 \sec ^2(\eta ) \xi
   _2^2 \xi _3^4 \xi _4^4 a'(t)^2 \csc ^6(\eta )}{\alpha ^{14} Q_{\alpha
   ,4}^8}+\frac{107520 \sec ^2(\eta ) \xi _1^2 \xi _3^6 \xi _4^2 a'(t)^2 \csc ^6(\eta
   )}{\alpha ^{12} Q_{\alpha ,4}^8}+\frac{107520 \sec ^2(\eta ) \xi _2^2 \xi _3^6 \xi
   _4^2 a'(t)^2 \csc ^6(\eta )}{\alpha ^{14} Q_{\alpha ,4}^8}-\frac{107520 \xi _2^4 \xi
   _3^4 \xi _4^2 a'(t)^2 \csc ^6(\eta )}{\alpha ^{14} Q_{\alpha ,4}^8}-\frac{107520 \sec
   ^2(\eta ) \xi _1^2 \xi _2^2 \xi _3^4 \xi _4^2 a'(t)^2 \csc ^6(\eta )}{\alpha ^{12}
   Q_{\alpha ,4}^8}-\frac{887040 \xi _1^2 \xi _2^2 \xi _3^4 \xi _4^2 a'(t)^2 \csc ^6(\eta
   )}{\alpha ^{12} Q_{\alpha ,4}^8}+\frac{1290240 \cot ^2(\eta ) \xi _1^2 \xi _2^4 \xi
   _3^6 a'(t)^2 \csc ^6(\eta )}{\alpha ^{14} Q_{\alpha ,4}^9}+\frac{645120 \sec ^2(\eta )
   \xi _1^2 \xi _2^2 \xi _3^4 \xi _4^4 a'(t)^2 \csc ^6(\eta )}{\alpha ^{14} Q_{\alpha
   ,4}^9}-\frac{1290240 \sec ^2(\eta ) \xi _1^2 \xi _2^2 \xi _3^6 \xi _4^2 a'(t)^2 \csc
   ^6(\eta )}{\alpha ^{14} Q_{\alpha ,4}^9}+\frac{1290240 \xi _1^2 \xi _2^4 \xi _3^4 \xi
   _4^2 a'(t)^2 \csc ^6(\eta )}{\alpha ^{14} Q_{\alpha ,4}^9}+\frac{1664 \xi _3^6 a'(t)^2
   a''(t) \csc ^6(\eta )}{\alpha ^9 Q_{\alpha ,4}^6}-\frac{30080 \xi _1^2 \xi _3^6
   a'(t)^2 a''(t) \csc ^6(\eta )}{\alpha ^9 Q_{\alpha ,4}^7}+\frac{53760 \xi _1^4 \xi
   _3^6 a'(t)^2 a''(t) \csc ^6(\eta )}{\alpha ^9 Q_{\alpha ,4}^8}-\frac{96 \xi _3^4
   a''(t) \csc ^6(\eta )}{\alpha ^7 Q_{\alpha ,4}^5}+\frac{640 \cot ^2(\eta ) \xi _3^6
   a''(t) \csc ^6(\eta )}{\alpha ^9 Q_{\alpha ,4}^6}+\frac{640 \xi _1^2 \xi _3^4 a''(t)
   \csc ^6(\eta )}{\alpha ^7 Q_{\alpha ,4}^6}+\frac{640 \xi _2^2 \xi _3^4 a''(t) \csc
   ^6(\eta )}{\alpha ^9 Q_{\alpha ,4}^6}+\frac{640 \xi _3^4 \xi _4^2 a''(t) \csc ^6(\eta
   )}{\alpha ^9 Q_{\alpha ,4}^6}-\frac{5120 \cot ^2(\eta ) \xi _1^2 \xi _3^6 a''(t) \csc
   ^6(\eta )}{\alpha ^9 Q_{\alpha ,4}^7}-\frac{5760 \cot ^2(\eta ) \xi _2^2 \xi _3^6
   a''(t) \csc ^6(\eta )}{\alpha ^{11} Q_{\alpha ,4}^7}-\frac{5120 \xi _1^2 \xi _2^2 \xi
   _3^4 a''(t) \csc ^6(\eta )}{\alpha ^9 Q_{\alpha ,4}^7}-\frac{5120 \xi _1^2 \xi _3^4
   \xi _4^2 a''(t) \csc ^6(\eta )}{\alpha ^9 Q_{\alpha ,4}^7}-\frac{5760 \xi _2^2 \xi
   _3^4 \xi _4^2 a''(t) \csc ^6(\eta )}{\alpha ^{11} Q_{\alpha ,4}^7}+\frac{53760 \cot
   ^2(\eta ) \xi _1^2 \xi _2^2 \xi _3^6 a''(t) \csc ^6(\eta )}{\alpha ^{11} Q_{\alpha
   ,4}^8}+\frac{53760 \xi _1^2 \xi _2^2 \xi _3^4 \xi _4^2 a''(t) \csc ^6(\eta )}{\alpha
   ^{11} Q_{\alpha ,4}^8}-\frac{50 \xi _3^2 \csc ^6(\eta )}{\alpha ^6 Q_{\alpha
   ,4}^4}+\frac{1520 \cot ^2(\eta ) \xi _3^4 \csc ^6(\eta )}{\alpha ^8 Q_{\alpha
   ,4}^5}-\frac{512 \cot (\eta ) \cot (2 \eta ) \xi _3^4 \csc ^6(\eta )}{\alpha ^8
   Q_{\alpha ,4}^5}-\frac{32 \xi _3^4 \csc ^6(\eta )}{\alpha ^8 Q_{\alpha
   ,4}^5}+\frac{688 \xi _2^2 \xi _3^2 \csc ^6(\eta )}{\alpha ^8 Q_{\alpha
   ,4}^5}-\frac{3008 \cot ^4(\eta ) \xi _3^6 \csc ^6(\eta )}{\alpha ^{10} Q_{\alpha
   ,4}^6}+\frac{1280 \cot ^3(\eta ) \cot (2 \eta ) \xi _3^6 \csc ^6(\eta )}{\alpha ^{10}
   Q_{\alpha ,4}^6}-\frac{24592 \cot ^2(\eta ) \xi _2^2 \xi _3^4 \csc ^6(\eta )}{\alpha
   ^{10} Q_{\alpha ,4}^6}+\frac{8960 \cot (\eta ) \cot (2 \eta ) \xi _2^2 \xi _3^4 \csc
   ^6(\eta )}{\alpha ^{10} Q_{\alpha ,4}^6}+\frac{240 \xi _2^2 \xi _3^4 \csc ^6(\eta
   )}{\alpha ^{10} Q_{\alpha ,4}^6}-\frac{1024 \xi _2^4 \xi _3^2 \csc ^6(\eta )}{\alpha
   ^{10} Q_{\alpha ,4}^6}+\frac{1664 \sec ^2(\eta ) \xi _3^4 \xi _4^2 \csc ^6(\eta
   )}{\alpha ^{10} Q_{\alpha ,4}^6}+\frac{6016 \xi _3^4 \xi _4^2 \csc ^6(\eta )}{\alpha
   ^{10} Q_{\alpha ,4}^6}+\frac{57280 \cot ^4(\eta ) \xi _2^2 \xi _3^6 \csc ^6(\eta
   )}{\alpha ^{12} Q_{\alpha ,4}^7}-\frac{26880 \cot ^3(\eta ) \cot (2 \eta ) \xi _2^2
   \xi _3^6 \csc ^6(\eta )}{\alpha ^{12} Q_{\alpha ,4}^7}+\frac{40960 \cot ^2(\eta ) \xi
   _2^4 \xi _3^4 \csc ^6(\eta )}{\alpha ^{12} Q_{\alpha ,4}^7}-\frac{15360 \cot (\eta )
   \cot (2 \eta ) \xi _2^4 \xi _3^4 \csc ^6(\eta )}{\alpha ^{12} Q_{\alpha
   ,4}^7}-\frac{30080 \sec ^2(\eta ) \xi _2^2 \xi _3^4 \xi _4^2 \csc ^6(\eta )}{\alpha
   ^{12} Q_{\alpha ,4}^7}-\frac{114560 \xi _2^2 \xi _3^4 \xi _4^2 \csc ^6(\eta )}{\alpha
   ^{12} Q_{\alpha ,4}^7}-\frac{107520 \cot ^4(\eta ) \xi _2^4 \xi _3^6 \csc ^6(\eta
   )}{\alpha ^{14} Q_{\alpha ,4}^8}+\frac{53760 \cot ^3(\eta ) \cot (2 \eta ) \xi _2^4
   \xi _3^6 \csc ^6(\eta )}{\alpha ^{14} Q_{\alpha ,4}^8}+\frac{53760 \sec ^2(\eta ) \xi
   _2^4 \xi _3^4 \xi _4^2 \csc ^6(\eta )}{\alpha ^{14} Q_{\alpha ,4}^8}+\frac{215040 \xi
   _2^4 \xi _3^4 \xi _4^2 \csc ^6(\eta )}{\alpha ^{14} Q_{\alpha ,4}^8}+\frac{2560 \cot
   (2 \eta ) \sec (\eta ) \xi _3^4 \xi _4^2 a'(t)^2 \csc ^5(\eta )}{\alpha ^{10}
   Q_{\alpha ,4}^6}-\frac{23040 \cot (2 \eta ) \sec (\eta ) \xi _1^2 \xi _3^4 \xi _4^2
   a'(t)^2 \csc ^5(\eta )}{\alpha ^{10} Q_{\alpha ,4}^7}-\frac{30720 \cot (2 \eta ) \sec
   (\eta ) \xi _2^2 \xi _3^4 \xi _4^2 a'(t)^2 \csc ^5(\eta )}{\alpha ^{12} Q_{\alpha
   ,4}^7}+\frac{322560 \cot (2 \eta ) \sec (\eta ) \xi _1^2 \xi _2^2 \xi _3^4 \xi _4^2
   a'(t)^2 \csc ^5(\eta )}{\alpha ^{12} Q_{\alpha ,4}^8}-\frac{3840 \cot (2 \eta ) \sec
   (\eta ) \xi _3^4 \xi _4^2 \csc ^5(\eta )}{\alpha ^{10} Q_{\alpha ,4}^6}+\frac{80640
   \cot (2 \eta ) \sec (\eta ) \xi _2^2 \xi _3^4 \xi _4^2 \csc ^5(\eta )}{\alpha ^{12}
   Q_{\alpha ,4}^7}-\frac{161280 \cot (2 \eta ) \sec (\eta ) \xi _2^4 \xi _3^4 \xi _4^2
   \csc ^5(\eta )}{\alpha ^{14} Q_{\alpha ,4}^8}+\frac{612 \xi _3^4 a'(t)^4 \csc ^4(\eta
   )}{\alpha ^8 Q_{\alpha ,4}^5}-\frac{6960 \xi _1^2 \xi _3^4 a'(t)^4 \csc ^4(\eta
   )}{\alpha ^8 Q_{\alpha ,4}^6}-\frac{8256 \xi _2^2 \xi _3^4 a'(t)^4 \csc ^4(\eta
   )}{\alpha ^{10} Q_{\alpha ,4}^6}-\frac{8256 \sec ^2(\eta ) \xi _3^4 \xi _4^2 a'(t)^4
   \csc ^4(\eta )}{\alpha ^{10} Q_{\alpha ,4}^6}+\frac{8640 \xi _1^4 \xi _3^4 a'(t)^4
   \csc ^4(\eta )}{\alpha ^8 Q_{\alpha ,4}^7}+\frac{15360 \xi _2^4 \xi _3^4 a'(t)^4 \csc
   ^4(\eta )}{\alpha ^{12} Q_{\alpha ,4}^7}+\frac{141120 \xi _1^2 \xi _2^2 \xi _3^4
   a'(t)^4 \csc ^4(\eta )}{\alpha ^{10} Q_{\alpha ,4}^7}+\frac{15360 \sec ^4(\eta ) \xi
   _3^4 \xi _4^4 a'(t)^4 \csc ^4(\eta )}{\alpha ^{12} Q_{\alpha ,4}^7}+\frac{141120 \sec
   ^2(\eta ) \xi _1^2 \xi _3^4 \xi _4^2 a'(t)^4 \csc ^4(\eta )}{\alpha ^{10} Q_{\alpha
   ,4}^7}+\frac{30720 \sec ^2(\eta ) \xi _2^2 \xi _3^4 \xi _4^2 a'(t)^4 \csc ^4(\eta
   )}{\alpha ^{12} Q_{\alpha ,4}^7}-\frac{322560 \xi _1^2 \xi _2^4 \xi _3^4 a'(t)^4 \csc
   ^4(\eta )}{\alpha ^{12} Q_{\alpha ,4}^8}-\frac{241920 \xi _1^4 \xi _2^2 \xi _3^4
   a'(t)^4 \csc ^4(\eta )}{\alpha ^{10} Q_{\alpha ,4}^8}-\frac{322560 \sec ^4(\eta ) \xi
   _1^2 \xi _3^4 \xi _4^4 a'(t)^4 \csc ^4(\eta )}{\alpha ^{12} Q_{\alpha
   ,4}^8}-\frac{241920 \sec ^2(\eta ) \xi _1^4 \xi _3^4 \xi _4^2 a'(t)^4 \csc ^4(\eta
   )}{\alpha ^{10} Q_{\alpha ,4}^8}-\frac{645120 \sec ^2(\eta ) \xi _1^2 \xi _2^2 \xi
   _3^4 \xi _4^2 a'(t)^4 \csc ^4(\eta )}{\alpha ^{12} Q_{\alpha ,4}^8}+\frac{645120 \xi
   _1^4 \xi _2^4 \xi _3^4 a'(t)^4 \csc ^4(\eta )}{\alpha ^{12} Q_{\alpha
   ,4}^9}+\frac{645120 \sec ^4(\eta ) \xi _1^4 \xi _3^4 \xi _4^4 a'(t)^4 \csc ^4(\eta
   )}{\alpha ^{12} Q_{\alpha ,4}^9}+\frac{1290240 \sec ^2(\eta ) \xi _1^4 \xi _2^2 \xi
   _3^4 \xi _4^2 a'(t)^4 \csc ^4(\eta )}{\alpha ^{12} Q_{\alpha ,4}^9}-\frac{45 \xi _3^2
   a'(t)^2 \csc ^4(\eta )}{\alpha ^6 Q_{\alpha ,4}^4}+\frac{1624 \cot ^2(\eta ) \xi _3^4
   a'(t)^2 \csc ^4(\eta )}{\alpha ^8 Q_{\alpha ,4}^5}-\frac{32 \sec ^2(\eta ) \xi _3^4
   a'(t)^2 \csc ^4(\eta )}{\alpha ^8 Q_{\alpha ,4}^5}-\frac{928 \cot (\eta ) \cot (2 \eta
   ) \xi _3^4 a'(t)^2 \csc ^4(\eta )}{\alpha ^8 Q_{\alpha ,4}^5}-\frac{128 \xi _3^4
   a'(t)^2 \csc ^4(\eta )}{\alpha ^8 Q_{\alpha ,4}^5}+\frac{96 \xi _1^2 \xi _3^2 a'(t)^2
   \csc ^4(\eta )}{\alpha ^6 Q_{\alpha ,4}^5}+\frac{1328 \xi _2^2 \xi _3^2 a'(t)^2 \csc
   ^4(\eta )}{\alpha ^8 Q_{\alpha ,4}^5}+\frac{400 \sec ^2(\eta ) \xi _3^2 \xi _4^2
   a'(t)^2 \csc ^4(\eta )}{\alpha ^8 Q_{\alpha ,4}^5}+\frac{696 \xi _3^2 \xi _4^2 a'(t)^2
   \csc ^4(\eta )}{\alpha ^8 Q_{\alpha ,4}^5}-\frac{9120 \cot ^2(\eta ) \xi _1^2 \xi _3^4
   a'(t)^2 \csc ^4(\eta )}{\alpha ^8 Q_{\alpha ,4}^6}+\frac{240 \sec ^2(\eta ) \xi _1^2
   \xi _3^4 a'(t)^2 \csc ^4(\eta )}{\alpha ^8 Q_{\alpha ,4}^6}+\frac{5760 \cot (\eta )
   \cot (2 \eta ) \xi _1^2 \xi _3^4 a'(t)^2 \csc ^4(\eta )}{\alpha ^8 Q_{\alpha
   ,4}^6}+\frac{960 \xi _1^2 \xi _3^4 a'(t)^2 \csc ^4(\eta )}{\alpha ^8 Q_{\alpha
   ,4}^6}-\frac{30480 \cot ^2(\eta ) \xi _2^2 \xi _3^4 a'(t)^2 \csc ^4(\eta )}{\alpha
   ^{10} Q_{\alpha ,4}^6}-\frac{1280 \cot ^2(2 \eta ) \xi _2^2 \xi _3^4 a'(t)^2 \csc
   ^4(\eta )}{\alpha ^{10} Q_{\alpha ,4}^6}+\frac{2560 \csc ^2(2 \eta ) \xi _2^2 \xi _3^4
   a'(t)^2 \csc ^4(\eta )}{\alpha ^{10} Q_{\alpha ,4}^6}+\frac{16960 \cot (\eta ) \cot (2
   \eta ) \xi _2^2 \xi _3^4 a'(t)^2 \csc ^4(\eta )}{\alpha ^{10} Q_{\alpha
   ,4}^6}-\frac{640 \sec ^4(\eta ) \xi _3^2 \xi _4^4 a'(t)^2 \csc ^4(\eta )}{\alpha ^{10}
   Q_{\alpha ,4}^6}-\frac{960 \sec ^2(\eta ) \xi _3^2 \xi _4^4 a'(t)^2 \csc ^4(\eta
   )}{\alpha ^{10} Q_{\alpha ,4}^6}-\frac{5440 \xi _2^4 \xi _3^2 a'(t)^2 \csc ^4(\eta
   )}{\alpha ^{10} Q_{\alpha ,4}^6}-\frac{7200 \xi _1^2 \xi _2^2 \xi _3^2 a'(t)^2 \csc
   ^4(\eta )}{\alpha ^8 Q_{\alpha ,4}^6}-\frac{640 \sec ^4(\eta ) \xi _3^4 \xi _4^2
   a'(t)^2 \csc ^4(\eta )}{\alpha ^{10} Q_{\alpha ,4}^6}+\frac{9600 \sec ^2(\eta ) \xi
   _3^4 \xi _4^2 a'(t)^2 \csc ^4(\eta )}{\alpha ^{10} Q_{\alpha ,4}^6}-\frac{2400 \sec
   ^2(\eta ) \xi _1^2 \xi _3^2 \xi _4^2 a'(t)^2 \csc ^4(\eta )}{\alpha ^8 Q_{\alpha
   ,4}^6}-\frac{4320 \xi _1^2 \xi _3^2 \xi _4^2 a'(t)^2 \csc ^4(\eta )}{\alpha ^8
   Q_{\alpha ,4}^6}-\frac{6080 \sec ^2(\eta ) \xi _2^2 \xi _3^2 \xi _4^2 a'(t)^2 \csc
   ^4(\eta )}{\alpha ^{10} Q_{\alpha ,4}^6}-\frac{11520 \xi _2^2 \xi _3^2 \xi _4^2
   a'(t)^2 \csc ^4(\eta )}{\alpha ^{10} Q_{\alpha ,4}^6}+\frac{84160 \cot ^2(\eta ) \xi
   _2^4 \xi _3^4 a'(t)^2 \csc ^4(\eta )}{\alpha ^{12} Q_{\alpha ,4}^7}-\frac{30720 \cot
   (\eta ) \cot (2 \eta ) \xi _2^4 \xi _3^4 a'(t)^2 \csc ^4(\eta )}{\alpha ^{12}
   Q_{\alpha ,4}^7}+\frac{233280 \cot ^2(\eta ) \xi _1^2 \xi _2^2 \xi _3^4 a'(t)^2 \csc
   ^4(\eta )}{\alpha ^{10} Q_{\alpha ,4}^7}+\frac{11520 \cot ^2(2 \eta ) \xi _1^2 \xi
   _2^2 \xi _3^4 a'(t)^2 \csc ^4(\eta )}{\alpha ^{10} Q_{\alpha ,4}^7}-\frac{23040 \csc
   ^2(2 \eta ) \xi _1^2 \xi _2^2 \xi _3^4 a'(t)^2 \csc ^4(\eta )}{\alpha ^{10} Q_{\alpha
   ,4}^7}-\frac{138240 \cot (\eta ) \cot (2 \eta ) \xi _1^2 \xi _2^2 \xi _3^4 a'(t)^2
   \csc ^4(\eta )}{\alpha ^{10} Q_{\alpha ,4}^7}-\frac{20480 \sec ^4(\eta ) \xi _3^4 \xi
   _4^4 a'(t)^2 \csc ^4(\eta )}{\alpha ^{12} Q_{\alpha ,4}^7}+\frac{5760 \sec ^4(\eta )
   \xi _1^2 \xi _3^2 \xi _4^4 a'(t)^2 \csc ^4(\eta )}{\alpha ^{10} Q_{\alpha
   ,4}^7}+\frac{8640 \sec ^2(\eta ) \xi _1^2 \xi _3^2 \xi _4^4 a'(t)^2 \csc ^4(\eta
   )}{\alpha ^{10} Q_{\alpha ,4}^7}+\frac{5120 \sec ^4(\eta ) \xi _2^2 \xi _3^2 \xi _4^4
   a'(t)^2 \csc ^4(\eta )}{\alpha ^{12} Q_{\alpha ,4}^7}+\frac{10240 \sec ^2(\eta ) \xi
   _2^2 \xi _3^2 \xi _4^4 a'(t)^2 \csc ^4(\eta )}{\alpha ^{12} Q_{\alpha
   ,4}^7}+\frac{5120 \xi _2^6 \xi _3^2 a'(t)^2 \csc ^4(\eta )}{\alpha ^{12} Q_{\alpha
   ,4}^7}+\frac{44160 \xi _1^2 \xi _2^4 \xi _3^2 a'(t)^2 \csc ^4(\eta )}{\alpha ^{10}
   Q_{\alpha ,4}^7}+\frac{5760 \sec ^4(\eta ) \xi _1^2 \xi _3^4 \xi _4^2 a'(t)^2 \csc
   ^4(\eta )}{\alpha ^{10} Q_{\alpha ,4}^7}-\frac{76800 \sec ^2(\eta ) \xi _1^2 \xi _3^4
   \xi _4^2 a'(t)^2 \csc ^4(\eta )}{\alpha ^{10} Q_{\alpha ,4}^7}+\frac{5120 \sec ^4(\eta
   ) \xi _2^2 \xi _3^4 \xi _4^2 a'(t)^2 \csc ^4(\eta )}{\alpha ^{12} Q_{\alpha
   ,4}^7}-\frac{137600 \sec ^2(\eta ) \xi _2^2 \xi _3^4 \xi _4^2 a'(t)^2 \csc ^4(\eta
   )}{\alpha ^{12} Q_{\alpha ,4}^7}+\frac{10240 \sec ^2(\eta ) \xi _2^4 \xi _3^2 \xi _4^2
   a'(t)^2 \csc ^4(\eta )}{\alpha ^{12} Q_{\alpha ,4}^7}+\frac{20480 \xi _2^4 \xi _3^2
   \xi _4^2 a'(t)^2 \csc ^4(\eta )}{\alpha ^{12} Q_{\alpha ,4}^7}+\frac{49920 \sec
   ^2(\eta ) \xi _1^2 \xi _2^2 \xi _3^2 \xi _4^2 a'(t)^2 \csc ^4(\eta )}{\alpha ^{10}
   Q_{\alpha ,4}^7}+\frac{94080 \xi _1^2 \xi _2^2 \xi _3^2 \xi _4^2 a'(t)^2 \csc ^4(\eta
   )}{\alpha ^{10} Q_{\alpha ,4}^7}-\frac{53760 \cot ^2(\eta ) \xi _2^6 \xi _3^4 a'(t)^2
   \csc ^4(\eta )}{\alpha ^{14} Q_{\alpha ,4}^8}-\frac{833280 \cot ^2(\eta ) \xi _1^2 \xi
   _2^4 \xi _3^4 a'(t)^2 \csc ^4(\eta )}{\alpha ^{12} Q_{\alpha ,4}^8}+\frac{322560 \cot
   (\eta ) \cot (2 \eta ) \xi _1^2 \xi _2^4 \xi _3^4 a'(t)^2 \csc ^4(\eta )}{\alpha ^{12}
   Q_{\alpha ,4}^8}+\frac{215040 \sec ^4(\eta ) \xi _1^2 \xi _3^4 \xi _4^4 a'(t)^2 \csc
   ^4(\eta )}{\alpha ^{12} Q_{\alpha ,4}^8}+\frac{215040 \sec ^4(\eta ) \xi _2^2 \xi _3^4
   \xi _4^4 a'(t)^2 \csc ^4(\eta )}{\alpha ^{14} Q_{\alpha ,4}^8}-\frac{53760 \sec
   ^4(\eta ) \xi _1^2 \xi _2^2 \xi _3^2 \xi _4^4 a'(t)^2 \csc ^4(\eta )}{\alpha ^{12}
   Q_{\alpha ,4}^8}-\frac{107520 \sec ^2(\eta ) \xi _1^2 \xi _2^2 \xi _3^2 \xi _4^4
   a'(t)^2 \csc ^4(\eta )}{\alpha ^{12} Q_{\alpha ,4}^8}-\frac{53760 \xi _1^2 \xi _2^6
   \xi _3^2 a'(t)^2 \csc ^4(\eta )}{\alpha ^{12} Q_{\alpha ,4}^8}+\frac{215040 \sec
   ^2(\eta ) \xi _2^4 \xi _3^4 \xi _4^2 a'(t)^2 \csc ^4(\eta )}{\alpha ^{14} Q_{\alpha
   ,4}^8}-\frac{53760 \sec ^4(\eta ) \xi _1^2 \xi _2^2 \xi _3^4 \xi _4^2 a'(t)^2 \csc
   ^4(\eta )}{\alpha ^{12} Q_{\alpha ,4}^8}+\frac{1344000 \sec ^2(\eta ) \xi _1^2 \xi
   _2^2 \xi _3^4 \xi _4^2 a'(t)^2 \csc ^4(\eta )}{\alpha ^{12} Q_{\alpha
   ,4}^8}-\frac{107520 \sec ^2(\eta ) \xi _1^2 \xi _2^4 \xi _3^2 \xi _4^2 a'(t)^2 \csc
   ^4(\eta )}{\alpha ^{12} Q_{\alpha ,4}^8}-\frac{215040 \xi _1^2 \xi _2^4 \xi _3^2 \xi
   _4^2 a'(t)^2 \csc ^4(\eta )}{\alpha ^{12} Q_{\alpha ,4}^8}+\frac{645120 \cot ^2(\eta )
   \xi _1^2 \xi _2^6 \xi _3^4 a'(t)^2 \csc ^4(\eta )}{\alpha ^{14} Q_{\alpha
   ,4}^9}-\frac{2580480 \sec ^4(\eta ) \xi _1^2 \xi _2^2 \xi _3^4 \xi _4^4 a'(t)^2 \csc
   ^4(\eta )}{\alpha ^{14} Q_{\alpha ,4}^9}-\frac{2580480 \sec ^2(\eta ) \xi _1^2 \xi
   _2^4 \xi _3^4 \xi _4^2 a'(t)^2 \csc ^4(\eta )}{\alpha ^{14} Q_{\alpha ,4}^9}+\frac{112
   \xi _3^4 a''(t)^2 \csc ^4(\eta )}{\alpha ^6 Q_{\alpha ,4}^5}-\frac{1664 \xi _1^2 \xi
   _3^4 a''(t)^2 \csc ^4(\eta )}{\alpha ^6 Q_{\alpha ,4}^6}+\frac{2560 \xi _1^4 \xi _3^4
   a''(t)^2 \csc ^4(\eta )}{\alpha ^6 Q_{\alpha ,4}^7}-\frac{688 \xi _3^4 a'(t)^2 a''(t)
   \csc ^4(\eta )}{\alpha ^7 Q_{\alpha ,4}^5}+\frac{8032 \xi _1^2 \xi _3^4 a'(t)^2 a''(t)
   \csc ^4(\eta )}{\alpha ^7 Q_{\alpha ,4}^6}+\frac{4992 \xi _2^2 \xi _3^4 a'(t)^2 a''(t)
   \csc ^4(\eta )}{\alpha ^9 Q_{\alpha ,4}^6}+\frac{4992 \sec ^2(\eta ) \xi _3^4 \xi _4^2
   a'(t)^2 a''(t) \csc ^4(\eta )}{\alpha ^9 Q_{\alpha ,4}^6}-\frac{9600 \xi _1^4 \xi _3^4
   a'(t)^2 a''(t) \csc ^4(\eta )}{\alpha ^7 Q_{\alpha ,4}^7}-\frac{90240 \xi _1^2 \xi
   _2^2 \xi _3^4 a'(t)^2 a''(t) \csc ^4(\eta )}{\alpha ^9 Q_{\alpha ,4}^7}-\frac{90240
   \sec ^2(\eta ) \xi _1^2 \xi _3^4 \xi _4^2 a'(t)^2 a''(t) \csc ^4(\eta )}{\alpha ^9
   Q_{\alpha ,4}^7}+\frac{161280 \xi _1^4 \xi _2^2 \xi _3^4 a'(t)^2 a''(t) \csc ^4(\eta
   )}{\alpha ^9 Q_{\alpha ,4}^8}+\frac{161280 \sec ^2(\eta ) \xi _1^4 \xi _3^4 \xi _4^2
   a'(t)^2 a''(t) \csc ^4(\eta )}{\alpha ^9 Q_{\alpha ,4}^8}+\frac{6 \xi _3^2 a''(t) \csc
   ^4(\eta )}{\alpha ^5 Q_{\alpha ,4}^4}-\frac{336 \cot ^2(\eta ) \xi _3^4 a''(t) \csc
   ^4(\eta )}{\alpha ^7 Q_{\alpha ,4}^5}+\frac{192 \cot (\eta ) \cot (2 \eta ) \xi _3^4
   a''(t) \csc ^4(\eta )}{\alpha ^7 Q_{\alpha ,4}^5}+\frac{64 \xi _1^2 \xi _3^2 a''(t)
   \csc ^4(\eta )}{\alpha ^5 Q_{\alpha ,4}^5}-\frac{288 \xi _2^2 \xi _3^2 a''(t) \csc
   ^4(\eta )}{\alpha ^7 Q_{\alpha ,4}^5}-\frac{96 \sec ^2(\eta ) \xi _3^2 \xi _4^2 a''(t)
   \csc ^4(\eta )}{\alpha ^7 Q_{\alpha ,4}^5}-\frac{144 \xi _3^2 \xi _4^2 a''(t) \csc
   ^4(\eta )}{\alpha ^7 Q_{\alpha ,4}^5}+\frac{1600 \cot ^2(\eta ) \xi _1^2 \xi _3^4
   a''(t) \csc ^4(\eta )}{\alpha ^7 Q_{\alpha ,4}^6}-\frac{1280 \cot (\eta ) \cot (2 \eta
   ) \xi _1^2 \xi _3^4 a''(t) \csc ^4(\eta )}{\alpha ^7 Q_{\alpha ,4}^6}+\frac{4320 \cot
   ^2(\eta ) \xi _2^2 \xi _3^4 a''(t) \csc ^4(\eta )}{\alpha ^9 Q_{\alpha
   ,4}^6}-\frac{1920 \cot (\eta ) \cot (2 \eta ) \xi _2^2 \xi _3^4 a''(t) \csc ^4(\eta
   )}{\alpha ^9 Q_{\alpha ,4}^6}+\frac{640 \xi _2^4 \xi _3^2 a''(t) \csc ^4(\eta
   )}{\alpha ^9 Q_{\alpha ,4}^6}+\frac{1280 \xi _1^2 \xi _2^2 \xi _3^2 a''(t) \csc
   ^4(\eta )}{\alpha ^7 Q_{\alpha ,4}^6}-\frac{1280 \sec ^2(\eta ) \xi _3^4 \xi _4^2
   a''(t) \csc ^4(\eta )}{\alpha ^9 Q_{\alpha ,4}^6}+\frac{640 \sec ^2(\eta ) \xi _1^2
   \xi _3^2 \xi _4^2 a''(t) \csc ^4(\eta )}{\alpha ^7 Q_{\alpha ,4}^6}+\frac{960 \xi _1^2
   \xi _3^2 \xi _4^2 a''(t) \csc ^4(\eta )}{\alpha ^7 Q_{\alpha ,4}^6}+\frac{640 \sec
   ^2(\eta ) \xi _2^2 \xi _3^2 \xi _4^2 a''(t) \csc ^4(\eta )}{\alpha ^9 Q_{\alpha
   ,4}^6}+\frac{1280 \xi _2^2 \xi _3^2 \xi _4^2 a''(t) \csc ^4(\eta )}{\alpha ^9
   Q_{\alpha ,4}^6}-\frac{5760 \cot ^2(\eta ) \xi _2^4 \xi _3^4 a''(t) \csc ^4(\eta
   )}{\alpha ^{11} Q_{\alpha ,4}^7}-\frac{28800 \cot ^2(\eta ) \xi _1^2 \xi _2^2 \xi _3^4
   a''(t) \csc ^4(\eta )}{\alpha ^9 Q_{\alpha ,4}^7}+\frac{15360 \cot (\eta ) \cot (2
   \eta ) \xi _1^2 \xi _2^2 \xi _3^4 a''(t) \csc ^4(\eta )}{\alpha ^9 Q_{\alpha
   ,4}^7}-\frac{5120 \xi _1^2 \xi _2^4 \xi _3^2 a''(t) \csc ^4(\eta )}{\alpha ^9
   Q_{\alpha ,4}^7}+\frac{10240 \sec ^2(\eta ) \xi _1^2 \xi _3^4 \xi _4^2 a''(t) \csc
   ^4(\eta )}{\alpha ^9 Q_{\alpha ,4}^7}+\frac{11520 \sec ^2(\eta ) \xi _2^2 \xi _3^4 \xi
   _4^2 a''(t) \csc ^4(\eta )}{\alpha ^{11} Q_{\alpha ,4}^7}-\frac{5120 \sec ^2(\eta )
   \xi _1^2 \xi _2^2 \xi _3^2 \xi _4^2 a''(t) \csc ^4(\eta )}{\alpha ^9 Q_{\alpha
   ,4}^7}-\frac{10240 \xi _1^2 \xi _2^2 \xi _3^2 \xi _4^2 a''(t) \csc ^4(\eta )}{\alpha
   ^9 Q_{\alpha ,4}^7}+\frac{53760 \cot ^2(\eta ) \xi _1^2 \xi _2^4 \xi _3^4 a''(t) \csc
   ^4(\eta )}{\alpha ^{11} Q_{\alpha ,4}^8}-\frac{107520 \sec ^2(\eta ) \xi _1^2 \xi _2^2
   \xi _3^4 \xi _4^2 a''(t) \csc ^4(\eta )}{\alpha ^{11} Q_{\alpha ,4}^8}+\frac{160 \xi
   _3^4 a'(t) a^{(3)}(t) \csc ^4(\eta )}{\alpha ^6 Q_{\alpha ,4}^5}-\frac{2432 \xi _1^2
   \xi _3^4 a'(t) a^{(3)}(t) \csc ^4(\eta )}{\alpha ^6 Q_{\alpha ,4}^6}+\frac{3840 \xi
   _1^4 \xi _3^4 a'(t) a^{(3)}(t) \csc ^4(\eta )}{\alpha ^6 Q_{\alpha ,4}^7}-\frac{7 \csc
   ^4(\eta )}{\alpha ^4 Q_{\alpha ,4}^3}+\frac{8 \xi _2^2 \csc ^4(\eta )}{\alpha ^6
   Q_{\alpha ,4}^4}-\frac{251 \cot ^2(\eta ) \xi _3^2 \csc ^4(\eta )}{\alpha ^6 Q_{\alpha
   ,4}^4}-\frac{32 \cot ^2(2 \eta ) \xi _3^2 \csc ^4(\eta )}{\alpha ^6 Q_{\alpha
   ,4}^4}+\frac{64 \csc ^2(2 \eta ) \xi _3^2 \csc ^4(\eta )}{\alpha ^6 Q_{\alpha
   ,4}^4}+\frac{6 \sec ^2(\eta ) \xi _3^2 \csc ^4(\eta )}{\alpha ^6 Q_{\alpha
   ,4}^4}+\frac{220 \cot (\eta ) \cot (2 \eta ) \xi _3^2 \csc ^4(\eta )}{\alpha ^6
   Q_{\alpha ,4}^4}+\frac{33 \xi _3^2 \csc ^4(\eta )}{\alpha ^6 Q_{\alpha
   ,4}^4}+\frac{868 \cot ^4(\eta ) \xi _3^4 \csc ^4(\eta )}{\alpha ^8 Q_{\alpha
   ,4}^5}-\frac{128 \cot ^2(\eta ) \xi _3^4 \csc ^4(\eta )}{\alpha ^8 Q_{\alpha
   ,4}^5}+\frac{192 \cot ^2(\eta ) \cot ^2(2 \eta ) \xi _3^4 \csc ^4(\eta )}{\alpha ^8
   Q_{\alpha ,4}^5}-\frac{384 \cot ^2(\eta ) \csc ^2(2 \eta ) \xi _3^4 \csc ^4(\eta
   )}{\alpha ^8 Q_{\alpha ,4}^5}-\frac{928 \cot ^3(\eta ) \cot (2 \eta ) \xi _3^4 \csc
   ^4(\eta )}{\alpha ^8 Q_{\alpha ,4}^5}+\frac{3648 \cot ^2(\eta ) \xi _2^2 \xi _3^2 \csc
   ^4(\eta )}{\alpha ^8 Q_{\alpha ,4}^5}+\frac{704 \cot ^2(2 \eta ) \xi _2^2 \xi _3^2
   \csc ^4(\eta )}{\alpha ^8 Q_{\alpha ,4}^5}-\frac{1408 \csc ^2(2 \eta ) \xi _2^2 \xi
   _3^2 \csc ^4(\eta )}{\alpha ^8 Q_{\alpha ,4}^5}-\frac{32 \sec ^2(\eta ) \xi _2^2 \xi
   _3^2 \csc ^4(\eta )}{\alpha ^8 Q_{\alpha ,4}^5}-\frac{3296 \cot (\eta ) \cot (2 \eta )
   \xi _2^2 \xi _3^2 \csc ^4(\eta )}{\alpha ^8 Q_{\alpha ,4}^5}-\frac{192 \xi _2^2 \xi
   _3^2 \csc ^4(\eta )}{\alpha ^8 Q_{\alpha ,4}^5}+\frac{224 \sec ^4(\eta ) \xi _3^2 \xi
   _4^2 \csc ^4(\eta )}{\alpha ^8 Q_{\alpha ,4}^5}-\frac{688 \sec ^2(\eta ) \xi _3^2 \xi
   _4^2 \csc ^4(\eta )}{\alpha ^8 Q_{\alpha ,4}^5}-\frac{512 \xi _3^2 \xi _4^2 \csc
   ^4(\eta )}{\alpha ^8 Q_{\alpha ,4}^5}-\frac{14784 \cot ^4(\eta ) \xi _2^2 \xi _3^4
   \csc ^4(\eta )}{\alpha ^{10} Q_{\alpha ,4}^6}+\frac{960 \cot ^2(\eta ) \xi _2^2 \xi
   _3^4 \csc ^4(\eta )}{\alpha ^{10} Q_{\alpha ,4}^6}-\frac{5120 \cot ^2(\eta ) \cot ^2(2
   \eta ) \xi _2^2 \xi _3^4 \csc ^4(\eta )}{\alpha ^{10} Q_{\alpha ,4}^6}+\frac{10240
   \cot ^2(\eta ) \csc ^2(2 \eta ) \xi _2^2 \xi _3^4 \csc ^4(\eta )}{\alpha ^{10}
   Q_{\alpha ,4}^6}+\frac{16960 \cot ^3(\eta ) \cot (2 \eta ) \xi _2^2 \xi _3^4 \csc
   ^4(\eta )}{\alpha ^{10} Q_{\alpha ,4}^6}-\frac{5632 \cot ^2(\eta ) \xi _2^4 \xi _3^2
   \csc ^4(\eta )}{\alpha ^{10} Q_{\alpha ,4}^6}-\frac{1280 \cot ^2(2 \eta ) \xi _2^4 \xi
   _3^2 \csc ^4(\eta )}{\alpha ^{10} Q_{\alpha ,4}^6}+\frac{2560 \csc ^2(2 \eta ) \xi
   _2^4 \xi _3^2 \csc ^4(\eta )}{\alpha ^{10} Q_{\alpha ,4}^6}+\frac{5120 \cot (\eta )
   \cot (2 \eta ) \xi _2^4 \xi _3^2 \csc ^4(\eta )}{\alpha ^{10} Q_{\alpha
   ,4}^6}-\frac{3008 \sec ^2(\eta ) \xi _3^4 \xi _4^2 \csc ^4(\eta )}{\alpha ^{10}
   Q_{\alpha ,4}^6}-\frac{3328 \sec ^4(\eta ) \xi _2^2 \xi _3^2 \xi _4^2 \csc ^4(\eta
   )}{\alpha ^{10} Q_{\alpha ,4}^6}+\frac{11680 \sec ^2(\eta ) \xi _2^2 \xi _3^2 \xi _4^2
   \csc ^4(\eta )}{\alpha ^{10} Q_{\alpha ,4}^6}+\frac{8768 \xi _2^2 \xi _3^2 \xi _4^2
   \csc ^4(\eta )}{\alpha ^{10} Q_{\alpha ,4}^6}+\frac{25600 \cot ^4(\eta ) \xi _2^4 \xi
   _3^4 \csc ^4(\eta )}{\alpha ^{12} Q_{\alpha ,4}^7}+\frac{11520 \cot ^2(\eta ) \cot
   ^2(2 \eta ) \xi _2^4 \xi _3^4 \csc ^4(\eta )}{\alpha ^{12} Q_{\alpha
   ,4}^7}-\frac{23040 \cot ^2(\eta ) \csc ^2(2 \eta ) \xi _2^4 \xi _3^4 \csc ^4(\eta
   )}{\alpha ^{12} Q_{\alpha ,4}^7}-\frac{30720 \cot ^3(\eta ) \cot (2 \eta ) \xi _2^4
   \xi _3^4 \csc ^4(\eta )}{\alpha ^{12} Q_{\alpha ,4}^7}+\frac{15360 \sec ^4(\eta ) \xi
   _3^4 \xi _4^4 \csc ^4(\eta )}{\alpha ^{12} Q_{\alpha ,4}^7}+\frac{57280 \sec ^2(\eta )
   \xi _2^2 \xi _3^4 \xi _4^2 \csc ^4(\eta )}{\alpha ^{12} Q_{\alpha ,4}^7}+\frac{5120
   \sec ^4(\eta ) \xi _2^4 \xi _3^2 \xi _4^2 \csc ^4(\eta )}{\alpha ^{12} Q_{\alpha
   ,4}^7}-\frac{20480 \sec ^2(\eta ) \xi _2^4 \xi _3^2 \xi _4^2 \csc ^4(\eta )}{\alpha
   ^{12} Q_{\alpha ,4}^7}-\frac{15360 \xi _2^4 \xi _3^2 \xi _4^2 \csc ^4(\eta )}{\alpha
   ^{12} Q_{\alpha ,4}^7}-\frac{322560 \sec ^4(\eta ) \xi _2^2 \xi _3^4 \xi _4^4 \csc
   ^4(\eta )}{\alpha ^{14} Q_{\alpha ,4}^8}-\frac{107520 \sec ^2(\eta ) \xi _2^4 \xi _3^4
   \xi _4^2 \csc ^4(\eta )}{\alpha ^{14} Q_{\alpha ,4}^8}+\frac{645120 \sec ^4(\eta ) \xi
   _2^4 \xi _3^4 \xi _4^4 \csc ^4(\eta )}{\alpha ^{16} Q_{\alpha ,4}^9}-\frac{928 \cot (2
   \eta ) \sec (\eta ) \xi _3^2 \xi _4^2 a'(t)^2 \csc ^3(\eta )}{\alpha ^8 Q_{\alpha
   ,4}^5}+\frac{1280 \cot (2 \eta ) \sec ^3(\eta ) \xi _3^2 \xi _4^4 a'(t)^2 \csc ^3(\eta
   )}{\alpha ^{10} Q_{\alpha ,4}^6}-\frac{1280 \cot (2 \eta ) \sec ^3(\eta ) \xi _3^4 \xi
   _4^2 a'(t)^2 \csc ^3(\eta )}{\alpha ^{10} Q_{\alpha ,4}^6}+\frac{5760 \cot (2 \eta )
   \sec (\eta ) \xi _1^2 \xi _3^2 \xi _4^2 a'(t)^2 \csc ^3(\eta )}{\alpha ^8 Q_{\alpha
   ,4}^6}+\frac{16960 \cot (2 \eta ) \sec (\eta ) \xi _2^2 \xi _3^2 \xi _4^2 a'(t)^2 \csc
   ^3(\eta )}{\alpha ^{10} Q_{\alpha ,4}^6}-\frac{11520 \cot (2 \eta ) \sec ^3(\eta ) \xi
   _1^2 \xi _3^2 \xi _4^4 a'(t)^2 \csc ^3(\eta )}{\alpha ^{10} Q_{\alpha
   ,4}^7}-\frac{15360 \cot (2 \eta ) \sec ^3(\eta ) \xi _2^2 \xi _3^2 \xi _4^4 a'(t)^2
   \csc ^3(\eta )}{\alpha ^{12} Q_{\alpha ,4}^7}+\frac{11520 \cot (2 \eta ) \sec ^3(\eta
   ) \xi _1^2 \xi _3^4 \xi _4^2 a'(t)^2 \csc ^3(\eta )}{\alpha ^{10} Q_{\alpha
   ,4}^7}+\frac{15360 \cot (2 \eta ) \sec ^3(\eta ) \xi _2^2 \xi _3^4 \xi _4^2 a'(t)^2
   \csc ^3(\eta )}{\alpha ^{12} Q_{\alpha ,4}^7}-\frac{30720 \cot (2 \eta ) \sec (\eta )
   \xi _2^4 \xi _3^2 \xi _4^2 a'(t)^2 \csc ^3(\eta )}{\alpha ^{12} Q_{\alpha
   ,4}^7}-\frac{138240 \cot (2 \eta ) \sec (\eta ) \xi _1^2 \xi _2^2 \xi _3^2 \xi _4^2
   a'(t)^2 \csc ^3(\eta )}{\alpha ^{10} Q_{\alpha ,4}^7}+\frac{161280 \cot (2 \eta ) \sec
   ^3(\eta ) \xi _1^2 \xi _2^2 \xi _3^2 \xi _4^4 a'(t)^2 \csc ^3(\eta )}{\alpha ^{12}
   Q_{\alpha ,4}^8}-\frac{161280 \cot (2 \eta ) \sec ^3(\eta ) \xi _1^2 \xi _2^2 \xi _3^4
   \xi _4^2 a'(t)^2 \csc ^3(\eta )}{\alpha ^{12} Q_{\alpha ,4}^8}+\frac{322560 \cot (2
   \eta ) \sec (\eta ) \xi _1^2 \xi _2^4 \xi _3^2 \xi _4^2 a'(t)^2 \csc ^3(\eta )}{\alpha
   ^{12} Q_{\alpha ,4}^8}+\frac{192 \cot (2 \eta ) \sec (\eta ) \xi _3^2 \xi _4^2 a''(t)
   \csc ^3(\eta )}{\alpha ^7 Q_{\alpha ,4}^5}-\frac{1280 \cot (2 \eta ) \sec (\eta ) \xi
   _1^2 \xi _3^2 \xi _4^2 a''(t) \csc ^3(\eta )}{\alpha ^7 Q_{\alpha ,4}^6}-\frac{1920
   \cot (2 \eta ) \sec (\eta ) \xi _2^2 \xi _3^2 \xi _4^2 a''(t) \csc ^3(\eta )}{\alpha
   ^9 Q_{\alpha ,4}^6}+\frac{15360 \cot (2 \eta ) \sec (\eta ) \xi _1^2 \xi _2^2 \xi _3^2
   \xi _4^2 a''(t) \csc ^3(\eta )}{\alpha ^9 Q_{\alpha ,4}^7}-\frac{12 \cot (2 \eta )
   \sec (\eta ) \xi _3^2 \csc ^3(\eta )}{\alpha ^6 Q_{\alpha ,4}^4}+\frac{96 \cot (2 \eta
   ) \sec (\eta ) \xi _2^2 \xi _3^2 \csc ^3(\eta )}{\alpha ^8 Q_{\alpha ,4}^5}+\frac{928
   \cot (2 \eta ) \sec (\eta ) \xi _3^2 \xi _4^2 \csc ^3(\eta )}{\alpha ^8 Q_{\alpha
   ,4}^5}-\frac{16960 \cot (2 \eta ) \sec (\eta ) \xi _2^2 \xi _3^2 \xi _4^2 \csc ^3(\eta
   )}{\alpha ^{10} Q_{\alpha ,4}^6}+\frac{30720 \cot (2 \eta ) \sec (\eta ) \xi _2^4 \xi
   _3^2 \xi _4^2 \csc ^3(\eta )}{\alpha ^{12} Q_{\alpha ,4}^7}-\frac{7 \xi _3^2 a'(t)^4
   \csc ^2(\eta )}{\alpha ^6 Q_{\alpha ,4}^4}-\frac{216 \xi _1^2 \xi _3^2 a'(t)^4 \csc
   ^2(\eta )}{\alpha ^6 Q_{\alpha ,4}^5}+\frac{1224 \xi _2^2 \xi _3^2 a'(t)^4 \csc
   ^2(\eta )}{\alpha ^8 Q_{\alpha ,4}^5}+\frac{1224 \sec ^2(\eta ) \xi _3^2 \xi _4^2
   a'(t)^4 \csc ^2(\eta )}{\alpha ^8 Q_{\alpha ,4}^5}-\frac{8256 \sec ^4(\eta ) \xi _3^2
   \xi _4^4 a'(t)^4 \csc ^2(\eta )}{\alpha ^{10} Q_{\alpha ,4}^6}+\frac{480 \xi _1^4 \xi
   _3^2 a'(t)^4 \csc ^2(\eta )}{\alpha ^6 Q_{\alpha ,4}^6}-\frac{8256 \xi _2^4 \xi _3^2
   a'(t)^4 \csc ^2(\eta )}{\alpha ^{10} Q_{\alpha ,4}^6}-\frac{13920 \xi _1^2 \xi _2^2
   \xi _3^2 a'(t)^4 \csc ^2(\eta )}{\alpha ^8 Q_{\alpha ,4}^6}-\frac{13920 \sec ^2(\eta )
   \xi _1^2 \xi _3^2 \xi _4^2 a'(t)^4 \csc ^2(\eta )}{\alpha ^8 Q_{\alpha
   ,4}^6}-\frac{16512 \sec ^2(\eta ) \xi _2^2 \xi _3^2 \xi _4^2 a'(t)^4 \csc ^2(\eta
   )}{\alpha ^{10} Q_{\alpha ,4}^6}+\frac{10240 \sec ^6(\eta ) \xi _3^2 \xi _4^6 a'(t)^4
   \csc ^2(\eta )}{\alpha ^{12} Q_{\alpha ,4}^7}+\frac{141120 \sec ^4(\eta ) \xi _1^2 \xi
   _3^2 \xi _4^4 a'(t)^4 \csc ^2(\eta )}{\alpha ^{10} Q_{\alpha ,4}^7}+\frac{30720 \sec
   ^4(\eta ) \xi _2^2 \xi _3^2 \xi _4^4 a'(t)^4 \csc ^2(\eta )}{\alpha ^{12} Q_{\alpha
   ,4}^7}+\frac{10240 \xi _2^6 \xi _3^2 a'(t)^4 \csc ^2(\eta )}{\alpha ^{12} Q_{\alpha
   ,4}^7}+\frac{141120 \xi _1^2 \xi _2^4 \xi _3^2 a'(t)^4 \csc ^2(\eta )}{\alpha ^{10}
   Q_{\alpha ,4}^7}+\frac{17280 \xi _1^4 \xi _2^2 \xi _3^2 a'(t)^4 \csc ^2(\eta )}{\alpha
   ^8 Q_{\alpha ,4}^7}+\frac{17280 \sec ^2(\eta ) \xi _1^4 \xi _3^2 \xi _4^2 a'(t)^4 \csc
   ^2(\eta )}{\alpha ^8 Q_{\alpha ,4}^7}+\frac{30720 \sec ^2(\eta ) \xi _2^4 \xi _3^2 \xi
   _4^2 a'(t)^4 \csc ^2(\eta )}{\alpha ^{12} Q_{\alpha ,4}^7}+\frac{282240 \sec ^2(\eta )
   \xi _1^2 \xi _2^2 \xi _3^2 \xi _4^2 a'(t)^4 \csc ^2(\eta )}{\alpha ^{10} Q_{\alpha
   ,4}^7}-\frac{215040 \sec ^6(\eta ) \xi _1^2 \xi _3^2 \xi _4^6 a'(t)^4 \csc ^2(\eta
   )}{\alpha ^{12} Q_{\alpha ,4}^8}-\frac{241920 \sec ^4(\eta ) \xi _1^4 \xi _3^2 \xi
   _4^4 a'(t)^4 \csc ^2(\eta )}{\alpha ^{10} Q_{\alpha ,4}^8}-\frac{645120 \sec ^4(\eta )
   \xi _1^2 \xi _2^2 \xi _3^2 \xi _4^4 a'(t)^4 \csc ^2(\eta )}{\alpha ^{12} Q_{\alpha
   ,4}^8}-\frac{215040 \xi _1^2 \xi _2^6 \xi _3^2 a'(t)^4 \csc ^2(\eta )}{\alpha ^{12}
   Q_{\alpha ,4}^8}-\frac{241920 \xi _1^4 \xi _2^4 \xi _3^2 a'(t)^4 \csc ^2(\eta
   )}{\alpha ^{10} Q_{\alpha ,4}^8}-\frac{645120 \sec ^2(\eta ) \xi _1^2 \xi _2^4 \xi
   _3^2 \xi _4^2 a'(t)^4 \csc ^2(\eta )}{\alpha ^{12} Q_{\alpha ,4}^8}-\frac{483840 \sec
   ^2(\eta ) \xi _1^4 \xi _2^2 \xi _3^2 \xi _4^2 a'(t)^4 \csc ^2(\eta )}{\alpha ^{10}
   Q_{\alpha ,4}^8}+\frac{430080 \sec ^6(\eta ) \xi _1^4 \xi _3^2 \xi _4^6 a'(t)^4 \csc
   ^2(\eta )}{\alpha ^{12} Q_{\alpha ,4}^9}+\frac{1290240 \sec ^4(\eta ) \xi _1^4 \xi
   _2^2 \xi _3^2 \xi _4^4 a'(t)^4 \csc ^2(\eta )}{\alpha ^{12} Q_{\alpha
   ,4}^9}+\frac{430080 \xi _1^4 \xi _2^6 \xi _3^2 a'(t)^4 \csc ^2(\eta )}{\alpha ^{12}
   Q_{\alpha ,4}^9}+\frac{1290240 \sec ^2(\eta ) \xi _1^4 \xi _2^4 \xi _3^2 \xi _4^2
   a'(t)^4 \csc ^2(\eta )}{\alpha ^{12} Q_{\alpha ,4}^9}-\frac{3 a'(t)^2 \csc ^2(\eta
   )}{\alpha ^4 Q_{\alpha ,4}^3}-\frac{3 \xi _1^2 a'(t)^2 \csc ^2(\eta )}{\alpha ^4
   Q_{\alpha ,4}^4}+\frac{13 \xi _2^2 a'(t)^2 \csc ^2(\eta )}{\alpha ^6 Q_{\alpha
   ,4}^4}-\frac{87 \cot ^2(\eta ) \xi _3^2 a'(t)^2 \csc ^2(\eta )}{\alpha ^6 Q_{\alpha
   ,4}^4}+\frac{13 \sec ^2(\eta ) \xi _3^2 a'(t)^2 \csc ^2(\eta )}{\alpha ^6 Q_{\alpha
   ,4}^4}+\frac{116 \cot (\eta ) \cot (2 \eta ) \xi _3^2 a'(t)^2 \csc ^2(\eta )}{\alpha
   ^6 Q_{\alpha ,4}^4}+\frac{52 \xi _3^2 a'(t)^2 \csc ^2(\eta )}{\alpha ^6 Q_{\alpha
   ,4}^4}+\frac{13 \sec ^2(\eta ) \xi _4^2 a'(t)^2 \csc ^2(\eta )}{\alpha ^6 Q_{\alpha
   ,4}^4}-\frac{32 \xi _2^4 a'(t)^2 \csc ^2(\eta )}{\alpha ^8 Q_{\alpha ,4}^5}-\frac{32
   \sec ^4(\eta ) \xi _4^4 a'(t)^2 \csc ^2(\eta )}{\alpha ^8 Q_{\alpha ,4}^5}-\frac{48
   \xi _1^2 \xi _2^2 a'(t)^2 \csc ^2(\eta )}{\alpha ^6 Q_{\alpha ,4}^5}+\frac{216 \cot
   ^2(\eta ) \xi _1^2 \xi _3^2 a'(t)^2 \csc ^2(\eta )}{\alpha ^6 Q_{\alpha
   ,4}^5}-\frac{48 \sec ^2(\eta ) \xi _1^2 \xi _3^2 a'(t)^2 \csc ^2(\eta )}{\alpha ^6
   Q_{\alpha ,4}^5}-\frac{288 \cot (\eta ) \cot (2 \eta ) \xi _1^2 \xi _3^2 a'(t)^2 \csc
   ^2(\eta )}{\alpha ^6 Q_{\alpha ,4}^5}-\frac{192 \xi _1^2 \xi _3^2 a'(t)^2 \csc ^2(\eta
   )}{\alpha ^6 Q_{\alpha ,4}^5}+\frac{2552 \cot ^2(\eta ) \xi _2^2 \xi _3^2 a'(t)^2 \csc
   ^2(\eta )}{\alpha ^8 Q_{\alpha ,4}^5}+\frac{928 \cot ^2(2 \eta ) \xi _2^2 \xi _3^2
   a'(t)^2 \csc ^2(\eta )}{\alpha ^8 Q_{\alpha ,4}^5}-\frac{1856 \csc ^2(2 \eta ) \xi
   _2^2 \xi _3^2 a'(t)^2 \csc ^2(\eta )}{\alpha ^8 Q_{\alpha ,4}^5}-\frac{64 \sec ^2(\eta
   ) \xi _2^2 \xi _3^2 a'(t)^2 \csc ^2(\eta )}{\alpha ^8 Q_{\alpha ,4}^5}-\frac{3712 \cot
   (\eta ) \cot (2 \eta ) \xi _2^2 \xi _3^2 a'(t)^2 \csc ^2(\eta )}{\alpha ^8 Q_{\alpha
   ,4}^5}-\frac{256 \xi _2^2 \xi _3^2 a'(t)^2 \csc ^2(\eta )}{\alpha ^8 Q_{\alpha
   ,4}^5}-\frac{48 \sec ^2(\eta ) \xi _1^2 \xi _4^2 a'(t)^2 \csc ^2(\eta )}{\alpha ^6
   Q_{\alpha ,4}^5}-\frac{64 \sec ^2(\eta ) \xi _2^2 \xi _4^2 a'(t)^2 \csc ^2(\eta
   )}{\alpha ^8 Q_{\alpha ,4}^5}+\frac{400 \sec ^4(\eta ) \xi _3^2 \xi _4^2 a'(t)^2 \csc
   ^2(\eta )}{\alpha ^8 Q_{\alpha ,4}^5}-\frac{2112 \sec ^2(\eta ) \xi _3^2 \xi _4^2
   a'(t)^2 \csc ^2(\eta )}{\alpha ^8 Q_{\alpha ,4}^5}+\frac{240 \xi _1^2 \xi _2^4 a'(t)^2
   \csc ^2(\eta )}{\alpha ^8 Q_{\alpha ,4}^6}+\frac{240 \sec ^4(\eta ) \xi _1^2 \xi _4^4
   a'(t)^2 \csc ^2(\eta )}{\alpha ^8 Q_{\alpha ,4}^6}-\frac{1280 \sec ^6(\eta ) \xi _3^2
   \xi _4^4 a'(t)^2 \csc ^2(\eta )}{\alpha ^{10} Q_{\alpha ,4}^6}+\frac{9600 \sec ^4(\eta
   ) \xi _3^2 \xi _4^4 a'(t)^2 \csc ^2(\eta )}{\alpha ^{10} Q_{\alpha ,4}^6}-\frac{10560
   \cot ^2(\eta ) \xi _2^4 \xi _3^2 a'(t)^2 \csc ^2(\eta )}{\alpha ^{10} Q_{\alpha
   ,4}^6}-\frac{2560 \cot ^2(2 \eta ) \xi _2^4 \xi _3^2 a'(t)^2 \csc ^2(\eta )}{\alpha
   ^{10} Q_{\alpha ,4}^6}+\frac{5120 \csc ^2(2 \eta ) \xi _2^4 \xi _3^2 a'(t)^2 \csc
   ^2(\eta )}{\alpha ^{10} Q_{\alpha ,4}^6}+\frac{15680 \cot (\eta ) \cot (2 \eta ) \xi
   _2^4 \xi _3^2 a'(t)^2 \csc ^2(\eta )}{\alpha ^{10} Q_{\alpha ,4}^6}-\frac{13920 \cot
   ^2(\eta ) \xi _1^2 \xi _2^2 \xi _3^2 a'(t)^2 \csc ^2(\eta )}{\alpha ^8 Q_{\alpha
   ,4}^6}-\frac{5760 \cot ^2(2 \eta ) \xi _1^2 \xi _2^2 \xi _3^2 a'(t)^2 \csc ^2(\eta
   )}{\alpha ^8 Q_{\alpha ,4}^6}+\frac{11520 \csc ^2(2 \eta ) \xi _1^2 \xi _2^2 \xi _3^2
   a'(t)^2 \csc ^2(\eta )}{\alpha ^8 Q_{\alpha ,4}^6}+\frac{480 \sec ^2(\eta ) \xi _1^2
   \xi _2^2 \xi _3^2 a'(t)^2 \csc ^2(\eta )}{\alpha ^8 Q_{\alpha ,4}^6}+\frac{20160 \cot
   (\eta ) \cot (2 \eta ) \xi _1^2 \xi _2^2 \xi _3^2 a'(t)^2 \csc ^2(\eta )}{\alpha ^8
   Q_{\alpha ,4}^6}+\frac{1920 \xi _1^2 \xi _2^2 \xi _3^2 a'(t)^2 \csc ^2(\eta )}{\alpha
   ^8 Q_{\alpha ,4}^6}-\frac{960 \sec ^4(\eta ) \xi _3^4 \xi _4^2 a'(t)^2 \csc ^2(\eta
   )}{\alpha ^{10} Q_{\alpha ,4}^6}+\frac{480 \sec ^2(\eta ) \xi _1^2 \xi _2^2 \xi _4^2
   a'(t)^2 \csc ^2(\eta )}{\alpha ^8 Q_{\alpha ,4}^6}-\frac{2400 \sec ^4(\eta ) \xi _1^2
   \xi _3^2 \xi _4^2 a'(t)^2 \csc ^2(\eta )}{\alpha ^8 Q_{\alpha ,4}^6}+\frac{11520 \sec
   ^2(\eta ) \xi _1^2 \xi _3^2 \xi _4^2 a'(t)^2 \csc ^2(\eta )}{\alpha ^8 Q_{\alpha
   ,4}^6}-\frac{6080 \sec ^4(\eta ) \xi _2^2 \xi _3^2 \xi _4^2 a'(t)^2 \csc ^2(\eta
   )}{\alpha ^{10} Q_{\alpha ,4}^6}-\frac{2560 \cot ^2(2 \eta ) \sec ^2(\eta ) \xi _2^2
   \xi _3^2 \xi _4^2 a'(t)^2 \csc ^2(\eta )}{\alpha ^{10} Q_{\alpha ,4}^6}+\frac{5120
   \csc ^2(2 \eta ) \sec ^2(\eta ) \xi _2^2 \xi _3^2 \xi _4^2 a'(t)^2 \csc ^2(\eta
   )}{\alpha ^{10} Q_{\alpha ,4}^6}+\frac{37920 \sec ^2(\eta ) \xi _2^2 \xi _3^2 \xi _4^2
   a'(t)^2 \csc ^2(\eta )}{\alpha ^{10} Q_{\alpha ,4}^6}-\frac{10240 \sec ^6(\eta ) \xi
   _3^2 \xi _4^6 a'(t)^2 \csc ^2(\eta )}{\alpha ^{12} Q_{\alpha ,4}^7}+\frac{5120 \sec
   ^6(\eta ) \xi _3^4 \xi _4^4 a'(t)^2 \csc ^2(\eta )}{\alpha ^{12} Q_{\alpha
   ,4}^7}+\frac{11520 \sec ^6(\eta ) \xi _1^2 \xi _3^2 \xi _4^4 a'(t)^2 \csc ^2(\eta
   )}{\alpha ^{10} Q_{\alpha ,4}^7}-\frac{76800 \sec ^4(\eta ) \xi _1^2 \xi _3^2 \xi _4^4
   a'(t)^2 \csc ^2(\eta )}{\alpha ^{10} Q_{\alpha ,4}^7}+\frac{10240 \sec ^6(\eta ) \xi
   _2^2 \xi _3^2 \xi _4^4 a'(t)^2 \csc ^2(\eta )}{\alpha ^{12} Q_{\alpha
   ,4}^7}-\frac{137600 \sec ^4(\eta ) \xi _2^2 \xi _3^2 \xi _4^4 a'(t)^2 \csc ^2(\eta
   )}{\alpha ^{12} Q_{\alpha ,4}^7}+\frac{10240 \cot ^2(\eta ) \xi _2^6 \xi _3^2 a'(t)^2
   \csc ^2(\eta )}{\alpha ^{12} Q_{\alpha ,4}^7}-\frac{15360 \cot (\eta ) \cot (2 \eta )
   \xi _2^6 \xi _3^2 a'(t)^2 \csc ^2(\eta )}{\alpha ^{12} Q_{\alpha ,4}^7}+\frac{85440
   \cot ^2(\eta ) \xi _1^2 \xi _2^4 \xi _3^2 a'(t)^2 \csc ^2(\eta )}{\alpha ^{10}
   Q_{\alpha ,4}^7}+\frac{23040 \cot ^2(2 \eta ) \xi _1^2 \xi _2^4 \xi _3^2 a'(t)^2 \csc
   ^2(\eta )}{\alpha ^{10} Q_{\alpha ,4}^7}-\frac{46080 \csc ^2(2 \eta ) \xi _1^2 \xi
   _2^4 \xi _3^2 a'(t)^2 \csc ^2(\eta )}{\alpha ^{10} Q_{\alpha ,4}^7}-\frac{126720 \cot
   (\eta ) \cot (2 \eta ) \xi _1^2 \xi _2^4 \xi _3^2 a'(t)^2 \csc ^2(\eta )}{\alpha ^{10}
   Q_{\alpha ,4}^7}+\frac{8640 \sec ^4(\eta ) \xi _1^2 \xi _3^4 \xi _4^2 a'(t)^2 \csc
   ^2(\eta )}{\alpha ^{10} Q_{\alpha ,4}^7}+\frac{10240 \sec ^4(\eta ) \xi _2^2 \xi _3^4
   \xi _4^2 a'(t)^2 \csc ^2(\eta )}{\alpha ^{12} Q_{\alpha ,4}^7}+\frac{10240 \sec
   ^4(\eta ) \xi _2^4 \xi _3^2 \xi _4^2 a'(t)^2 \csc ^2(\eta )}{\alpha ^{12} Q_{\alpha
   ,4}^7}-\frac{127360 \sec ^2(\eta ) \xi _2^4 \xi _3^2 \xi _4^2 a'(t)^2 \csc ^2(\eta
   )}{\alpha ^{12} Q_{\alpha ,4}^7}+\frac{49920 \sec ^4(\eta ) \xi _1^2 \xi _2^2 \xi _3^2
   \xi _4^2 a'(t)^2 \csc ^2(\eta )}{\alpha ^{10} Q_{\alpha ,4}^7}+\frac{23040 \cot ^2(2
   \eta ) \sec ^2(\eta ) \xi _1^2 \xi _2^2 \xi _3^2 \xi _4^2 a'(t)^2 \csc ^2(\eta
   )}{\alpha ^{10} Q_{\alpha ,4}^7}-\frac{46080 \csc ^2(2 \eta ) \sec ^2(\eta ) \xi _1^2
   \xi _2^2 \xi _3^2 \xi _4^2 a'(t)^2 \csc ^2(\eta )}{\alpha ^{10} Q_{\alpha
   ,4}^7}-\frac{278400 \sec ^2(\eta ) \xi _1^2 \xi _2^2 \xi _3^2 \xi _4^2 a'(t)^2 \csc
   ^2(\eta )}{\alpha ^{10} Q_{\alpha ,4}^7}+\frac{107520 \sec ^6(\eta ) \xi _1^2 \xi _3^2
   \xi _4^6 a'(t)^2 \csc ^2(\eta )}{\alpha ^{12} Q_{\alpha ,4}^8}+\frac{107520 \sec
   ^6(\eta ) \xi _2^2 \xi _3^2 \xi _4^6 a'(t)^2 \csc ^2(\eta )}{\alpha ^{14} Q_{\alpha
   ,4}^8}-\frac{53760 \sec ^6(\eta ) \xi _1^2 \xi _3^4 \xi _4^4 a'(t)^2 \csc ^2(\eta
   )}{\alpha ^{12} Q_{\alpha ,4}^8}-\frac{53760 \sec ^6(\eta ) \xi _2^2 \xi _3^4 \xi _4^4
   a'(t)^2 \csc ^2(\eta )}{\alpha ^{14} Q_{\alpha ,4}^8}+\frac{215040 \sec ^4(\eta ) \xi
   _2^4 \xi _3^2 \xi _4^4 a'(t)^2 \csc ^2(\eta )}{\alpha ^{14} Q_{\alpha
   ,4}^8}-\frac{107520 \sec ^6(\eta ) \xi _1^2 \xi _2^2 \xi _3^2 \xi _4^4 a'(t)^2 \csc
   ^2(\eta )}{\alpha ^{12} Q_{\alpha ,4}^8}+\frac{1344000 \sec ^4(\eta ) \xi _1^2 \xi
   _2^2 \xi _3^2 \xi _4^4 a'(t)^2 \csc ^2(\eta )}{\alpha ^{12} Q_{\alpha
   ,4}^8}-\frac{107520 \cot ^2(\eta ) \xi _1^2 \xi _2^6 \xi _3^2 a'(t)^2 \csc ^2(\eta
   )}{\alpha ^{12} Q_{\alpha ,4}^8}+\frac{161280 \cot (\eta ) \cot (2 \eta ) \xi _1^2 \xi
   _2^6 \xi _3^2 a'(t)^2 \csc ^2(\eta )}{\alpha ^{12} Q_{\alpha ,4}^8}-\frac{107520 \sec
   ^4(\eta ) \xi _1^2 \xi _2^2 \xi _3^4 \xi _4^2 a'(t)^2 \csc ^2(\eta )}{\alpha ^{12}
   Q_{\alpha ,4}^8}+\frac{107520 \sec ^2(\eta ) \xi _2^6 \xi _3^2 \xi _4^2 a'(t)^2 \csc
   ^2(\eta )}{\alpha ^{14} Q_{\alpha ,4}^8}-\frac{107520 \sec ^4(\eta ) \xi _1^2 \xi _2^4
   \xi _3^2 \xi _4^2 a'(t)^2 \csc ^2(\eta )}{\alpha ^{12} Q_{\alpha ,4}^8}+\frac{1236480
   \sec ^2(\eta ) \xi _1^2 \xi _2^4 \xi _3^2 \xi _4^2 a'(t)^2 \csc ^2(\eta )}{\alpha
   ^{12} Q_{\alpha ,4}^8}-\frac{1290240 \sec ^6(\eta ) \xi _1^2 \xi _2^2 \xi _3^2 \xi
   _4^6 a'(t)^2 \csc ^2(\eta )}{\alpha ^{14} Q_{\alpha ,4}^9}+\frac{645120 \sec ^6(\eta )
   \xi _1^2 \xi _2^2 \xi _3^4 \xi _4^4 a'(t)^2 \csc ^2(\eta )}{\alpha ^{14} Q_{\alpha
   ,4}^9}-\frac{2580480 \sec ^4(\eta ) \xi _1^2 \xi _2^4 \xi _3^2 \xi _4^4 a'(t)^2 \csc
   ^2(\eta )}{\alpha ^{14} Q_{\alpha ,4}^9}-\frac{1290240 \sec ^2(\eta ) \xi _1^2 \xi
   _2^6 \xi _3^2 \xi _4^2 a'(t)^2 \csc ^2(\eta )}{\alpha ^{14} Q_{\alpha ,4}^9}+\frac{20
   \xi _3^2 a''(t)^2 \csc ^2(\eta )}{\alpha ^4 Q_{\alpha ,4}^4}-\frac{432 \xi _1^2 \xi
   _3^2 a''(t)^2 \csc ^2(\eta )}{\alpha ^4 Q_{\alpha ,4}^5}+\frac{224 \xi _2^2 \xi _3^2
   a''(t)^2 \csc ^2(\eta )}{\alpha ^6 Q_{\alpha ,4}^5}+\frac{224 \sec ^2(\eta ) \xi _3^2
   \xi _4^2 a''(t)^2 \csc ^2(\eta )}{\alpha ^6 Q_{\alpha ,4}^5}+\frac{768 \xi _1^4 \xi
   _3^2 a''(t)^2 \csc ^2(\eta )}{\alpha ^4 Q_{\alpha ,4}^6}-\frac{3328 \xi _1^2 \xi _2^2
   \xi _3^2 a''(t)^2 \csc ^2(\eta )}{\alpha ^6 Q_{\alpha ,4}^6}-\frac{3328 \sec ^2(\eta )
   \xi _1^2 \xi _3^2 \xi _4^2 a''(t)^2 \csc ^2(\eta )}{\alpha ^6 Q_{\alpha
   ,4}^6}+\frac{5120 \xi _1^4 \xi _2^2 \xi _3^2 a''(t)^2 \csc ^2(\eta )}{\alpha ^6
   Q_{\alpha ,4}^7}+\frac{5120 \sec ^2(\eta ) \xi _1^4 \xi _3^2 \xi _4^2 a''(t)^2 \csc
   ^2(\eta )}{\alpha ^6 Q_{\alpha ,4}^7}-\frac{8 \xi _3^2 a'(t)^2 a''(t) \csc ^2(\eta
   )}{\alpha ^5 Q_{\alpha ,4}^4}+\frac{672 \xi _1^2 \xi _3^2 a'(t)^2 a''(t) \csc ^2(\eta
   )}{\alpha ^5 Q_{\alpha ,4}^5}-\frac{1376 \xi _2^2 \xi _3^2 a'(t)^2 a''(t) \csc ^2(\eta
   )}{\alpha ^7 Q_{\alpha ,4}^5}-\frac{1376 \sec ^2(\eta ) \xi _3^2 \xi _4^2 a'(t)^2
   a''(t) \csc ^2(\eta )}{\alpha ^7 Q_{\alpha ,4}^5}+\frac{4992 \sec ^4(\eta ) \xi _3^2
   \xi _4^4 a'(t)^2 a''(t) \csc ^2(\eta )}{\alpha ^9 Q_{\alpha ,4}^6}-\frac{1344 \xi _1^4
   \xi _3^2 a'(t)^2 a''(t) \csc ^2(\eta )}{\alpha ^5 Q_{\alpha ,4}^6}+\frac{4992 \xi _2^4
   \xi _3^2 a'(t)^2 a''(t) \csc ^2(\eta )}{\alpha ^9 Q_{\alpha ,4}^6}+\frac{16064 \xi
   _1^2 \xi _2^2 \xi _3^2 a'(t)^2 a''(t) \csc ^2(\eta )}{\alpha ^7 Q_{\alpha
   ,4}^6}+\frac{16064 \sec ^2(\eta ) \xi _1^2 \xi _3^2 \xi _4^2 a'(t)^2 a''(t) \csc
   ^2(\eta )}{\alpha ^7 Q_{\alpha ,4}^6}+\frac{9984 \sec ^2(\eta ) \xi _2^2 \xi _3^2 \xi
   _4^2 a'(t)^2 a''(t) \csc ^2(\eta )}{\alpha ^9 Q_{\alpha ,4}^6}-\frac{90240 \sec
   ^4(\eta ) \xi _1^2 \xi _3^2 \xi _4^4 a'(t)^2 a''(t) \csc ^2(\eta )}{\alpha ^9
   Q_{\alpha ,4}^7}-\frac{90240 \xi _1^2 \xi _2^4 \xi _3^2 a'(t)^2 a''(t) \csc ^2(\eta
   )}{\alpha ^9 Q_{\alpha ,4}^7}-\frac{19200 \xi _1^4 \xi _2^2 \xi _3^2 a'(t)^2 a''(t)
   \csc ^2(\eta )}{\alpha ^7 Q_{\alpha ,4}^7}-\frac{19200 \sec ^2(\eta ) \xi _1^4 \xi
   _3^2 \xi _4^2 a'(t)^2 a''(t) \csc ^2(\eta )}{\alpha ^7 Q_{\alpha ,4}^7}-\frac{180480
   \sec ^2(\eta ) \xi _1^2 \xi _2^2 \xi _3^2 \xi _4^2 a'(t)^2 a''(t) \csc ^2(\eta
   )}{\alpha ^9 Q_{\alpha ,4}^7}+\frac{161280 \sec ^4(\eta ) \xi _1^4 \xi _3^2 \xi _4^4
   a'(t)^2 a''(t) \csc ^2(\eta )}{\alpha ^9 Q_{\alpha ,4}^8}+\frac{161280 \xi _1^4 \xi
   _2^4 \xi _3^2 a'(t)^2 a''(t) \csc ^2(\eta )}{\alpha ^9 Q_{\alpha ,4}^8}+\frac{322560
   \sec ^2(\eta ) \xi _1^4 \xi _2^2 \xi _3^2 \xi _4^2 a'(t)^2 a''(t) \csc ^2(\eta
   )}{\alpha ^9 Q_{\alpha ,4}^8}-\frac{a''(t) \csc ^2(\eta )}{\alpha ^3 Q_{\alpha
   ,4}^3}+\frac{10 \xi _1^2 a''(t) \csc ^2(\eta )}{\alpha ^3 Q_{\alpha ,4}^4}-\frac{6 \xi
   _2^2 a''(t) \csc ^2(\eta )}{\alpha ^5 Q_{\alpha ,4}^4}+\frac{18 \cot ^2(\eta ) \xi
   _3^2 a''(t) \csc ^2(\eta )}{\alpha ^5 Q_{\alpha ,4}^4}-\frac{6 \sec ^2(\eta ) \xi _3^2
   a''(t) \csc ^2(\eta )}{\alpha ^5 Q_{\alpha ,4}^4}-\frac{24 \cot (\eta ) \cot (2 \eta )
   \xi _3^2 a''(t) \csc ^2(\eta )}{\alpha ^5 Q_{\alpha ,4}^4}-\frac{24 \xi _3^2 a''(t)
   \csc ^2(\eta )}{\alpha ^5 Q_{\alpha ,4}^4}-\frac{6 \sec ^2(\eta ) \xi _4^2 a''(t) \csc
   ^2(\eta )}{\alpha ^5 Q_{\alpha ,4}^4}+\frac{32 \xi _1^2 \xi _2^2 a''(t) \csc ^2(\eta
   )}{\alpha ^5 Q_{\alpha ,4}^5}+\frac{48 \cot ^2(\eta ) \xi _1^2 \xi _3^2 a''(t) \csc
   ^2(\eta )}{\alpha ^5 Q_{\alpha ,4}^5}+\frac{32 \sec ^2(\eta ) \xi _1^2 \xi _3^2 a''(t)
   \csc ^2(\eta )}{\alpha ^5 Q_{\alpha ,4}^5}-\frac{64 \cot (\eta ) \cot (2 \eta ) \xi
   _1^2 \xi _3^2 a''(t) \csc ^2(\eta )}{\alpha ^5 Q_{\alpha ,4}^5}+\frac{128 \xi _1^2 \xi
   _3^2 a''(t) \csc ^2(\eta )}{\alpha ^5 Q_{\alpha ,4}^5}-\frac{528 \cot ^2(\eta ) \xi
   _2^2 \xi _3^2 a''(t) \csc ^2(\eta )}{\alpha ^7 Q_{\alpha ,4}^5}-\frac{192 \cot ^2(2
   \eta ) \xi _2^2 \xi _3^2 a''(t) \csc ^2(\eta )}{\alpha ^7 Q_{\alpha ,4}^5}+\frac{384
   \csc ^2(2 \eta ) \xi _2^2 \xi _3^2 a''(t) \csc ^2(\eta )}{\alpha ^7 Q_{\alpha
   ,4}^5}+\frac{768 \cot (\eta ) \cot (2 \eta ) \xi _2^2 \xi _3^2 a''(t) \csc ^2(\eta
   )}{\alpha ^7 Q_{\alpha ,4}^5}+\frac{32 \sec ^2(\eta ) \xi _1^2 \xi _4^2 a''(t) \csc
   ^2(\eta )}{\alpha ^5 Q_{\alpha ,4}^5}-\frac{96 \sec ^4(\eta ) \xi _3^2 \xi _4^2 a''(t)
   \csc ^2(\eta )}{\alpha ^7 Q_{\alpha ,4}^5}+\frac{384 \sec ^2(\eta ) \xi _3^2 \xi _4^2
   a''(t) \csc ^2(\eta )}{\alpha ^7 Q_{\alpha ,4}^5}-\frac{1280 \sec ^4(\eta ) \xi _3^2
   \xi _4^4 a''(t) \csc ^2(\eta )}{\alpha ^9 Q_{\alpha ,4}^6}+\frac{1280 \cot ^2(\eta )
   \xi _2^4 \xi _3^2 a''(t) \csc ^2(\eta )}{\alpha ^9 Q_{\alpha ,4}^6}-\frac{1920 \cot
   (\eta ) \cot (2 \eta ) \xi _2^4 \xi _3^2 a''(t) \csc ^2(\eta )}{\alpha ^9 Q_{\alpha
   ,4}^6}+\frac{2240 \cot ^2(\eta ) \xi _1^2 \xi _2^2 \xi _3^2 a''(t) \csc ^2(\eta
   )}{\alpha ^7 Q_{\alpha ,4}^6}+\frac{1280 \cot ^2(2 \eta ) \xi _1^2 \xi _2^2 \xi _3^2
   a''(t) \csc ^2(\eta )}{\alpha ^7 Q_{\alpha ,4}^6}-\frac{2560 \csc ^2(2 \eta ) \xi _1^2
   \xi _2^2 \xi _3^2 a''(t) \csc ^2(\eta )}{\alpha ^7 Q_{\alpha ,4}^6}-\frac{3200 \cot
   (\eta ) \cot (2 \eta ) \xi _1^2 \xi _2^2 \xi _3^2 a''(t) \csc ^2(\eta )}{\alpha ^7
   Q_{\alpha ,4}^6}+\frac{640 \sec ^4(\eta ) \xi _1^2 \xi _3^2 \xi _4^2 a''(t) \csc
   ^2(\eta )}{\alpha ^7 Q_{\alpha ,4}^6}-\frac{1280 \sec ^2(\eta ) \xi _1^2 \xi _3^2 \xi
   _4^2 a''(t) \csc ^2(\eta )}{\alpha ^7 Q_{\alpha ,4}^6}+\frac{640 \sec ^4(\eta ) \xi
   _2^2 \xi _3^2 \xi _4^2 a''(t) \csc ^2(\eta )}{\alpha ^9 Q_{\alpha ,4}^6}-\frac{6080
   \sec ^2(\eta ) \xi _2^2 \xi _3^2 \xi _4^2 a''(t) \csc ^2(\eta )}{\alpha ^9 Q_{\alpha
   ,4}^6}+\frac{10240 \sec ^4(\eta ) \xi _1^2 \xi _3^2 \xi _4^4 a''(t) \csc ^2(\eta
   )}{\alpha ^9 Q_{\alpha ,4}^7}+\frac{11520 \sec ^4(\eta ) \xi _2^2 \xi _3^2 \xi _4^4
   a''(t) \csc ^2(\eta )}{\alpha ^{11} Q_{\alpha ,4}^7}-\frac{10240 \cot ^2(\eta ) \xi
   _1^2 \xi _2^4 \xi _3^2 a''(t) \csc ^2(\eta )}{\alpha ^9 Q_{\alpha ,4}^7}+\frac{15360
   \cot (\eta ) \cot (2 \eta ) \xi _1^2 \xi _2^4 \xi _3^2 a''(t) \csc ^2(\eta )}{\alpha
   ^9 Q_{\alpha ,4}^7}+\frac{11520 \sec ^2(\eta ) \xi _2^4 \xi _3^2 \xi _4^2 a''(t) \csc
   ^2(\eta )}{\alpha ^{11} Q_{\alpha ,4}^7}-\frac{5120 \sec ^4(\eta ) \xi _1^2 \xi _2^2
   \xi _3^2 \xi _4^2 a''(t) \csc ^2(\eta )}{\alpha ^9 Q_{\alpha ,4}^7}+\frac{37120 \sec
   ^2(\eta ) \xi _1^2 \xi _2^2 \xi _3^2 \xi _4^2 a''(t) \csc ^2(\eta )}{\alpha ^9
   Q_{\alpha ,4}^7}-\frac{107520 \sec ^4(\eta ) \xi _1^2 \xi _2^2 \xi _3^2 \xi _4^4
   a''(t) \csc ^2(\eta )}{\alpha ^{11} Q_{\alpha ,4}^8}-\frac{107520 \sec ^2(\eta ) \xi
   _1^2 \xi _2^4 \xi _3^2 \xi _4^2 a''(t) \csc ^2(\eta )}{\alpha ^{11} Q_{\alpha
   ,4}^8}+\frac{36 \xi _3^2 a'(t) a^{(3)}(t) \csc ^2(\eta )}{\alpha ^4 Q_{\alpha
   ,4}^4}-\frac{736 \xi _1^2 \xi _3^2 a'(t) a^{(3)}(t) \csc ^2(\eta )}{\alpha ^4
   Q_{\alpha ,4}^5}+\frac{320 \xi _2^2 \xi _3^2 a'(t) a^{(3)}(t) \csc ^2(\eta )}{\alpha
   ^6 Q_{\alpha ,4}^5}+\frac{320 \sec ^2(\eta ) \xi _3^2 \xi _4^2 a'(t) a^{(3)}(t) \csc
   ^2(\eta )}{\alpha ^6 Q_{\alpha ,4}^5}+\frac{1344 \xi _1^4 \xi _3^2 a'(t) a^{(3)}(t)
   \csc ^2(\eta )}{\alpha ^4 Q_{\alpha ,4}^6}-\frac{4864 \xi _1^2 \xi _2^2 \xi _3^2 a'(t)
   a^{(3)}(t) \csc ^2(\eta )}{\alpha ^6 Q_{\alpha ,4}^6}-\frac{4864 \sec ^2(\eta ) \xi
   _1^2 \xi _3^2 \xi _4^2 a'(t) a^{(3)}(t) \csc ^2(\eta )}{\alpha ^6 Q_{\alpha
   ,4}^6}+\frac{7680 \xi _1^4 \xi _2^2 \xi _3^2 a'(t) a^{(3)}(t) \csc ^2(\eta )}{\alpha
   ^6 Q_{\alpha ,4}^7}+\frac{7680 \sec ^2(\eta ) \xi _1^4 \xi _3^2 \xi _4^2 a'(t)
   a^{(3)}(t) \csc ^2(\eta )}{\alpha ^6 Q_{\alpha ,4}^7}+\frac{8 \xi _3^2 a^{(4)}(t) \csc
   ^2(\eta )}{\alpha ^3 Q_{\alpha ,4}^4}-\frac{96 \xi _1^2 \xi _3^2 a^{(4)}(t) \csc
   ^2(\eta )}{\alpha ^3 Q_{\alpha ,4}^5}+\frac{128 \xi _1^4 \xi _3^2 a^{(4)}(t) \csc
   ^2(\eta )}{\alpha ^3 Q_{\alpha ,4}^6}-\frac{4 \cot ^2(\eta ) \csc ^2(\eta )}{\alpha ^4
   Q_{\alpha ,4}^3}+\frac{\sec ^2(\eta ) \csc ^2(\eta )}{\alpha ^4 Q_{\alpha
   ,4}^3}+\frac{4 \cot (\eta ) \cot (2 \eta ) \csc ^2(\eta )}{\alpha ^4 Q_{\alpha
   ,4}^3}+\frac{2 \csc ^2(\eta )}{\alpha ^4 Q_{\alpha ,4}^3}+\frac{16 \cot ^2(\eta ) \xi
   _2^2 \csc ^2(\eta )}{\alpha ^6 Q_{\alpha ,4}^4}+\frac{12 \cot ^2(2 \eta ) \xi _2^2
   \csc ^2(\eta )}{\alpha ^6 Q_{\alpha ,4}^4}-\frac{24 \csc ^2(2 \eta ) \xi _2^2 \csc
   ^2(\eta )}{\alpha ^6 Q_{\alpha ,4}^4}-\frac{24 \cot (\eta ) \cot (2 \eta ) \xi _2^2
   \csc ^2(\eta )}{\alpha ^6 Q_{\alpha ,4}^4}-\frac{44 \cot ^4(\eta ) \xi _3^2 \csc
   ^2(\eta )}{\alpha ^6 Q_{\alpha ,4}^4}+\frac{36 \cot ^2(\eta ) \xi _3^2 \csc ^2(\eta
   )}{\alpha ^6 Q_{\alpha ,4}^4}-\frac{64 \cot ^2(\eta ) \cot ^2(2 \eta ) \xi _3^2 \csc
   ^2(\eta )}{\alpha ^6 Q_{\alpha ,4}^4}+\frac{128 \cot ^2(\eta ) \csc ^2(2 \eta ) \xi
   _3^2 \csc ^2(\eta )}{\alpha ^6 Q_{\alpha ,4}^4}+\frac{64 \cot (\eta ) \cot (2 \eta )
   \csc ^2(2 \eta ) \xi _3^2 \csc ^2(\eta )}{\alpha ^6 Q_{\alpha ,4}^4}-\frac{16 \sec
   ^2(\eta ) \xi _3^2 \csc ^2(\eta )}{\alpha ^6 Q_{\alpha ,4}^4}+\frac{104 \cot ^3(\eta )
   \cot (2 \eta ) \xi _3^2 \csc ^2(\eta )}{\alpha ^6 Q_{\alpha ,4}^4}-\frac{48 \cot (\eta
   ) \cot (2 \eta ) \xi _3^2 \csc ^2(\eta )}{\alpha ^6 Q_{\alpha ,4}^4}+\frac{6 \sec
   ^4(\eta ) \xi _4^2 \csc ^2(\eta )}{\alpha ^6 Q_{\alpha ,4}^4}-\frac{16 \sec ^2(\eta )
   \xi _4^2 \csc ^2(\eta )}{\alpha ^6 Q_{\alpha ,4}^4}+\frac{656 \cot ^4(\eta ) \xi _2^2
   \xi _3^2 \csc ^2(\eta )}{\alpha ^8 Q_{\alpha ,4}^5}-\frac{384 \cot (\eta ) \cot ^3(2
   \eta ) \xi _2^2 \xi _3^2 \csc ^2(\eta )}{\alpha ^8 Q_{\alpha ,4}^5}-\frac{256 \cot
   ^2(\eta ) \xi _2^2 \xi _3^2 \csc ^2(\eta )}{\alpha ^8 Q_{\alpha ,4}^5}+\frac{1312 \cot
   ^2(\eta ) \cot ^2(2 \eta ) \xi _2^2 \xi _3^2 \csc ^2(\eta )}{\alpha ^8 Q_{\alpha
   ,4}^5}-\frac{2624 \cot ^2(\eta ) \csc ^2(2 \eta ) \xi _2^2 \xi _3^2 \csc ^2(\eta
   )}{\alpha ^8 Q_{\alpha ,4}^5}-\frac{1536 \cot (\eta ) \cot (2 \eta ) \csc ^2(2 \eta )
   \xi _2^2 \xi _3^2 \csc ^2(\eta )}{\alpha ^8 Q_{\alpha ,4}^5}+\frac{96 \sec ^2(\eta )
   \xi _2^2 \xi _3^2 \csc ^2(\eta )}{\alpha ^8 Q_{\alpha ,4}^5}-\frac{1600 \cot ^3(\eta )
   \cot (2 \eta ) \xi _2^2 \xi _3^2 \csc ^2(\eta )}{\alpha ^8 Q_{\alpha ,4}^5}+\frac{384
   \cot (\eta ) \cot (2 \eta ) \xi _2^2 \xi _3^2 \csc ^2(\eta )}{\alpha ^8 Q_{\alpha
   ,4}^5}-\frac{32 \sec ^4(\eta ) \xi _2^2 \xi _4^2 \csc ^2(\eta )}{\alpha ^8 Q_{\alpha
   ,4}^5}+\frac{96 \sec ^2(\eta ) \xi _2^2 \xi _4^2 \csc ^2(\eta )}{\alpha ^8 Q_{\alpha
   ,4}^5}-\frac{688 \sec ^4(\eta ) \xi _3^2 \xi _4^2 \csc ^2(\eta )}{\alpha ^8 Q_{\alpha
   ,4}^5}-\frac{384 \cot ^2(2 \eta ) \sec ^2(\eta ) \xi _3^2 \xi _4^2 \csc ^2(\eta
   )}{\alpha ^8 Q_{\alpha ,4}^5}+\frac{768 \csc ^2(2 \eta ) \sec ^2(\eta ) \xi _3^2 \xi
   _4^2 \csc ^2(\eta )}{\alpha ^8 Q_{\alpha ,4}^5}+\frac{968 \sec ^2(\eta ) \xi _3^2 \xi
   _4^2 \csc ^2(\eta )}{\alpha ^8 Q_{\alpha ,4}^5}+\frac{1664 \sec ^6(\eta ) \xi _3^2 \xi
   _4^4 \csc ^2(\eta )}{\alpha ^{10} Q_{\alpha ,4}^6}-\frac{3008 \sec ^4(\eta ) \xi _3^2
   \xi _4^4 \csc ^2(\eta )}{\alpha ^{10} Q_{\alpha ,4}^6}-\frac{1024 \cot ^4(\eta ) \xi
   _2^4 \xi _3^2 \csc ^2(\eta )}{\alpha ^{10} Q_{\alpha ,4}^6}+\frac{1280 \cot (\eta )
   \cot ^3(2 \eta ) \xi _2^4 \xi _3^2 \csc ^2(\eta )}{\alpha ^{10} Q_{\alpha
   ,4}^6}-\frac{2560 \cot ^2(\eta ) \cot ^2(2 \eta ) \xi _2^4 \xi _3^2 \csc ^2(\eta
   )}{\alpha ^{10} Q_{\alpha ,4}^6}+\frac{5120 \cot ^2(\eta ) \csc ^2(2 \eta ) \xi _2^4
   \xi _3^2 \csc ^2(\eta )}{\alpha ^{10} Q_{\alpha ,4}^6}+\frac{2560 \cot (\eta ) \cot (2
   \eta ) \csc ^2(2 \eta ) \xi _2^4 \xi _3^2 \csc ^2(\eta )}{\alpha ^{10} Q_{\alpha
   ,4}^6}+\frac{2560 \cot ^3(\eta ) \cot (2 \eta ) \xi _2^4 \xi _3^2 \csc ^2(\eta
   )}{\alpha ^{10} Q_{\alpha ,4}^6}+\frac{11680 \sec ^4(\eta ) \xi _2^2 \xi _3^2 \xi _4^2
   \csc ^2(\eta )}{\alpha ^{10} Q_{\alpha ,4}^6}+\frac{10240 \cot ^2(2 \eta ) \sec
   ^2(\eta ) \xi _2^2 \xi _3^2 \xi _4^2 \csc ^2(\eta )}{\alpha ^{10} Q_{\alpha
   ,4}^6}-\frac{20480 \csc ^2(2 \eta ) \sec ^2(\eta ) \xi _2^2 \xi _3^2 \xi _4^2 \csc
   ^2(\eta )}{\alpha ^{10} Q_{\alpha ,4}^6}-\frac{13952 \sec ^2(\eta ) \xi _2^2 \xi _3^2
   \xi _4^2 \csc ^2(\eta )}{\alpha ^{10} Q_{\alpha ,4}^6}-\frac{30080 \sec ^6(\eta ) \xi
   _2^2 \xi _3^2 \xi _4^4 \csc ^2(\eta )}{\alpha ^{12} Q_{\alpha ,4}^7}+\frac{57280 \sec
   ^4(\eta ) \xi _2^2 \xi _3^2 \xi _4^4 \csc ^2(\eta )}{\alpha ^{12} Q_{\alpha
   ,4}^7}-\frac{20480 \sec ^4(\eta ) \xi _2^4 \xi _3^2 \xi _4^2 \csc ^2(\eta )}{\alpha
   ^{12} Q_{\alpha ,4}^7}-\frac{23040 \cot ^2(2 \eta ) \sec ^2(\eta ) \xi _2^4 \xi _3^2
   \xi _4^2 \csc ^2(\eta )}{\alpha ^{12} Q_{\alpha ,4}^7}+\frac{46080 \csc ^2(2 \eta )
   \sec ^2(\eta ) \xi _2^4 \xi _3^2 \xi _4^2 \csc ^2(\eta )}{\alpha ^{12} Q_{\alpha
   ,4}^7}+\frac{20480 \sec ^2(\eta ) \xi _2^4 \xi _3^2 \xi _4^2 \csc ^2(\eta )}{\alpha
   ^{12} Q_{\alpha ,4}^7}+\frac{53760 \sec ^6(\eta ) \xi _2^4 \xi _3^2 \xi _4^4 \csc
   ^2(\eta )}{\alpha ^{14} Q_{\alpha ,4}^8}-\frac{107520 \sec ^4(\eta ) \xi _2^4 \xi _3^2
   \xi _4^4 \csc ^2(\eta )}{\alpha ^{14} Q_{\alpha ,4}^8}+\frac{928 \cot (2 \eta ) \sec
   ^3(\eta ) \xi _3^2 \xi _4^2 a'(t)^2 \csc (\eta )}{\alpha ^8 Q_{\alpha
   ,4}^5}-\frac{2560 \cot (2 \eta ) \sec ^5(\eta ) \xi _3^2 \xi _4^4 a'(t)^2 \csc (\eta
   )}{\alpha ^{10} Q_{\alpha ,4}^6}-\frac{5760 \cot (2 \eta ) \sec ^3(\eta ) \xi _1^2 \xi
   _3^2 \xi _4^2 a'(t)^2 \csc (\eta )}{\alpha ^8 Q_{\alpha ,4}^6}-\frac{16960 \cot (2
   \eta ) \sec ^3(\eta ) \xi _2^2 \xi _3^2 \xi _4^2 a'(t)^2 \csc (\eta )}{\alpha ^{10}
   Q_{\alpha ,4}^6}+\frac{23040 \cot (2 \eta ) \sec ^5(\eta ) \xi _1^2 \xi _3^2 \xi _4^4
   a'(t)^2 \csc (\eta )}{\alpha ^{10} Q_{\alpha ,4}^7}+\frac{30720 \cot (2 \eta ) \sec
   ^5(\eta ) \xi _2^2 \xi _3^2 \xi _4^4 a'(t)^2 \csc (\eta )}{\alpha ^{12} Q_{\alpha
   ,4}^7}+\frac{30720 \cot (2 \eta ) \sec ^3(\eta ) \xi _2^4 \xi _3^2 \xi _4^2 a'(t)^2
   \csc (\eta )}{\alpha ^{12} Q_{\alpha ,4}^7}+\frac{138240 \cot (2 \eta ) \sec ^3(\eta )
   \xi _1^2 \xi _2^2 \xi _3^2 \xi _4^2 a'(t)^2 \csc (\eta )}{\alpha ^{10} Q_{\alpha
   ,4}^7}-\frac{322560 \cot (2 \eta ) \sec ^5(\eta ) \xi _1^2 \xi _2^2 \xi _3^2 \xi _4^4
   a'(t)^2 \csc (\eta )}{\alpha ^{12} Q_{\alpha ,4}^8}-\frac{322560 \cot (2 \eta ) \sec
   ^3(\eta ) \xi _1^2 \xi _2^4 \xi _3^2 \xi _4^2 a'(t)^2 \csc (\eta )}{\alpha ^{12}
   Q_{\alpha ,4}^8}-\frac{192 \cot (2 \eta ) \sec ^3(\eta ) \xi _3^2 \xi _4^2 a''(t) \csc
   (\eta )}{\alpha ^7 Q_{\alpha ,4}^5}+\frac{1280 \cot (2 \eta ) \sec ^3(\eta ) \xi _1^2
   \xi _3^2 \xi _4^2 a''(t) \csc (\eta )}{\alpha ^7 Q_{\alpha ,4}^6}+\frac{1920 \cot (2
   \eta ) \sec ^3(\eta ) \xi _2^2 \xi _3^2 \xi _4^2 a''(t) \csc (\eta )}{\alpha ^9
   Q_{\alpha ,4}^6}-\frac{15360 \cot (2 \eta ) \sec ^3(\eta ) \xi _1^2 \xi _2^2 \xi _3^2
   \xi _4^2 a''(t) \csc (\eta )}{\alpha ^9 Q_{\alpha ,4}^7}+\frac{12 \cot (2 \eta ) \sec
   ^3(\eta ) \xi _4^2 \csc (\eta )}{\alpha ^6 Q_{\alpha ,4}^4}-\frac{96 \cot (2 \eta )
   \sec ^3(\eta ) \xi _2^2 \xi _4^2 \csc (\eta )}{\alpha ^8 Q_{\alpha ,4}^5}-\frac{928
   \cot (2 \eta ) \sec ^3(\eta ) \xi _3^2 \xi _4^2 \csc (\eta )}{\alpha ^8 Q_{\alpha
   ,4}^5}+\frac{3840 \cot (2 \eta ) \sec ^5(\eta ) \xi _3^2 \xi _4^4 \csc (\eta )}{\alpha
   ^{10} Q_{\alpha ,4}^6}+\frac{16960 \cot (2 \eta ) \sec ^3(\eta ) \xi _2^2 \xi _3^2 \xi
   _4^2 \csc (\eta )}{\alpha ^{10} Q_{\alpha ,4}^6}-\frac{80640 \cot (2 \eta ) \sec
   ^5(\eta ) \xi _2^2 \xi _3^2 \xi _4^4 \csc (\eta )}{\alpha ^{12} Q_{\alpha
   ,4}^7}-\frac{30720 \cot (2 \eta ) \sec ^3(\eta ) \xi _2^4 \xi _3^2 \xi _4^2 \csc (\eta
   )}{\alpha ^{12} Q_{\alpha ,4}^7}+\frac{161280 \cot (2 \eta ) \sec ^5(\eta ) \xi _2^4
   \xi _3^2 \xi _4^4 \csc (\eta )}{\alpha ^{14} Q_{\alpha ,4}^8}-\frac{44 \sec ^2(\eta )
   \xi _4^2 \tan ^4(\eta )}{\alpha ^6 Q_{\alpha ,4}^4}+\frac{868 \sec ^4(\eta ) \xi _4^4
   \tan ^4(\eta )}{\alpha ^8 Q_{\alpha ,4}^5}+\frac{656 \sec ^2(\eta ) \xi _2^2 \xi _4^2
   \tan ^4(\eta )}{\alpha ^8 Q_{\alpha ,4}^5}-\frac{3008 \sec ^6(\eta ) \xi _4^6 \tan
   ^4(\eta )}{\alpha ^{10} Q_{\alpha ,4}^6}-\frac{14784 \sec ^4(\eta ) \xi _2^2 \xi _4^4
   \tan ^4(\eta )}{\alpha ^{10} Q_{\alpha ,4}^6}-\frac{1024 \sec ^2(\eta ) \xi _2^4 \xi
   _4^2 \tan ^4(\eta )}{\alpha ^{10} Q_{\alpha ,4}^6}+\frac{2560 \sec ^8(\eta ) \xi _4^8
   \tan ^4(\eta )}{\alpha ^{12} Q_{\alpha ,4}^7}+\frac{57280 \sec ^6(\eta ) \xi _2^2 \xi
   _4^6 \tan ^4(\eta )}{\alpha ^{12} Q_{\alpha ,4}^7}+\frac{25600 \sec ^4(\eta ) \xi _2^4
   \xi _4^4 \tan ^4(\eta )}{\alpha ^{12} Q_{\alpha ,4}^7}-\frac{53760 \sec ^8(\eta ) \xi
   _2^2 \xi _4^8 \tan ^4(\eta )}{\alpha ^{14} Q_{\alpha ,4}^8}-\frac{107520 \sec ^6(\eta
   ) \xi _2^4 \xi _4^6 \tan ^4(\eta )}{\alpha ^{14} Q_{\alpha ,4}^8}+\frac{107520 \sec
   ^8(\eta ) \xi _2^4 \xi _4^8 \tan ^4(\eta )}{\alpha ^{16} Q_{\alpha ,4}^9}+\frac{9
   a'(t)^4}{\alpha ^4 Q_{\alpha ,4}^3}-\frac{27 \xi _1^2 a'(t)^4}{\alpha ^4 Q_{\alpha
   ,4}^4}-\frac{7 \xi _2^2 a'(t)^4}{\alpha ^6 Q_{\alpha ,4}^4}-\frac{7 \sec ^2(\eta ) \xi
   _4^2 a'(t)^4}{\alpha ^6 Q_{\alpha ,4}^4}+\frac{36 \xi _1^4 a'(t)^4}{\alpha ^4
   Q_{\alpha ,4}^5}+\frac{612 \xi _2^4 a'(t)^4}{\alpha ^8 Q_{\alpha ,4}^5}+\frac{612 \sec
   ^4(\eta ) \xi _4^4 a'(t)^4}{\alpha ^8 Q_{\alpha ,4}^5}-\frac{216 \xi _1^2 \xi _2^2
   a'(t)^4}{\alpha ^6 Q_{\alpha ,4}^5}-\frac{216 \sec ^2(\eta ) \xi _1^2 \xi _4^2
   a'(t)^4}{\alpha ^6 Q_{\alpha ,4}^5}+\frac{1224 \sec ^2(\eta ) \xi _2^2 \xi _4^2
   a'(t)^4}{\alpha ^8 Q_{\alpha ,4}^5}-\frac{2752 \xi _2^6 a'(t)^4}{\alpha ^{10}
   Q_{\alpha ,4}^6}-\frac{2752 \sec ^6(\eta ) \xi _4^6 a'(t)^4}{\alpha ^{10} Q_{\alpha
   ,4}^6}-\frac{6960 \xi _1^2 \xi _2^4 a'(t)^4}{\alpha ^8 Q_{\alpha ,4}^6}-\frac{6960
   \sec ^4(\eta ) \xi _1^2 \xi _4^4 a'(t)^4}{\alpha ^8 Q_{\alpha ,4}^6}-\frac{8256 \sec
   ^4(\eta ) \xi _2^2 \xi _4^4 a'(t)^4}{\alpha ^{10} Q_{\alpha ,4}^6}+\frac{480 \xi _1^4
   \xi _2^2 a'(t)^4}{\alpha ^6 Q_{\alpha ,4}^6}+\frac{480 \sec ^2(\eta ) \xi _1^4 \xi
   _4^2 a'(t)^4}{\alpha ^6 Q_{\alpha ,4}^6}-\frac{8256 \sec ^2(\eta ) \xi _2^4 \xi _4^2
   a'(t)^4}{\alpha ^{10} Q_{\alpha ,4}^6}-\frac{13920 \sec ^2(\eta ) \xi _1^2 \xi _2^2
   \xi _4^2 a'(t)^4}{\alpha ^8 Q_{\alpha ,4}^6}+\frac{2560 \xi _2^8 a'(t)^4}{\alpha ^{12}
   Q_{\alpha ,4}^7}+\frac{2560 \sec ^8(\eta ) \xi _4^8 a'(t)^4}{\alpha ^{12} Q_{\alpha
   ,4}^7}+\frac{47040 \xi _1^2 \xi _2^6 a'(t)^4}{\alpha ^{10} Q_{\alpha
   ,4}^7}+\frac{47040 \sec ^6(\eta ) \xi _1^2 \xi _4^6 a'(t)^4}{\alpha ^{10} Q_{\alpha
   ,4}^7}+\frac{10240 \sec ^6(\eta ) \xi _2^2 \xi _4^6 a'(t)^4}{\alpha ^{12} Q_{\alpha
   ,4}^7}+\frac{8640 \xi _1^4 \xi _2^4 a'(t)^4}{\alpha ^8 Q_{\alpha ,4}^7}+\frac{8640
   \sec ^4(\eta ) \xi _1^4 \xi _4^4 a'(t)^4}{\alpha ^8 Q_{\alpha ,4}^7}+\frac{15360 \sec
   ^4(\eta ) \xi _2^4 \xi _4^4 a'(t)^4}{\alpha ^{12} Q_{\alpha ,4}^7}+\frac{141120 \sec
   ^4(\eta ) \xi _1^2 \xi _2^2 \xi _4^4 a'(t)^4}{\alpha ^{10} Q_{\alpha
   ,4}^7}+\frac{10240 \sec ^2(\eta ) \xi _2^6 \xi _4^2 a'(t)^4}{\alpha ^{12} Q_{\alpha
   ,4}^7}+\frac{141120 \sec ^2(\eta ) \xi _1^2 \xi _2^4 \xi _4^2 a'(t)^4}{\alpha ^{10}
   Q_{\alpha ,4}^7}+\frac{17280 \sec ^2(\eta ) \xi _1^4 \xi _2^2 \xi _4^2 a'(t)^4}{\alpha
   ^8 Q_{\alpha ,4}^7}-\frac{53760 \xi _1^2 \xi _2^8 a'(t)^4}{\alpha ^{12} Q_{\alpha
   ,4}^8}-\frac{53760 \sec ^8(\eta ) \xi _1^2 \xi _4^8 a'(t)^4}{\alpha ^{12} Q_{\alpha
   ,4}^8}-\frac{80640 \xi _1^4 \xi _2^6 a'(t)^4}{\alpha ^{10} Q_{\alpha
   ,4}^8}-\frac{80640 \sec ^6(\eta ) \xi _1^4 \xi _4^6 a'(t)^4}{\alpha ^{10} Q_{\alpha
   ,4}^8}-\frac{215040 \sec ^6(\eta ) \xi _1^2 \xi _2^2 \xi _4^6 a'(t)^4}{\alpha ^{12}
   Q_{\alpha ,4}^8}-\frac{322560 \sec ^4(\eta ) \xi _1^2 \xi _2^4 \xi _4^4
   a'(t)^4}{\alpha ^{12} Q_{\alpha ,4}^8}-\frac{241920 \sec ^4(\eta ) \xi _1^4 \xi _2^2
   \xi _4^4 a'(t)^4}{\alpha ^{10} Q_{\alpha ,4}^8}-\frac{215040 \sec ^2(\eta ) \xi _1^2
   \xi _2^6 \xi _4^2 a'(t)^4}{\alpha ^{12} Q_{\alpha ,4}^8}-\frac{241920 \sec ^2(\eta )
   \xi _1^4 \xi _2^4 \xi _4^2 a'(t)^4}{\alpha ^{10} Q_{\alpha ,4}^8}+\frac{107520 \xi
   _1^4 \xi _2^8 a'(t)^4}{\alpha ^{12} Q_{\alpha ,4}^9}+\frac{107520 \sec ^8(\eta ) \xi
   _1^4 \xi _4^8 a'(t)^4}{\alpha ^{12} Q_{\alpha ,4}^9}+\frac{430080 \sec ^6(\eta ) \xi
   _1^4 \xi _2^2 \xi _4^6 a'(t)^4}{\alpha ^{12} Q_{\alpha ,4}^9}+\frac{645120 \sec
   ^4(\eta ) \xi _1^4 \xi _2^4 \xi _4^4 a'(t)^4}{\alpha ^{12} Q_{\alpha
   ,4}^9}+\frac{430080 \sec ^2(\eta ) \xi _1^4 \xi _2^6 \xi _4^2 a'(t)^4}{\alpha ^{12}
   Q_{\alpha ,4}^9}-\frac{104 \cot (2 \eta ) \sec ^2(\eta ) \xi _4^2 \tan ^3(\eta
   )}{\alpha ^6 Q_{\alpha ,4}^4}+\frac{928 \cot (2 \eta ) \sec ^4(\eta ) \xi _4^4 \tan
   ^3(\eta )}{\alpha ^8 Q_{\alpha ,4}^5}+\frac{1600 \cot (2 \eta ) \sec ^2(\eta ) \xi
   _2^2 \xi _4^2 \tan ^3(\eta )}{\alpha ^8 Q_{\alpha ,4}^5}-\frac{1280 \cot (2 \eta )
   \sec ^6(\eta ) \xi _4^6 \tan ^3(\eta )}{\alpha ^{10} Q_{\alpha ,4}^6}-\frac{16960 \cot
   (2 \eta ) \sec ^4(\eta ) \xi _2^2 \xi _4^4 \tan ^3(\eta )}{\alpha ^{10} Q_{\alpha
   ,4}^6}-\frac{2560 \cot (2 \eta ) \sec ^2(\eta ) \xi _2^4 \xi _4^2 \tan ^3(\eta
   )}{\alpha ^{10} Q_{\alpha ,4}^6}+\frac{26880 \cot (2 \eta ) \sec ^6(\eta ) \xi _2^2
   \xi _4^6 \tan ^3(\eta )}{\alpha ^{12} Q_{\alpha ,4}^7}+\frac{30720 \cot (2 \eta ) \sec
   ^4(\eta ) \xi _2^4 \xi _4^4 \tan ^3(\eta )}{\alpha ^{12} Q_{\alpha ,4}^7}-\frac{53760
   \cot (2 \eta ) \sec ^6(\eta ) \xi _2^4 \xi _4^6 \tan ^3(\eta )}{\alpha ^{14} Q_{\alpha
   ,4}^8}-\frac{4 \sec ^2(\eta ) \tan ^2(\eta )}{\alpha ^4 Q_{\alpha ,4}^3}+\frac{16 \sec
   ^2(\eta ) \xi _2^2 \tan ^2(\eta )}{\alpha ^6 Q_{\alpha ,4}^4}-\frac{251 \sec ^4(\eta )
   \xi _4^2 \tan ^2(\eta )}{\alpha ^6 Q_{\alpha ,4}^4}-\frac{64 \cot ^2(2 \eta ) \sec
   ^2(\eta ) \xi _4^2 \tan ^2(\eta )}{\alpha ^6 Q_{\alpha ,4}^4}+\frac{128 \csc ^2(2 \eta
   ) \sec ^2(\eta ) \xi _4^2 \tan ^2(\eta )}{\alpha ^6 Q_{\alpha ,4}^4}+\frac{36 \sec
   ^2(\eta ) \xi _4^2 \tan ^2(\eta )}{\alpha ^6 Q_{\alpha ,4}^4}+\frac{1520 \sec ^6(\eta
   ) \xi _4^4 \tan ^2(\eta )}{\alpha ^8 Q_{\alpha ,4}^5}+\frac{192 \cot ^2(2 \eta ) \sec
   ^4(\eta ) \xi _4^4 \tan ^2(\eta )}{\alpha ^8 Q_{\alpha ,4}^5}-\frac{384 \csc ^2(2 \eta
   ) \sec ^4(\eta ) \xi _4^4 \tan ^2(\eta )}{\alpha ^8 Q_{\alpha ,4}^5}-\frac{128 \sec
   ^4(\eta ) \xi _4^4 \tan ^2(\eta )}{\alpha ^8 Q_{\alpha ,4}^5}+\frac{3648 \sec ^4(\eta
   ) \xi _2^2 \xi _4^2 \tan ^2(\eta )}{\alpha ^8 Q_{\alpha ,4}^5}+\frac{1312 \cot ^2(2
   \eta ) \sec ^2(\eta ) \xi _2^2 \xi _4^2 \tan ^2(\eta )}{\alpha ^8 Q_{\alpha
   ,4}^5}-\frac{2624 \csc ^2(2 \eta ) \sec ^2(\eta ) \xi _2^2 \xi _4^2 \tan ^2(\eta
   )}{\alpha ^8 Q_{\alpha ,4}^5}-\frac{256 \sec ^2(\eta ) \xi _2^2 \xi _4^2 \tan ^2(\eta
   )}{\alpha ^8 Q_{\alpha ,4}^5}-\frac{1664 \sec ^8(\eta ) \xi _4^6 \tan ^2(\eta
   )}{\alpha ^{10} Q_{\alpha ,4}^6}-\frac{24592 \sec ^6(\eta ) \xi _2^2 \xi _4^4 \tan
   ^2(\eta )}{\alpha ^{10} Q_{\alpha ,4}^6}-\frac{5120 \cot ^2(2 \eta ) \sec ^4(\eta )
   \xi _2^2 \xi _4^4 \tan ^2(\eta )}{\alpha ^{10} Q_{\alpha ,4}^6}+\frac{10240 \csc ^2(2
   \eta ) \sec ^4(\eta ) \xi _2^2 \xi _4^4 \tan ^2(\eta )}{\alpha ^{10} Q_{\alpha
   ,4}^6}+\frac{960 \sec ^4(\eta ) \xi _2^2 \xi _4^4 \tan ^2(\eta )}{\alpha ^{10}
   Q_{\alpha ,4}^6}-\frac{5632 \sec ^4(\eta ) \xi _2^4 \xi _4^2 \tan ^2(\eta )}{\alpha
   ^{10} Q_{\alpha ,4}^6}-\frac{2560 \cot ^2(2 \eta ) \sec ^2(\eta ) \xi _2^4 \xi _4^2
   \tan ^2(\eta )}{\alpha ^{10} Q_{\alpha ,4}^6}+\frac{5120 \csc ^2(2 \eta ) \sec ^2(\eta
   ) \xi _2^4 \xi _4^2 \tan ^2(\eta )}{\alpha ^{10} Q_{\alpha ,4}^6}+\frac{30080 \sec
   ^8(\eta ) \xi _2^2 \xi _4^6 \tan ^2(\eta )}{\alpha ^{12} Q_{\alpha ,4}^7}+\frac{40960
   \sec ^6(\eta ) \xi _2^4 \xi _4^4 \tan ^2(\eta )}{\alpha ^{12} Q_{\alpha
   ,4}^7}+\frac{11520 \cot ^2(2 \eta ) \sec ^4(\eta ) \xi _2^4 \xi _4^4 \tan ^2(\eta
   )}{\alpha ^{12} Q_{\alpha ,4}^7}-\frac{23040 \csc ^2(2 \eta ) \sec ^4(\eta ) \xi _2^4
   \xi _4^4 \tan ^2(\eta )}{\alpha ^{12} Q_{\alpha ,4}^7}-\frac{53760 \sec ^8(\eta ) \xi
   _2^4 \xi _4^6 \tan ^2(\eta )}{\alpha ^{14} Q_{\alpha ,4}^8}-\frac{87 \sec ^2(\eta )
   \xi _4^2 \tan ^2(\eta ) a'(t)^2}{\alpha ^6 Q_{\alpha ,4}^4}+\frac{1624 \sec ^4(\eta )
   \xi _4^4 \tan ^2(\eta ) a'(t)^2}{\alpha ^8 Q_{\alpha ,4}^5}+\frac{216 \sec ^2(\eta )
   \xi _1^2 \xi _4^2 \tan ^2(\eta ) a'(t)^2}{\alpha ^6 Q_{\alpha ,4}^5}+\frac{2552 \sec
   ^2(\eta ) \xi _2^2 \xi _4^2 \tan ^2(\eta ) a'(t)^2}{\alpha ^8 Q_{\alpha
   ,4}^5}-\frac{5760 \sec ^6(\eta ) \xi _4^6 \tan ^2(\eta ) a'(t)^2}{\alpha ^{10}
   Q_{\alpha ,4}^6}-\frac{9120 \sec ^4(\eta ) \xi _1^2 \xi _4^4 \tan ^2(\eta )
   a'(t)^2}{\alpha ^8 Q_{\alpha ,4}^6}-\frac{30480 \sec ^4(\eta ) \xi _2^2 \xi _4^4 \tan
   ^2(\eta ) a'(t)^2}{\alpha ^{10} Q_{\alpha ,4}^6}-\frac{10560 \sec ^2(\eta ) \xi _2^4
   \xi _4^2 \tan ^2(\eta ) a'(t)^2}{\alpha ^{10} Q_{\alpha ,4}^6}-\frac{13920 \sec
   ^2(\eta ) \xi _1^2 \xi _2^2 \xi _4^2 \tan ^2(\eta ) a'(t)^2}{\alpha ^8 Q_{\alpha
   ,4}^6}+\frac{5120 \sec ^8(\eta ) \xi _4^8 \tan ^2(\eta ) a'(t)^2}{\alpha ^{12}
   Q_{\alpha ,4}^7}+\frac{47040 \sec ^6(\eta ) \xi _1^2 \xi _4^6 \tan ^2(\eta )
   a'(t)^2}{\alpha ^{10} Q_{\alpha ,4}^7}+\frac{79040 \sec ^6(\eta ) \xi _2^2 \xi _4^6
   \tan ^2(\eta ) a'(t)^2}{\alpha ^{12} Q_{\alpha ,4}^7}+\frac{84160 \sec ^4(\eta ) \xi
   _2^4 \xi _4^4 \tan ^2(\eta ) a'(t)^2}{\alpha ^{12} Q_{\alpha ,4}^7}+\frac{233280 \sec
   ^4(\eta ) \xi _1^2 \xi _2^2 \xi _4^4 \tan ^2(\eta ) a'(t)^2}{\alpha ^{10} Q_{\alpha
   ,4}^7}+\frac{10240 \sec ^2(\eta ) \xi _2^6 \xi _4^2 \tan ^2(\eta ) a'(t)^2}{\alpha
   ^{12} Q_{\alpha ,4}^7}+\frac{85440 \sec ^2(\eta ) \xi _1^2 \xi _2^4 \xi _4^2 \tan
   ^2(\eta ) a'(t)^2}{\alpha ^{10} Q_{\alpha ,4}^7}-\frac{53760 \sec ^8(\eta ) \xi _1^2
   \xi _4^8 \tan ^2(\eta ) a'(t)^2}{\alpha ^{12} Q_{\alpha ,4}^8}-\frac{53760 \sec
   ^8(\eta ) \xi _2^2 \xi _4^8 \tan ^2(\eta ) a'(t)^2}{\alpha ^{14} Q_{\alpha
   ,4}^8}-\frac{107520 \sec ^6(\eta ) \xi _2^4 \xi _4^6 \tan ^2(\eta ) a'(t)^2}{\alpha
   ^{14} Q_{\alpha ,4}^8}-\frac{779520 \sec ^6(\eta ) \xi _1^2 \xi _2^2 \xi _4^6 \tan
   ^2(\eta ) a'(t)^2}{\alpha ^{12} Q_{\alpha ,4}^8}-\frac{53760 \sec ^4(\eta ) \xi _2^6
   \xi _4^4 \tan ^2(\eta ) a'(t)^2}{\alpha ^{14} Q_{\alpha ,4}^8}-\frac{833280 \sec
   ^4(\eta ) \xi _1^2 \xi _2^4 \xi _4^4 \tan ^2(\eta ) a'(t)^2}{\alpha ^{12} Q_{\alpha
   ,4}^8}-\frac{107520 \sec ^2(\eta ) \xi _1^2 \xi _2^6 \xi _4^2 \tan ^2(\eta )
   a'(t)^2}{\alpha ^{12} Q_{\alpha ,4}^8}+\frac{645120 \sec ^8(\eta ) \xi _1^2 \xi _2^2
   \xi _4^8 \tan ^2(\eta ) a'(t)^2}{\alpha ^{14} Q_{\alpha ,4}^9}+\frac{1290240 \sec
   ^6(\eta ) \xi _1^2 \xi _2^4 \xi _4^6 \tan ^2(\eta ) a'(t)^2}{\alpha ^{14} Q_{\alpha
   ,4}^9}+\frac{645120 \sec ^4(\eta ) \xi _1^2 \xi _2^6 \xi _4^4 \tan ^2(\eta )
   a'(t)^2}{\alpha ^{14} Q_{\alpha ,4}^9}-\frac{116 \cot (2 \eta ) \sec ^2(\eta ) \xi
   _4^2 \tan (\eta ) a'(t)^2}{\alpha ^6 Q_{\alpha ,4}^4}+\frac{928 \cot (2 \eta ) \sec
   ^4(\eta ) \xi _4^4 \tan (\eta ) a'(t)^2}{\alpha ^8 Q_{\alpha ,4}^5}+\frac{288 \cot (2
   \eta ) \sec ^2(\eta ) \xi _1^2 \xi _4^2 \tan (\eta ) a'(t)^2}{\alpha ^6 Q_{\alpha
   ,4}^5}+\frac{3712 \cot (2 \eta ) \sec ^2(\eta ) \xi _2^2 \xi _4^2 \tan (\eta )
   a'(t)^2}{\alpha ^8 Q_{\alpha ,4}^5}-\frac{1280 \cot (2 \eta ) \sec ^6(\eta ) \xi _4^6
   \tan (\eta ) a'(t)^2}{\alpha ^{10} Q_{\alpha ,4}^6}-\frac{5760 \cot (2 \eta ) \sec
   ^4(\eta ) \xi _1^2 \xi _4^4 \tan (\eta ) a'(t)^2}{\alpha ^8 Q_{\alpha
   ,4}^6}-\frac{16960 \cot (2 \eta ) \sec ^4(\eta ) \xi _2^2 \xi _4^4 \tan (\eta )
   a'(t)^2}{\alpha ^{10} Q_{\alpha ,4}^6}-\frac{15680 \cot (2 \eta ) \sec ^2(\eta ) \xi
   _2^4 \xi _4^2 \tan (\eta ) a'(t)^2}{\alpha ^{10} Q_{\alpha ,4}^6}-\frac{20160 \cot (2
   \eta ) \sec ^2(\eta ) \xi _1^2 \xi _2^2 \xi _4^2 \tan (\eta ) a'(t)^2}{\alpha ^8
   Q_{\alpha ,4}^6}+\frac{11520 \cot (2 \eta ) \sec ^6(\eta ) \xi _1^2 \xi _4^6 \tan
   (\eta ) a'(t)^2}{\alpha ^{10} Q_{\alpha ,4}^7}+\frac{15360 \cot (2 \eta ) \sec ^6(\eta
   ) \xi _2^2 \xi _4^6 \tan (\eta ) a'(t)^2}{\alpha ^{12} Q_{\alpha ,4}^7}+\frac{30720
   \cot (2 \eta ) \sec ^4(\eta ) \xi _2^4 \xi _4^4 \tan (\eta ) a'(t)^2}{\alpha ^{12}
   Q_{\alpha ,4}^7}+\frac{138240 \cot (2 \eta ) \sec ^4(\eta ) \xi _1^2 \xi _2^2 \xi _4^4
   \tan (\eta ) a'(t)^2}{\alpha ^{10} Q_{\alpha ,4}^7}+\frac{15360 \cot (2 \eta ) \sec
   ^2(\eta ) \xi _2^6 \xi _4^2 \tan (\eta ) a'(t)^2}{\alpha ^{12} Q_{\alpha
   ,4}^7}+\frac{126720 \cot (2 \eta ) \sec ^2(\eta ) \xi _1^2 \xi _2^4 \xi _4^2 \tan
   (\eta ) a'(t)^2}{\alpha ^{10} Q_{\alpha ,4}^7}-\frac{161280 \cot (2 \eta ) \sec
   ^6(\eta ) \xi _1^2 \xi _2^2 \xi _4^6 \tan (\eta ) a'(t)^2}{\alpha ^{12} Q_{\alpha
   ,4}^8}-\frac{322560 \cot (2 \eta ) \sec ^4(\eta ) \xi _1^2 \xi _2^4 \xi _4^4 \tan
   (\eta ) a'(t)^2}{\alpha ^{12} Q_{\alpha ,4}^8}-\frac{161280 \cot (2 \eta ) \sec
   ^2(\eta ) \xi _1^2 \xi _2^6 \xi _4^2 \tan (\eta ) a'(t)^2}{\alpha ^{12} Q_{\alpha
   ,4}^8}-\frac{4 \cot ^2(2 \eta ) a'(t)^2}{\alpha ^4 Q_{\alpha ,4}^3}+\frac{8 \csc ^2(2
   \eta ) a'(t)^2}{\alpha ^4 Q_{\alpha ,4}^3}-\frac{3 \sec ^2(\eta ) a'(t)^2}{\alpha ^4
   Q_{\alpha ,4}^3}-\frac{6 a'(t)^2}{\alpha ^4 Q_{\alpha ,4}^3}-\frac{3 \sec ^2(\eta )
   \xi _1^2 a'(t)^2}{\alpha ^4 Q_{\alpha ,4}^4}-\frac{12 \xi _1^2 a'(t)^2}{\alpha ^4
   Q_{\alpha ,4}^4}-\frac{116 \cot ^2(2 \eta ) \xi _2^2 a'(t)^2}{\alpha ^6 Q_{\alpha
   ,4}^4}+\frac{232 \csc ^2(2 \eta ) \xi _2^2 a'(t)^2}{\alpha ^6 Q_{\alpha
   ,4}^4}+\frac{13 \sec ^2(\eta ) \xi _2^2 a'(t)^2}{\alpha ^6 Q_{\alpha ,4}^4}+\frac{52
   \xi _2^2 a'(t)^2}{\alpha ^6 Q_{\alpha ,4}^4}-\frac{45 \sec ^4(\eta ) \xi _4^2
   a'(t)^2}{\alpha ^6 Q_{\alpha ,4}^4}+\frac{52 \sec ^2(\eta ) \xi _4^2 a'(t)^2}{\alpha
   ^6 Q_{\alpha ,4}^4}+\frac{928 \cot ^2(2 \eta ) \xi _2^4 a'(t)^2}{\alpha ^8 Q_{\alpha
   ,4}^5}-\frac{1856 \csc ^2(2 \eta ) \xi _2^4 a'(t)^2}{\alpha ^8 Q_{\alpha
   ,4}^5}-\frac{32 \sec ^2(\eta ) \xi _2^4 a'(t)^2}{\alpha ^8 Q_{\alpha ,4}^5}-\frac{128
   \xi _2^4 a'(t)^2}{\alpha ^8 Q_{\alpha ,4}^5}+\frac{432 \sec ^6(\eta ) \xi _4^4
   a'(t)^2}{\alpha ^8 Q_{\alpha ,4}^5}-\frac{128 \sec ^4(\eta ) \xi _4^4 a'(t)^2}{\alpha
   ^8 Q_{\alpha ,4}^5}+\frac{288 \cot ^2(2 \eta ) \xi _1^2 \xi _2^2 a'(t)^2}{\alpha ^6
   Q_{\alpha ,4}^5}-\frac{576 \csc ^2(2 \eta ) \xi _1^2 \xi _2^2 a'(t)^2}{\alpha ^6
   Q_{\alpha ,4}^5}-\frac{48 \sec ^2(\eta ) \xi _1^2 \xi _2^2 a'(t)^2}{\alpha ^6
   Q_{\alpha ,4}^5}-\frac{192 \xi _1^2 \xi _2^2 a'(t)^2}{\alpha ^6 Q_{\alpha
   ,4}^5}+\frac{96 \sec ^4(\eta ) \xi _1^2 \xi _4^2 a'(t)^2}{\alpha ^6 Q_{\alpha
   ,4}^5}-\frac{192 \sec ^2(\eta ) \xi _1^2 \xi _4^2 a'(t)^2}{\alpha ^6 Q_{\alpha
   ,4}^5}+\frac{1328 \sec ^4(\eta ) \xi _2^2 \xi _4^2 a'(t)^2}{\alpha ^8 Q_{\alpha
   ,4}^5}+\frac{928 \cot ^2(2 \eta ) \sec ^2(\eta ) \xi _2^2 \xi _4^2 a'(t)^2}{\alpha ^8
   Q_{\alpha ,4}^5}-\frac{1856 \csc ^2(2 \eta ) \sec ^2(\eta ) \xi _2^2 \xi _4^2
   a'(t)^2}{\alpha ^8 Q_{\alpha ,4}^5}-\frac{256 \sec ^2(\eta ) \xi _2^2 \xi _4^2
   a'(t)^2}{\alpha ^8 Q_{\alpha ,4}^5}+\frac{696 \sec ^4(\eta ) \xi _3^2 \xi _4^2
   a'(t)^2}{\alpha ^8 Q_{\alpha ,4}^5}-\frac{1280 \cot ^2(2 \eta ) \xi _2^6
   a'(t)^2}{\alpha ^{10} Q_{\alpha ,4}^6}+\frac{2560 \csc ^2(2 \eta ) \xi _2^6
   a'(t)^2}{\alpha ^{10} Q_{\alpha ,4}^6}-\frac{640 \sec ^8(\eta ) \xi _4^6
   a'(t)^2}{\alpha ^{10} Q_{\alpha ,4}^6}-\frac{5760 \cot ^2(2 \eta ) \xi _1^2 \xi _2^4
   a'(t)^2}{\alpha ^8 Q_{\alpha ,4}^6}+\frac{11520 \csc ^2(2 \eta ) \xi _1^2 \xi _2^4
   a'(t)^2}{\alpha ^8 Q_{\alpha ,4}^6}+\frac{240 \sec ^2(\eta ) \xi _1^2 \xi _2^4
   a'(t)^2}{\alpha ^8 Q_{\alpha ,4}^6}+\frac{960 \xi _1^2 \xi _2^4 a'(t)^2}{\alpha ^8
   Q_{\alpha ,4}^6}-\frac{2640 \sec ^6(\eta ) \xi _1^2 \xi _4^4 a'(t)^2}{\alpha ^8
   Q_{\alpha ,4}^6}+\frac{960 \sec ^4(\eta ) \xi _1^2 \xi _4^4 a'(t)^2}{\alpha ^8
   Q_{\alpha ,4}^6}-\frac{6080 \sec ^6(\eta ) \xi _2^2 \xi _4^4 a'(t)^2}{\alpha ^{10}
   Q_{\alpha ,4}^6}-\frac{1280 \cot ^2(2 \eta ) \sec ^4(\eta ) \xi _2^2 \xi _4^4
   a'(t)^2}{\alpha ^{10} Q_{\alpha ,4}^6}+\frac{2560 \csc ^2(2 \eta ) \sec ^4(\eta ) \xi
   _2^2 \xi _4^4 a'(t)^2}{\alpha ^{10} Q_{\alpha ,4}^6}-\frac{6720 \sec ^6(\eta ) \xi
   _3^2 \xi _4^4 a'(t)^2}{\alpha ^{10} Q_{\alpha ,4}^6}-\frac{5440 \sec ^4(\eta ) \xi
   _2^4 \xi _4^2 a'(t)^2}{\alpha ^{10} Q_{\alpha ,4}^6}-\frac{2560 \cot ^2(2 \eta ) \sec
   ^2(\eta ) \xi _2^4 \xi _4^2 a'(t)^2}{\alpha ^{10} Q_{\alpha ,4}^6}+\frac{5120 \csc
   ^2(2 \eta ) \sec ^2(\eta ) \xi _2^4 \xi _4^2 a'(t)^2}{\alpha ^{10} Q_{\alpha
   ,4}^6}-\frac{7200 \sec ^4(\eta ) \xi _1^2 \xi _2^2 \xi _4^2 a'(t)^2}{\alpha ^8
   Q_{\alpha ,4}^6}-\frac{5760 \cot ^2(2 \eta ) \sec ^2(\eta ) \xi _1^2 \xi _2^2 \xi _4^2
   a'(t)^2}{\alpha ^8 Q_{\alpha ,4}^6}+\frac{11520 \csc ^2(2 \eta ) \sec ^2(\eta ) \xi
   _1^2 \xi _2^2 \xi _4^2 a'(t)^2}{\alpha ^8 Q_{\alpha ,4}^6}+\frac{1920 \sec ^2(\eta )
   \xi _1^2 \xi _2^2 \xi _4^2 a'(t)^2}{\alpha ^8 Q_{\alpha ,4}^6}-\frac{4320 \sec ^4(\eta
   ) \xi _1^2 \xi _3^2 \xi _4^2 a'(t)^2}{\alpha ^8 Q_{\alpha ,4}^6}-\frac{11520 \sec
   ^4(\eta ) \xi _2^2 \xi _3^2 \xi _4^2 a'(t)^2}{\alpha ^{10} Q_{\alpha
   ,4}^6}+\frac{11520 \cot ^2(2 \eta ) \xi _1^2 \xi _2^6 a'(t)^2}{\alpha ^{10} Q_{\alpha
   ,4}^7}-\frac{23040 \csc ^2(2 \eta ) \xi _1^2 \xi _2^6 a'(t)^2}{\alpha ^{10} Q_{\alpha
   ,4}^7}+\frac{5760 \sec ^8(\eta ) \xi _1^2 \xi _4^6 a'(t)^2}{\alpha ^{10} Q_{\alpha
   ,4}^7}+\frac{5120 \sec ^8(\eta ) \xi _2^2 \xi _4^6 a'(t)^2}{\alpha ^{12} Q_{\alpha
   ,4}^7}+\frac{10240 \sec ^8(\eta ) \xi _3^2 \xi _4^6 a'(t)^2}{\alpha ^{12} Q_{\alpha
   ,4}^7}+\frac{10240 \sec ^6(\eta ) \xi _2^4 \xi _4^4 a'(t)^2}{\alpha ^{12} Q_{\alpha
   ,4}^7}+\frac{49920 \sec ^6(\eta ) \xi _1^2 \xi _2^2 \xi _4^4 a'(t)^2}{\alpha ^{10}
   Q_{\alpha ,4}^7}+\frac{11520 \cot ^2(2 \eta ) \sec ^4(\eta ) \xi _1^2 \xi _2^2 \xi
   _4^4 a'(t)^2}{\alpha ^{10} Q_{\alpha ,4}^7}-\frac{23040 \csc ^2(2 \eta ) \sec ^4(\eta
   ) \xi _1^2 \xi _2^2 \xi _4^4 a'(t)^2}{\alpha ^{10} Q_{\alpha ,4}^7}+\frac{55680 \sec
   ^6(\eta ) \xi _1^2 \xi _3^2 \xi _4^4 a'(t)^2}{\alpha ^{10} Q_{\alpha
   ,4}^7}+\frac{89280 \sec ^6(\eta ) \xi _2^2 \xi _3^2 \xi _4^4 a'(t)^2}{\alpha ^{12}
   Q_{\alpha ,4}^7}+\frac{5120 \sec ^4(\eta ) \xi _2^6 \xi _4^2 a'(t)^2}{\alpha ^{12}
   Q_{\alpha ,4}^7}+\frac{44160 \sec ^4(\eta ) \xi _1^2 \xi _2^4 \xi _4^2 a'(t)^2}{\alpha
   ^{10} Q_{\alpha ,4}^7}+\frac{23040 \cot ^2(2 \eta ) \sec ^2(\eta ) \xi _1^2 \xi _2^4
   \xi _4^2 a'(t)^2}{\alpha ^{10} Q_{\alpha ,4}^7}-\frac{46080 \csc ^2(2 \eta ) \sec
   ^2(\eta ) \xi _1^2 \xi _2^4 \xi _4^2 a'(t)^2}{\alpha ^{10} Q_{\alpha
   ,4}^7}+\frac{20480 \sec ^4(\eta ) \xi _2^4 \xi _3^2 \xi _4^2 a'(t)^2}{\alpha ^{12}
   Q_{\alpha ,4}^7}+\frac{94080 \sec ^4(\eta ) \xi _1^2 \xi _2^2 \xi _3^2 \xi _4^2
   a'(t)^2}{\alpha ^{10} Q_{\alpha ,4}^7}-\frac{53760 \sec ^8(\eta ) \xi _1^2 \xi _2^2
   \xi _4^6 a'(t)^2}{\alpha ^{12} Q_{\alpha ,4}^8}-\frac{107520 \sec ^8(\eta ) \xi _1^2
   \xi _3^2 \xi _4^6 a'(t)^2}{\alpha ^{12} Q_{\alpha ,4}^8}-\frac{107520 \sec ^8(\eta )
   \xi _2^2 \xi _3^2 \xi _4^6 a'(t)^2}{\alpha ^{14} Q_{\alpha ,4}^8}-\frac{107520 \sec
   ^6(\eta ) \xi _1^2 \xi _2^4 \xi _4^4 a'(t)^2}{\alpha ^{12} Q_{\alpha
   ,4}^8}-\frac{107520 \sec ^6(\eta ) \xi _2^4 \xi _3^2 \xi _4^4 a'(t)^2}{\alpha ^{14}
   Q_{\alpha ,4}^8}-\frac{887040 \sec ^6(\eta ) \xi _1^2 \xi _2^2 \xi _3^2 \xi _4^4
   a'(t)^2}{\alpha ^{12} Q_{\alpha ,4}^8}-\frac{53760 \sec ^4(\eta ) \xi _1^2 \xi _2^6
   \xi _4^2 a'(t)^2}{\alpha ^{12} Q_{\alpha ,4}^8}-\frac{215040 \sec ^4(\eta ) \xi _1^2
   \xi _2^4 \xi _3^2 \xi _4^2 a'(t)^2}{\alpha ^{12} Q_{\alpha ,4}^8}+\frac{1290240 \sec
   ^8(\eta ) \xi _1^2 \xi _2^2 \xi _3^2 \xi _4^6 a'(t)^2}{\alpha ^{14} Q_{\alpha
   ,4}^9}+\frac{1290240 \sec ^6(\eta ) \xi _1^2 \xi _2^4 \xi _3^2 \xi _4^4
   a'(t)^2}{\alpha ^{14} Q_{\alpha ,4}^9}+\frac{9 a''(t)^2}{\alpha ^2 Q_{\alpha
   ,4}^3}-\frac{108 \xi _1^2 a''(t)^2}{\alpha ^2 Q_{\alpha ,4}^4}+\frac{20 \xi _2^2
   a''(t)^2}{\alpha ^4 Q_{\alpha ,4}^4}+\frac{20 \sec ^2(\eta ) \xi _4^2 a''(t)^2}{\alpha
   ^4 Q_{\alpha ,4}^4}+\frac{144 \xi _1^4 a''(t)^2}{\alpha ^2 Q_{\alpha ,4}^5}+\frac{112
   \xi _2^4 a''(t)^2}{\alpha ^6 Q_{\alpha ,4}^5}+\frac{112 \sec ^4(\eta ) \xi _4^4
   a''(t)^2}{\alpha ^6 Q_{\alpha ,4}^5}-\frac{432 \xi _1^2 \xi _2^2 a''(t)^2}{\alpha ^4
   Q_{\alpha ,4}^5}-\frac{432 \sec ^2(\eta ) \xi _1^2 \xi _4^2 a''(t)^2}{\alpha ^4
   Q_{\alpha ,4}^5}+\frac{224 \sec ^2(\eta ) \xi _2^2 \xi _4^2 a''(t)^2}{\alpha ^6
   Q_{\alpha ,4}^5}-\frac{1664 \xi _1^2 \xi _2^4 a''(t)^2}{\alpha ^6 Q_{\alpha
   ,4}^6}-\frac{1664 \sec ^4(\eta ) \xi _1^2 \xi _4^4 a''(t)^2}{\alpha ^6 Q_{\alpha
   ,4}^6}+\frac{768 \xi _1^4 \xi _2^2 a''(t)^2}{\alpha ^4 Q_{\alpha ,4}^6}+\frac{768 \sec
   ^2(\eta ) \xi _1^4 \xi _4^2 a''(t)^2}{\alpha ^4 Q_{\alpha ,4}^6}-\frac{3328 \sec
   ^2(\eta ) \xi _1^2 \xi _2^2 \xi _4^2 a''(t)^2}{\alpha ^6 Q_{\alpha ,4}^6}+\frac{2560
   \xi _1^4 \xi _2^4 a''(t)^2}{\alpha ^6 Q_{\alpha ,4}^7}+\frac{2560 \sec ^4(\eta ) \xi
   _1^4 \xi _4^4 a''(t)^2}{\alpha ^6 Q_{\alpha ,4}^7}+\frac{5120 \sec ^2(\eta ) \xi _1^4
   \xi _2^2 \xi _4^2 a''(t)^2}{\alpha ^6 Q_{\alpha ,4}^7}-\frac{4 \cot (2 \eta ) \sec
   ^2(\eta ) \tan (\eta )}{\alpha ^4 Q_{\alpha ,4}^3}+\frac{24 \cot (2 \eta ) \sec
   ^2(\eta ) \xi _2^2 \tan (\eta )}{\alpha ^6 Q_{\alpha ,4}^4}-\frac{220 \cot (2 \eta )
   \sec ^4(\eta ) \xi _4^2 \tan (\eta )}{\alpha ^6 Q_{\alpha ,4}^4}-\frac{64 \cot (2 \eta
   ) \csc ^2(2 \eta ) \sec ^2(\eta ) \xi _4^2 \tan (\eta )}{\alpha ^6 Q_{\alpha
   ,4}^4}+\frac{48 \cot (2 \eta ) \sec ^2(\eta ) \xi _4^2 \tan (\eta )}{\alpha ^6
   Q_{\alpha ,4}^4}+\frac{512 \cot (2 \eta ) \sec ^6(\eta ) \xi _4^4 \tan (\eta )}{\alpha
   ^8 Q_{\alpha ,4}^5}+\frac{3296 \cot (2 \eta ) \sec ^4(\eta ) \xi _2^2 \xi _4^2 \tan
   (\eta )}{\alpha ^8 Q_{\alpha ,4}^5}+\frac{384 \cot ^3(2 \eta ) \sec ^2(\eta ) \xi _2^2
   \xi _4^2 \tan (\eta )}{\alpha ^8 Q_{\alpha ,4}^5}+\frac{1536 \cot (2 \eta ) \csc ^2(2
   \eta ) \sec ^2(\eta ) \xi _2^2 \xi _4^2 \tan (\eta )}{\alpha ^8 Q_{\alpha
   ,4}^5}-\frac{384 \cot (2 \eta ) \sec ^2(\eta ) \xi _2^2 \xi _4^2 \tan (\eta )}{\alpha
   ^8 Q_{\alpha ,4}^5}-\frac{8960 \cot (2 \eta ) \sec ^6(\eta ) \xi _2^2 \xi _4^4 \tan
   (\eta )}{\alpha ^{10} Q_{\alpha ,4}^6}-\frac{5120 \cot (2 \eta ) \sec ^4(\eta ) \xi
   _2^4 \xi _4^2 \tan (\eta )}{\alpha ^{10} Q_{\alpha ,4}^6}-\frac{1280 \cot ^3(2 \eta )
   \sec ^2(\eta ) \xi _2^4 \xi _4^2 \tan (\eta )}{\alpha ^{10} Q_{\alpha
   ,4}^6}-\frac{2560 \cot (2 \eta ) \csc ^2(2 \eta ) \sec ^2(\eta ) \xi _2^4 \xi _4^2
   \tan (\eta )}{\alpha ^{10} Q_{\alpha ,4}^6}+\frac{15360 \cot (2 \eta ) \sec ^6(\eta )
   \xi _2^4 \xi _4^4 \tan (\eta )}{\alpha ^{12} Q_{\alpha ,4}^7}+\frac{18 \sec ^2(\eta )
   \xi _4^2 \tan ^2(\eta ) a''(t)}{\alpha ^5 Q_{\alpha ,4}^4}-\frac{336 \sec ^4(\eta )
   \xi _4^4 \tan ^2(\eta ) a''(t)}{\alpha ^7 Q_{\alpha ,4}^5}+\frac{48 \sec ^2(\eta ) \xi
   _1^2 \xi _4^2 \tan ^2(\eta ) a''(t)}{\alpha ^5 Q_{\alpha ,4}^5}-\frac{528 \sec ^2(\eta
   ) \xi _2^2 \xi _4^2 \tan ^2(\eta ) a''(t)}{\alpha ^7 Q_{\alpha ,4}^5}+\frac{640 \sec
   ^6(\eta ) \xi _4^6 \tan ^2(\eta ) a''(t)}{\alpha ^9 Q_{\alpha ,4}^6}+\frac{1600 \sec
   ^4(\eta ) \xi _1^2 \xi _4^4 \tan ^2(\eta ) a''(t)}{\alpha ^7 Q_{\alpha
   ,4}^6}+\frac{4320 \sec ^4(\eta ) \xi _2^2 \xi _4^4 \tan ^2(\eta ) a''(t)}{\alpha ^9
   Q_{\alpha ,4}^6}+\frac{1280 \sec ^2(\eta ) \xi _2^4 \xi _4^2 \tan ^2(\eta )
   a''(t)}{\alpha ^9 Q_{\alpha ,4}^6}+\frac{2240 \sec ^2(\eta ) \xi _1^2 \xi _2^2 \xi
   _4^2 \tan ^2(\eta ) a''(t)}{\alpha ^7 Q_{\alpha ,4}^6}-\frac{5120 \sec ^6(\eta ) \xi
   _1^2 \xi _4^6 \tan ^2(\eta ) a''(t)}{\alpha ^9 Q_{\alpha ,4}^7}-\frac{5760 \sec
   ^6(\eta ) \xi _2^2 \xi _4^6 \tan ^2(\eta ) a''(t)}{\alpha ^{11} Q_{\alpha
   ,4}^7}-\frac{5760 \sec ^4(\eta ) \xi _2^4 \xi _4^4 \tan ^2(\eta ) a''(t)}{\alpha ^{11}
   Q_{\alpha ,4}^7}-\frac{28800 \sec ^4(\eta ) \xi _1^2 \xi _2^2 \xi _4^4 \tan ^2(\eta )
   a''(t)}{\alpha ^9 Q_{\alpha ,4}^7}-\frac{10240 \sec ^2(\eta ) \xi _1^2 \xi _2^4 \xi
   _4^2 \tan ^2(\eta ) a''(t)}{\alpha ^9 Q_{\alpha ,4}^7}+\frac{53760 \sec ^6(\eta ) \xi
   _1^2 \xi _2^2 \xi _4^6 \tan ^2(\eta ) a''(t)}{\alpha ^{11} Q_{\alpha
   ,4}^8}+\frac{53760 \sec ^4(\eta ) \xi _1^2 \xi _2^4 \xi _4^4 \tan ^2(\eta )
   a''(t)}{\alpha ^{11} Q_{\alpha ,4}^8}-\frac{9 a'(t)^2 a''(t)}{\alpha ^3 Q_{\alpha
   ,4}^3}+\frac{108 \xi _1^2 a'(t)^2 a''(t)}{\alpha ^3 Q_{\alpha ,4}^4}-\frac{8 \xi _2^2
   a'(t)^2 a''(t)}{\alpha ^5 Q_{\alpha ,4}^4}-\frac{8 \sec ^2(\eta ) \xi _4^2 a'(t)^2
   a''(t)}{\alpha ^5 Q_{\alpha ,4}^4}-\frac{144 \xi _1^4 a'(t)^2 a''(t)}{\alpha ^3
   Q_{\alpha ,4}^5}-\frac{688 \xi _2^4 a'(t)^2 a''(t)}{\alpha ^7 Q_{\alpha
   ,4}^5}-\frac{688 \sec ^4(\eta ) \xi _4^4 a'(t)^2 a''(t)}{\alpha ^7 Q_{\alpha
   ,4}^5}+\frac{672 \xi _1^2 \xi _2^2 a'(t)^2 a''(t)}{\alpha ^5 Q_{\alpha
   ,4}^5}+\frac{672 \sec ^2(\eta ) \xi _1^2 \xi _4^2 a'(t)^2 a''(t)}{\alpha ^5 Q_{\alpha
   ,4}^5}-\frac{1376 \sec ^2(\eta ) \xi _2^2 \xi _4^2 a'(t)^2 a''(t)}{\alpha ^7 Q_{\alpha
   ,4}^5}+\frac{1664 \xi _2^6 a'(t)^2 a''(t)}{\alpha ^9 Q_{\alpha ,4}^6}+\frac{1664 \sec
   ^6(\eta ) \xi _4^6 a'(t)^2 a''(t)}{\alpha ^9 Q_{\alpha ,4}^6}+\frac{8032 \xi _1^2 \xi
   _2^4 a'(t)^2 a''(t)}{\alpha ^7 Q_{\alpha ,4}^6}+\frac{8032 \sec ^4(\eta ) \xi _1^2 \xi
   _4^4 a'(t)^2 a''(t)}{\alpha ^7 Q_{\alpha ,4}^6}+\frac{4992 \sec ^4(\eta ) \xi _2^2 \xi
   _4^4 a'(t)^2 a''(t)}{\alpha ^9 Q_{\alpha ,4}^6}-\frac{1344 \xi _1^4 \xi _2^2 a'(t)^2
   a''(t)}{\alpha ^5 Q_{\alpha ,4}^6}-\frac{1344 \sec ^2(\eta ) \xi _1^4 \xi _4^2 a'(t)^2
   a''(t)}{\alpha ^5 Q_{\alpha ,4}^6}+\frac{4992 \sec ^2(\eta ) \xi _2^4 \xi _4^2 a'(t)^2
   a''(t)}{\alpha ^9 Q_{\alpha ,4}^6}+\frac{16064 \sec ^2(\eta ) \xi _1^2 \xi _2^2 \xi
   _4^2 a'(t)^2 a''(t)}{\alpha ^7 Q_{\alpha ,4}^6}-\frac{30080 \xi _1^2 \xi _2^6 a'(t)^2
   a''(t)}{\alpha ^9 Q_{\alpha ,4}^7}-\frac{30080 \sec ^6(\eta ) \xi _1^2 \xi _4^6
   a'(t)^2 a''(t)}{\alpha ^9 Q_{\alpha ,4}^7}-\frac{9600 \xi _1^4 \xi _2^4 a'(t)^2
   a''(t)}{\alpha ^7 Q_{\alpha ,4}^7}-\frac{9600 \sec ^4(\eta ) \xi _1^4 \xi _4^4 a'(t)^2
   a''(t)}{\alpha ^7 Q_{\alpha ,4}^7}-\frac{90240 \sec ^4(\eta ) \xi _1^2 \xi _2^2 \xi
   _4^4 a'(t)^2 a''(t)}{\alpha ^9 Q_{\alpha ,4}^7}-\frac{90240 \sec ^2(\eta ) \xi _1^2
   \xi _2^4 \xi _4^2 a'(t)^2 a''(t)}{\alpha ^9 Q_{\alpha ,4}^7}-\frac{19200 \sec ^2(\eta
   ) \xi _1^4 \xi _2^2 \xi _4^2 a'(t)^2 a''(t)}{\alpha ^7 Q_{\alpha ,4}^7}+\frac{53760
   \xi _1^4 \xi _2^6 a'(t)^2 a''(t)}{\alpha ^9 Q_{\alpha ,4}^8}+\frac{53760 \sec ^6(\eta
   ) \xi _1^4 \xi _4^6 a'(t)^2 a''(t)}{\alpha ^9 Q_{\alpha ,4}^8}+\frac{161280 \sec
   ^4(\eta ) \xi _1^4 \xi _2^2 \xi _4^4 a'(t)^2 a''(t)}{\alpha ^9 Q_{\alpha
   ,4}^8}+\frac{161280 \sec ^2(\eta ) \xi _1^4 \xi _2^4 \xi _4^2 a'(t)^2 a''(t)}{\alpha
   ^9 Q_{\alpha ,4}^8}+\frac{24 \cot (2 \eta ) \sec ^2(\eta ) \xi _4^2 \tan (\eta )
   a''(t)}{\alpha ^5 Q_{\alpha ,4}^4}-\frac{192 \cot (2 \eta ) \sec ^4(\eta ) \xi _4^4
   \tan (\eta ) a''(t)}{\alpha ^7 Q_{\alpha ,4}^5}+\frac{64 \cot (2 \eta ) \sec ^2(\eta )
   \xi _1^2 \xi _4^2 \tan (\eta ) a''(t)}{\alpha ^5 Q_{\alpha ,4}^5}-\frac{768 \cot (2
   \eta ) \sec ^2(\eta ) \xi _2^2 \xi _4^2 \tan (\eta ) a''(t)}{\alpha ^7 Q_{\alpha
   ,4}^5}+\frac{1280 \cot (2 \eta ) \sec ^4(\eta ) \xi _1^2 \xi _4^4 \tan (\eta )
   a''(t)}{\alpha ^7 Q_{\alpha ,4}^6}+\frac{1920 \cot (2 \eta ) \sec ^4(\eta ) \xi _2^2
   \xi _4^4 \tan (\eta ) a''(t)}{\alpha ^9 Q_{\alpha ,4}^6}+\frac{1920 \cot (2 \eta )
   \sec ^2(\eta ) \xi _2^4 \xi _4^2 \tan (\eta ) a''(t)}{\alpha ^9 Q_{\alpha
   ,4}^6}+\frac{3200 \cot (2 \eta ) \sec ^2(\eta ) \xi _1^2 \xi _2^2 \xi _4^2 \tan (\eta
   ) a''(t)}{\alpha ^7 Q_{\alpha ,4}^6}-\frac{15360 \cot (2 \eta ) \sec ^4(\eta ) \xi
   _1^2 \xi _2^2 \xi _4^4 \tan (\eta ) a''(t)}{\alpha ^9 Q_{\alpha ,4}^7}-\frac{15360
   \cot (2 \eta ) \sec ^2(\eta ) \xi _1^2 \xi _2^4 \xi _4^2 \tan (\eta ) a''(t)}{\alpha
   ^9 Q_{\alpha ,4}^7}-\frac{\sec ^2(\eta ) a''(t)}{\alpha ^3 Q_{\alpha ,4}^3}-\frac{4
   a''(t)}{\alpha ^3 Q_{\alpha ,4}^3}+\frac{10 \sec ^2(\eta ) \xi _1^2 a''(t)}{\alpha ^3
   Q_{\alpha ,4}^4}+\frac{40 \xi _1^2 a''(t)}{\alpha ^3 Q_{\alpha ,4}^4}+\frac{24 \cot
   ^2(2 \eta ) \xi _2^2 a''(t)}{\alpha ^5 Q_{\alpha ,4}^4}-\frac{48 \csc ^2(2 \eta ) \xi
   _2^2 a''(t)}{\alpha ^5 Q_{\alpha ,4}^4}-\frac{6 \sec ^2(\eta ) \xi _2^2 a''(t)}{\alpha
   ^5 Q_{\alpha ,4}^4}-\frac{24 \xi _2^2 a''(t)}{\alpha ^5 Q_{\alpha ,4}^4}+\frac{6 \sec
   ^4(\eta ) \xi _4^2 a''(t)}{\alpha ^5 Q_{\alpha ,4}^4}-\frac{24 \sec ^2(\eta ) \xi _4^2
   a''(t)}{\alpha ^5 Q_{\alpha ,4}^4}-\frac{192 \cot ^2(2 \eta ) \xi _2^4 a''(t)}{\alpha
   ^7 Q_{\alpha ,4}^5}+\frac{384 \csc ^2(2 \eta ) \xi _2^4 a''(t)}{\alpha ^7 Q_{\alpha
   ,4}^5}-\frac{96 \sec ^6(\eta ) \xi _4^4 a''(t)}{\alpha ^7 Q_{\alpha ,4}^5}+\frac{64
   \cot ^2(2 \eta ) \xi _1^2 \xi _2^2 a''(t)}{\alpha ^5 Q_{\alpha ,4}^5}-\frac{128 \csc
   ^2(2 \eta ) \xi _1^2 \xi _2^2 a''(t)}{\alpha ^5 Q_{\alpha ,4}^5}+\frac{32 \sec ^2(\eta
   ) \xi _1^2 \xi _2^2 a''(t)}{\alpha ^5 Q_{\alpha ,4}^5}+\frac{128 \xi _1^2 \xi _2^2
   a''(t)}{\alpha ^5 Q_{\alpha ,4}^5}+\frac{64 \sec ^4(\eta ) \xi _1^2 \xi _4^2
   a''(t)}{\alpha ^5 Q_{\alpha ,4}^5}+\frac{128 \sec ^2(\eta ) \xi _1^2 \xi _4^2
   a''(t)}{\alpha ^5 Q_{\alpha ,4}^5}-\frac{288 \sec ^4(\eta ) \xi _2^2 \xi _4^2
   a''(t)}{\alpha ^7 Q_{\alpha ,4}^5}-\frac{192 \cot ^2(2 \eta ) \sec ^2(\eta ) \xi _2^2
   \xi _4^2 a''(t)}{\alpha ^7 Q_{\alpha ,4}^5}+\frac{384 \csc ^2(2 \eta ) \sec ^2(\eta )
   \xi _2^2 \xi _4^2 a''(t)}{\alpha ^7 Q_{\alpha ,4}^5}-\frac{144 \sec ^4(\eta ) \xi _3^2
   \xi _4^2 a''(t)}{\alpha ^7 Q_{\alpha ,4}^5}+\frac{1280 \cot ^2(2 \eta ) \xi _1^2 \xi
   _2^4 a''(t)}{\alpha ^7 Q_{\alpha ,4}^6}-\frac{2560 \csc ^2(2 \eta ) \xi _1^2 \xi _2^4
   a''(t)}{\alpha ^7 Q_{\alpha ,4}^6}+\frac{640 \sec ^6(\eta ) \xi _1^2 \xi _4^4
   a''(t)}{\alpha ^7 Q_{\alpha ,4}^6}+\frac{640 \sec ^6(\eta ) \xi _2^2 \xi _4^4
   a''(t)}{\alpha ^9 Q_{\alpha ,4}^6}+\frac{640 \sec ^6(\eta ) \xi _3^2 \xi _4^4
   a''(t)}{\alpha ^9 Q_{\alpha ,4}^6}+\frac{640 \sec ^4(\eta ) \xi _2^4 \xi _4^2
   a''(t)}{\alpha ^9 Q_{\alpha ,4}^6}+\frac{1280 \sec ^4(\eta ) \xi _1^2 \xi _2^2 \xi
   _4^2 a''(t)}{\alpha ^7 Q_{\alpha ,4}^6}+\frac{1280 \cot ^2(2 \eta ) \sec ^2(\eta ) \xi
   _1^2 \xi _2^2 \xi _4^2 a''(t)}{\alpha ^7 Q_{\alpha ,4}^6}-\frac{2560 \csc ^2(2 \eta )
   \sec ^2(\eta ) \xi _1^2 \xi _2^2 \xi _4^2 a''(t)}{\alpha ^7 Q_{\alpha ,4}^6}+\frac{960
   \sec ^4(\eta ) \xi _1^2 \xi _3^2 \xi _4^2 a''(t)}{\alpha ^7 Q_{\alpha
   ,4}^6}+\frac{1280 \sec ^4(\eta ) \xi _2^2 \xi _3^2 \xi _4^2 a''(t)}{\alpha ^9
   Q_{\alpha ,4}^6}-\frac{5120 \sec ^6(\eta ) \xi _1^2 \xi _2^2 \xi _4^4 a''(t)}{\alpha
   ^9 Q_{\alpha ,4}^7}-\frac{5120 \sec ^6(\eta ) \xi _1^2 \xi _3^2 \xi _4^4
   a''(t)}{\alpha ^9 Q_{\alpha ,4}^7}-\frac{5760 \sec ^6(\eta ) \xi _2^2 \xi _3^2 \xi
   _4^4 a''(t)}{\alpha ^{11} Q_{\alpha ,4}^7}-\frac{5120 \sec ^4(\eta ) \xi _1^2 \xi _2^4
   \xi _4^2 a''(t)}{\alpha ^9 Q_{\alpha ,4}^7}-\frac{10240 \sec ^4(\eta ) \xi _1^2 \xi
   _2^2 \xi _3^2 \xi _4^2 a''(t)}{\alpha ^9 Q_{\alpha ,4}^7}+\frac{53760 \sec ^6(\eta )
   \xi _1^2 \xi _2^2 \xi _3^2 \xi _4^4 a''(t)}{\alpha ^{11} Q_{\alpha ,4}^8}+\frac{12
   a'(t) a^{(3)}(t)}{\alpha ^2 Q_{\alpha ,4}^3}-\frac{144 \xi _1^2 a'(t)
   a^{(3)}(t)}{\alpha ^2 Q_{\alpha ,4}^4}+\frac{36 \xi _2^2 a'(t) a^{(3)}(t)}{\alpha ^4
   Q_{\alpha ,4}^4}+\frac{36 \sec ^2(\eta ) \xi _4^2 a'(t) a^{(3)}(t)}{\alpha ^4
   Q_{\alpha ,4}^4}+\frac{192 \xi _1^4 a'(t) a^{(3)}(t)}{\alpha ^2 Q_{\alpha
   ,4}^5}+\frac{160 \xi _2^4 a'(t) a^{(3)}(t)}{\alpha ^6 Q_{\alpha ,4}^5}+\frac{160 \sec
   ^4(\eta ) \xi _4^4 a'(t) a^{(3)}(t)}{\alpha ^6 Q_{\alpha ,4}^5}-\frac{736 \xi _1^2 \xi
   _2^2 a'(t) a^{(3)}(t)}{\alpha ^4 Q_{\alpha ,4}^5}-\frac{736 \sec ^2(\eta ) \xi _1^2
   \xi _4^2 a'(t) a^{(3)}(t)}{\alpha ^4 Q_{\alpha ,4}^5}+\frac{320 \sec ^2(\eta ) \xi
   _2^2 \xi _4^2 a'(t) a^{(3)}(t)}{\alpha ^6 Q_{\alpha ,4}^5}-\frac{2432 \xi _1^2 \xi
   _2^4 a'(t) a^{(3)}(t)}{\alpha ^6 Q_{\alpha ,4}^6}-\frac{2432 \sec ^4(\eta ) \xi _1^2
   \xi _4^4 a'(t) a^{(3)}(t)}{\alpha ^6 Q_{\alpha ,4}^6}+\frac{1344 \xi _1^4 \xi _2^2
   a'(t) a^{(3)}(t)}{\alpha ^4 Q_{\alpha ,4}^6}+\frac{1344 \sec ^2(\eta ) \xi _1^4 \xi
   _4^2 a'(t) a^{(3)}(t)}{\alpha ^4 Q_{\alpha ,4}^6}-\frac{4864 \sec ^2(\eta ) \xi _1^2
   \xi _2^2 \xi _4^2 a'(t) a^{(3)}(t)}{\alpha ^6 Q_{\alpha ,4}^6}+\frac{3840 \xi _1^4 \xi
   _2^4 a'(t) a^{(3)}(t)}{\alpha ^6 Q_{\alpha ,4}^7}+\frac{3840 \sec ^4(\eta ) \xi _1^4
   \xi _4^4 a'(t) a^{(3)}(t)}{\alpha ^6 Q_{\alpha ,4}^7}+\frac{7680 \sec ^2(\eta ) \xi
   _1^4 \xi _2^2 \xi _4^2 a'(t) a^{(3)}(t)}{\alpha ^6 Q_{\alpha ,4}^7}+\frac{6
   a^{(4)}(t)}{\alpha  Q_{\alpha ,4}^3}-\frac{72 \xi _1^2 a^{(4)}(t)}{\alpha  Q_{\alpha
   ,4}^4}+\frac{8 \xi _2^2 a^{(4)}(t)}{\alpha ^3 Q_{\alpha ,4}^4}+\frac{8 \sec ^2(\eta )
   \xi _4^2 a^{(4)}(t)}{\alpha ^3 Q_{\alpha ,4}^4}+\frac{96 \xi _1^4 a^{(4)}(t)}{\alpha 
   Q_{\alpha ,4}^5}-\frac{96 \xi _1^2 \xi _2^2 a^{(4)}(t)}{\alpha ^3 Q_{\alpha
   ,4}^5}-\frac{96 \sec ^2(\eta ) \xi _1^2 \xi _4^2 a^{(4)}(t)}{\alpha ^3 Q_{\alpha
   ,4}^5}+\frac{128 \xi _1^4 \xi _2^2 a^{(4)}(t)}{\alpha ^3 Q_{\alpha ,4}^6}+\frac{128
   \sec ^2(\eta ) \xi _1^4 \xi _4^2 a^{(4)}(t)}{\alpha ^3 Q_{\alpha ,4}^6}-\frac{7 \sec
   ^4(\eta )}{\alpha ^4 Q_{\alpha ,4}^3}+\frac{2 \sec ^2(\eta )}{\alpha ^4 Q_{\alpha
   ,4}^3}+\frac{4}{\alpha ^4 Q_{\alpha ,4}^3}+\frac{256 \csc ^4(2 \eta ) \xi _2^2}{\alpha
   ^6 Q_{\alpha ,4}^4}+\frac{8 \sec ^4(\eta ) \xi _2^2}{\alpha ^6 Q_{\alpha
   ,4}^4}+\frac{48 \cot ^2(2 \eta ) \xi _2^2}{\alpha ^6 Q_{\alpha ,4}^4}+\frac{320 \cot
   ^2(2 \eta ) \csc ^2(2 \eta ) \xi _2^2}{\alpha ^6 Q_{\alpha ,4}^4}-\frac{96 \csc ^2(2
   \eta ) \xi _2^2}{\alpha ^6 Q_{\alpha ,4}^4}+\frac{12 \cot ^2(2 \eta ) \sec ^2(\eta )
   \xi _2^2}{\alpha ^6 Q_{\alpha ,4}^4}-\frac{24 \csc ^2(2 \eta ) \sec ^2(\eta ) \xi
   _2^2}{\alpha ^6 Q_{\alpha ,4}^4}-\frac{50 \sec ^6(\eta ) \xi _4^2}{\alpha ^6 Q_{\alpha
   ,4}^4}-\frac{32 \cot ^2(2 \eta ) \sec ^4(\eta ) \xi _4^2}{\alpha ^6 Q_{\alpha
   ,4}^4}+\frac{64 \csc ^2(2 \eta ) \sec ^4(\eta ) \xi _4^2}{\alpha ^6 Q_{\alpha
   ,4}^4}+\frac{33 \sec ^4(\eta ) \xi _4^2}{\alpha ^6 Q_{\alpha ,4}^4}+\frac{64 \cot ^4(2
   \eta ) \xi _2^4}{\alpha ^8 Q_{\alpha ,4}^5}-\frac{256 \csc ^4(2 \eta ) \xi
   _2^4}{\alpha ^8 Q_{\alpha ,4}^5}-\frac{768 \cot ^2(2 \eta ) \csc ^2(2 \eta ) \xi
   _2^4}{\alpha ^8 Q_{\alpha ,4}^5}+\frac{112 \sec ^8(\eta ) \xi _4^4}{\alpha ^8
   Q_{\alpha ,4}^5}-\frac{32 \sec ^6(\eta ) \xi _4^4}{\alpha ^8 Q_{\alpha
   ,4}^5}+\frac{688 \sec ^6(\eta ) \xi _2^2 \xi _4^2}{\alpha ^8 Q_{\alpha
   ,4}^5}+\frac{704 \cot ^2(2 \eta ) \sec ^4(\eta ) \xi _2^2 \xi _4^2}{\alpha ^8
   Q_{\alpha ,4}^5}-\frac{1408 \csc ^2(2 \eta ) \sec ^4(\eta ) \xi _2^2 \xi _4^2}{\alpha
   ^8 Q_{\alpha ,4}^5}-\frac{192 \sec ^4(\eta ) \xi _2^2 \xi _4^2}{\alpha ^8 Q_{\alpha
   ,4}^5}-\frac{512 \sec ^4(\eta ) \xi _3^2 \xi _4^2}{\alpha ^8 Q_{\alpha
   ,4}^5}-\frac{1664 \sec ^8(\eta ) \xi _2^2 \xi _4^4}{\alpha ^{10} Q_{\alpha
   ,4}^6}+\frac{240 \sec ^6(\eta ) \xi _2^2 \xi _4^4}{\alpha ^{10} Q_{\alpha
   ,4}^6}+\frac{6016 \sec ^6(\eta ) \xi _3^2 \xi _4^4}{\alpha ^{10} Q_{\alpha
   ,4}^6}-\frac{1024 \sec ^6(\eta ) \xi _2^4 \xi _4^2}{\alpha ^{10} Q_{\alpha
   ,4}^6}-\frac{1280 \cot ^2(2 \eta ) \sec ^4(\eta ) \xi _2^4 \xi _4^2}{\alpha ^{10}
   Q_{\alpha ,4}^6}+\frac{2560 \csc ^2(2 \eta ) \sec ^4(\eta ) \xi _2^4 \xi _4^2}{\alpha
   ^{10} Q_{\alpha ,4}^6}+\frac{8768 \sec ^4(\eta ) \xi _2^2 \xi _3^2 \xi _4^2}{\alpha
   ^{10} Q_{\alpha ,4}^6}-\frac{10240 \sec ^8(\eta ) \xi _3^2 \xi _4^6}{\alpha ^{12}
   Q_{\alpha ,4}^7}+\frac{2560 \sec ^8(\eta ) \xi _2^4 \xi _4^4}{\alpha ^{12} Q_{\alpha
   ,4}^7}-\frac{114560 \sec ^6(\eta ) \xi _2^2 \xi _3^2 \xi _4^4}{\alpha ^{12} Q_{\alpha
   ,4}^7}-\frac{15360 \sec ^4(\eta ) \xi _2^4 \xi _3^2 \xi _4^2}{\alpha ^{12} Q_{\alpha
   ,4}^7}+\frac{215040 \sec ^8(\eta ) \xi _2^2 \xi _3^2 \xi _4^6}{\alpha ^{14} Q_{\alpha
   ,4}^8}+\frac{215040 \sec ^6(\eta ) \xi _2^4 \xi _3^2 \xi _4^4}{\alpha ^{14} Q_{\alpha
   ,4}^8}-\frac{430080 \sec ^8(\eta ) \xi _2^4 \xi _3^2 \xi _4^6}{\alpha ^{16} Q_{\alpha
   ,4}^9}. 
   \end{math}
\end{center}
}

\subsection*{Acknowledgement} The second author acknowledges support from NSF grants 
DMS-1201512 and PHY-1205440. Part of this work was done at the Perimeter Institute for Theoretical Physics, 
supported by the Government of Canada through Industry Canada and by the
Province of Ontario through the Ministry of Economic Development and Innovation.


\begin{thebibliography}{99}

\bibitem{Ampl} N.~Arkani-Hamed, J.~Bourjaily, F.~Cachazo, A.~Goncharov, A.~Postnikov, J.~Trnka,
{\em Grassmannian Geometry of Scattering Amplitudes}, Cambridge University Press, 2016.

\bibitem{BEK} S.~Bloch, H.~Esnault, D.~Kreimer, {\em On motives associated to graph polynomials}, 
Comm. Math. Phys. 267 (2006), no. 1, 181--225.

\bibitem{BrSch}  F.~Brown, O.~Schnetz, {\em A K3 in $\phi^4$}, Duke Math. J. Vol.161 (2012) no. 10, 1817--1862. 

\bibitem{CC}  A.H.~Chamseddine, A.~Connes, {\em Spectral action for Robertson--Walker metrics}, 
J. High Energy Phys. (2012) no. 10, 101, 29 pages

\bibitem{CCact} A.H.~Chamseddine, A.~Connes, {\em The spectral action principle}, 
Comm. Math. Phys. 186 (1997), no. 3, 731--750. 

\bibitem{CoS3} A.~Connes, {\em Geometry from the spectral point of view}, Lett. Math. Phys. 34 (1995), no. 3, 203--238.

\bibitem{CoMa} A.~Connes, M.~Marcolli, {\em Renormalization and motivic Galois theory}, 
Int. Math. Res. Not. 2004, no. 76, 4073--4091. 

\bibitem{FFMRationality}
W. Fan, F. Fathizadeh, M. Marcolli,
{\it Spectral Action for Bianchi Type-IX Cosmological Models,}
J. High Energy Phys. 10 (2015) 085. 

\bibitem{FFM2} W. Fan, F. Fathizadeh, M. Marcolli, {\em Modular forms in the spectral action of Bianchi IX gravitational instantons}, arXiv:1511.05321

\bibitem{FGK}
 F.~Fathizadeh, A.~Ghorbanpour, M.~Khalkhali, {\em Rationality of spectral action for 
 Robertson--Walker metrics}, J. High Energy Phys. (2014) no. 12, 064, f21 pages
  
   \bibitem{Gon1} J.~Golden, A.B.~Goncharov, M.~Spradlin, C.~Vergu, A.~Volovich,
 {\em Motivic Amplitudes and Cluster Coordinates},  arXiv:1305.1617
 
 \bibitem{Gon2} A.B.~Goncharov, M.~Spradlin, C.~Vergu, A.~Volovich, {\em 
 Classical Polylogarithms for Amplitudes and Wilson Loops},  Phys. Rev. Lett. 105 (2010) no. 15, 151605, 4 pp. 
  
 \bibitem{GVFbook}
J.M. ~Gracia-Bondia, J.C. ~Varilly, H. Figueroa, 
{\em Elements of noncommutative geometry,} 
Birkh\"auser, 2001. 
 
\bibitem{Mar} M.~Marcolli, {\em Feynman motives}, World Scientific, 2010.  

\bibitem{Mar2} M.~Marcolli, {\em Noncommutative Cosmology}, World Scientific, to appear.

\bibitem{Rost} M.~Rost, {\em 
The motive of a Pfister form}, Preprint (1998)  \newline
{\tt www.physik.uni-regensburg.de/$\sim$rom03516/motive.html}

\bibitem{WvS} W.~van Suijlekom, {\em Noncommutative Geometry and Particle Physics},
Springer, 2014.

\bibitem{Vishik1} A.~Vishik, {\em Integral motives of quadrics}, Max-Planck-Institut f\"ur Mathematik Bonn, Preprint MPI-1998-13, 1--82 (1998).

\bibitem{Vishik2} A.~Vishik, {\em Motives of quadrics with applications to the theory of quadratic forms}, in ``Geometric methods in the algebraic theory of quadratic forms", pp.~25--101, Lecture Notes in Math., Vol.1835, Springer, 2004. 

\bibitem{Voe} V.~Voevodsly, A.~Suslin, E.M.~Friedlander, {\em Cyles, transfers, and motivic
homology theories}, Princeton University Press, 2000.

\end{thebibliography}
\end{document}